\documentclass[a4paper, 12pt, titlepage, draft=false]{scrartcl} 
\makeatletter
\DeclareOldFontCommand{\rm}{\normalfont\rmfamily}{\mathrm}
\DeclareOldFontCommand{\sf}{\normalfont\sffamily}{\mathsf}
\DeclareOldFontCommand{\tt}{\normalfont\ttfamily}{\mathtt}
\DeclareOldFontCommand{\bf}{\normalfont\bfseries}{\mathbf}
\DeclareOldFontCommand{\it}{\normalfont\itshape}{\mathit}
\DeclareOldFontCommand{\sl}{\normalfont\slshape}{\@nomath\sl}
\DeclareOldFontCommand{\sc}{\normalfont\scshape}{\@nomath\sc}
\makeatother

\usepackage{ifdraft} 

\ifdraft{
\usepackage{comment}
}

\usepackage{color}
\usepackage[usenames,dvipsnames]{xcolor}
\usepackage{verbatim}
\usepackage{comment}

\makeatletter
\def\@xobeysp{\mbox{}\space}
\def\verbatim@font{\normalfont\ttfamily\raggedright\leftskip\@totalleftmargin}
\makeatother

\usepackage{csvsimple}
\usepackage{amsmath}
\usepackage{amssymb}
\usepackage{amsfonts}
\usepackage{amsthm}
\usepackage{todonotes}
\usepackage{import}
\usepackage{booktabs}
\usepackage{tabularx}
\usepackage{longtable}
\usepackage{hhline}
\usepackage{float}
\usepackage{lscape}
\usepackage[export]{adjustbox}
\usepackage{changepage}
\newcommand{\ra}[1]{\renewcommand{\arraystretch}{#1}}

\usepackage[utf8]{inputenc}
\usepackage[T1]{fontenc}
\usepackage[ngerman,english]{babel}

\usepackage{geometry}

\usepackage{listings}

\usepackage{inconsolata}

\usepackage{color}
\definecolor{pblue}{rgb}{0.13,0.13,1}
\definecolor{pgreen}{rgb}{0,0.5,0}
\definecolor{pred}{rgb}{0.9,0,0}
\definecolor{pgrey}{rgb}{0.46,0.45,0.48}
\usepackage{caption}

\usepackage{algorithm}
\usepackage{algorithmicx}
\usepackage[noend]{algpseudocode}

\usepackage{graphicx}
\graphicspath{{}{./figures/}}

\usepackage{caption}
\usepackage{subcaption}
\usepackage[hyphenbreaks]{breakurl}
\usepackage[hyphens]{url}
\usepackage{hyperref}
\usepackage{etoolbox}
\hypersetup{colorlinks=true}
\hypersetup{
	urlcolor=black,
	citecolor=black,
	linkcolor=black,
}
\makeatletter
\def\@footnotecolor{black}
\define@key{Hyp}{footnotecolor}{%
	\HyColor@HyperrefColor{#1}\@footnotecolor%
}
\patchcmd{\@footnotemark}{\hyper@linkstart{link}}{\hyper@linkstart{footnote}}{}{}
\makeatother

\usepackage[multiple]{footmisc}

\usepackage{indentfirst}
\usepackage{soul}

\usepackage{enumitem}
\usepackage[sort&compress,nameinlink,noabbrev]{cleveref} %
\usepackage{nameref} 
\usepackage[backend=bibtex,style=numeric,
hyperref=true,natbib=true,
backref=true,backrefstyle=three,
sortcites=false,
maxbibnames=50,
maxcitenames=3,
firstinits=true, 
isbn=true,url=true,doi=true]
{biblatex} 
\patchcmd{\thebibliography}{\clubpenalty4000}{\clubpenalty10000}{}{}
\patchcmd{\thebibliography}{\widowpenalty4000}{\widowpenalty10000}{}{}
\patchcmd{\bibsetup}{\interlinepenalty=5000}{\interlinepenalty=10000}{}{}
\bibliography{bibliography}

\patchcmd{\url}{\clubpenalty4000}{\clubpenalty10000}{}{}
\patchcmd{\url}{\widowpenalty4000}{\widowpenalty10000}{}{}
\patchcmd{\verb}{\clubpenalty4000}{\clubpenalty10000}{}{}
\patchcmd{\verb}{\widowpenalty4000}{\widowpenalty10000}{}{}

\usepackage{array}
\newcolumntype{x}[1]{>{\centering\arraybackslash\hspace{0pt}}m{#1}}

\theoremstyle{plain}

\usepackage{tikz}
\usepackage{pgfplots}
\usepackage{pgfplotstable}
\usepgfplotslibrary{groupplots}
\pgfplotsset{compat=1.14}
\usepackage{etoolbox}

\def\figurepath#1{../figures/#1}

\newenvironment{smallitemize}
{ \begin{itemize}
		\setlength{\itemsep}{0pt}
		\setlength{\parskip}{0pt}
		\setlength{\parsep}{0pt}     
}
{ \end{itemize}                  }

\usepackage{fancyhdr}
\makeatletter
\newcommand{\rightorleftmark}{%
	\begingroup\protected@edef\x{\rightmark}%
	\ifx\x\@empty
	\endgroup\leftmark
	\else
	\endgroup\rightmark
	\fi}
\makeatother
\usepackage{setspace}
\usepackage{units}
\usepackage[a4paper]{anysize}
\marginsize{3.5cm}{2.75cm}{2cm}{3cm} %
\usepackage[many]{tcolorbox}
\newtcolorbox{myboxi}[1][]{
	breakable,
	title=#1,
	colback=white,
	colbacktitle=white,
	coltitle=black,
	fonttitle=\bfseries,
	bottomrule=1pt,
	toprule=1pt,
	leftrule=1pt,
	rightrule=1pt,
	titlerule=0pt,
	arc=2pt,
	outer arc=2pt,
	colframe=black,
	enlarge top by=12pt,
}

\usepackage[section]{placeins}
\makeatletter
\AtBeginDocument{%
	\expandafter\renewcommand\expandafter\subsection\expandafter{%
		\expandafter\@fb@secFB\subsection
	}%
}
\makeatother

\begin{document}
\pagenumbering{Roman}
\newgeometry{left=3cm,bottom=2cm,right=3cm,top=3cm}

\begin{titlepage}
\begin{center}

\includegraphics[height=0.13\textheight]{\figurepath{logo_dima.pdf}}
\hfill
\includegraphics[height=0.13\textheight]{\figurepath{logo_tub.png}}

\vspace*{1.0cm}

\LARGE
\textsc{Implementation and Evaluation of a Framework to calculate Impact Measures for Wikipedia Authors}

\vspace{0.5cm}

\Large \textsc{Bachelor Thesis}

\vspace{0.1cm}

by

\vspace{0.2cm}

\textbf{Sebastian \textsc{Neef}}\\
s.neef@campus.tu-berlin.de

\vspace{0.5cm}

\vfill

\large
Submitted to the Faculty IV, Electrical Engineering and Computer Science
Database Systems and Information Management Group
in partial fulfillment of the requirements for the degree of

\textbf{Bachelor of Science in Computer Science}

at the

\textsc{Technische Universit\"{a}t Berlin} \\

June 29, 2017 

\vfill

\begin{flushright}
\normalsize
\emph{Thesis Advisors:}\\
Moritz \textsc{Schubotz} \\

\ \\

\emph{Thesis Supervisor:}\\
Prof. Dr. Volker \textsc{Markl}\\
Prof. Dr. Odej \textsc{Kao}\\
\end{flushright}

\end{center}
\end{titlepage}

\restoregeometry

\clearpage
\thispagestyle{empty}
\mbox{}
\newpage

\section*{Eidesstattliche Erkl\"arung}

Hiermit erkläre ich, dass ich die vorliegende Arbeit selbstständig und eigenhändig sowie ohne unerlaubte fremde Hilfe und ausschließlich unter Verwendung der aufgeführten Quellen und Hilfsmittel angefertigt habe.
\ \\
\ \\
Die selbständige und eigenständige Anfertigung versichert an Eides statt:
\begin{flushright}
Berlin, den 29. Juni 2017

\ \\
\ \\
Sebastian \textsc{Neef}

\end{flushright}

\newpage


\section*{Zusammenfassung} \label{sec:zusammenfassung}
\selectlanguage{ngerman}
Die Wikipedia ist eine offene, kollaborative Webseite, welche von jedermann, auch anonym, bearbeitet und deswegen Opfer von unerwünschten Änderungen werden kann. Anhand der kontinuierlichen Versionierung aller Änderungen und mit Ranglisten, basierend auf der Berechnung von \textit{impact measures}, können unerwünschte Änderungen erkannt sowie angesehene Nutzer identifiziert werden \cite{Adler:2008:MAC:1822258.1822279}. Allerdings benötigt das Verarbeiten vieler Millionen Revisionen auf einem einzelnen System viel Zeit. Der Autor implementiert ein quelloffenes Framework, um solche Ranglisten auf verteilten Systemen im MapReduce-Stil zu berechnen, und evaluiert dessen Performance auf verschieden großen Datensätzen. Die Nachimplementierung der \textit{contribution measures} von \citeauthor{Adler:2008:MAC:1822258.1822279} sollen die Erweiterbarkeit und Nutzbarkeit, als auch die Probleme beim Handhaben von riesigen Datensätzen und deren mögliche Lösungsansätze demonstrieren. In den Ergebnissen werden die verschiedenen Optimierungen diskutiert und gezeigt, dass horizontale Skalierung die gesamte Verarbeitungsdauer reduzieren kann. 

\section*{Abstract} \label{sec:abstract}
\selectlanguage{english}
Wikipedia, an open collaborative website, can be edited by anyone, even anonymously, thus becoming victim to ill-intentioned changes. 
Therefore, ranking Wikipedia authors by calculating impact measures based on the edit history can help to identify reputational users or harmful activity such as vandalism \cite{Adler:2008:MAC:1822258.1822279}. However, processing millions of edits on one system can take a long time. The author implements an open source framework to calculate such rankings in a distributed way (MapReduce) and evaluates its performance on various sized datasets. A reimplementation of the contribution measures by \citeauthor{Adler:2008:MAC:1822258.1822279} demonstrates its extensibility and usability, as well as problems of handling huge datasets and their possible resolutions. The results put different performance optimizations into perspective and show that horizontal scaling can decrease the total processing time.

\newpage
\setcounter{tocdepth}{2}
\tableofcontents
\newpage
\pagestyle{fancy}
\pagenumbering{arabic}
\section{Introduction} \label{sec:introduction}
In this first section the thesis' topic is introduced. Starting with the background and related work then continuing with the problem statement and goal. Important keywords, formulas and abbreviations are defined, followed by general information about the used software and hardware setup.

\subsection{Background} \label{sec:introduction-background}
People publish articles or contribute to open source projects without an immediate reward, but hope for indirect rewards like extending their skill set, own marketability or peer recognition \citep[p. 253]{Rafaeli08onlinemotivational}. 
This may be one of the reasons for using platforms like ResearchGate\footnote{\url{https://www.researchgate.net/}}. 
It uses an unknown algorithm \citep{kraker2015critical} to calculate an individual's \verb|RG Score|\footnote{\url{https://www.researchgate.net/publicprofile.RGScoreFAQ.html}} and position her in the community, leading to said indirect rewards. 
Such impact measures have been used to accept or reject applicants \cite[p. 391]{lbormann}.

Another well known measure of the quality of a scientist's work is the \verb|h-index| (see \ref{sec:introduction-definitions-hindex}) \cite[p. 392]{lbormann}.

Unlike ResearchGate, the free encyclopedia Wikipedia\footnote{\url{https://wikipedia.org}} does not display author performance measures besides edit counts. 
Thus making it harder for them to turn their contributions into indirect rewards. 
In fact, Wikimedia relied on the edit count and votes to nominate new moderators \cite{wmelecgl}. 
Another point which makes researching this topic interesting are the high user and edit counts. 
In 2008, Wikipedia had more than 300.000 authors with at least ten edits and the numbers have been growing by 5 - 10 percent per month since then \cite[p.  243 - 244]{Rafaeli08onlinemotivational}. Even at the time of writing in March 2017, Wikipedia has more than 10.000 active authors with more than 100 edits in that month \cite{wmusertable}.

\subsection{Related work} \label{sec:introduction-relatedwork}
Previous research focused only on the edit count \cite{Wilkinson:2007:CQW:1296951.1296968, kittur2007power, Suh:2008:LVI:1357054.1357214} or length of changed text \cite{asw,Adler:2007:CRS:1242572.1242608}. 
Schwartz \citeyearpar{asw} discovered high discrepancies between some users' edit  and text count. He noticed that top contributors by edit count are not necessarily on the top by text count \cite{asw}.

\citeauthor{Adler:2008:MAC:1822258.1822279} conclude that the quantitative measures edit or text count can be manipulated easily.
Therefor, more weight should be put on the content's quality with qualitative measures. 
One measure they introduced was the longevity of a change \cite[p. 1f]{Adler:2008:MAC:1822258.1822279}.
Their concepts have been used in more recent publications like the WikiTrust program \cite[p. 38]{DeAlfaro:2011:RSO:1978542.1978560} or as a consideration for another rating system \cite[p. 75]{Pantola:2010:RRR:1854099.1854116}. 

A service that makes use of Wikipedia's history trying to detect vandalism by classifying an edit's quality, is the ``Objective Revision Evaluation Service (ORES)`` by MediaWiki \citep{mwores}. Its approach focuses on machine learning, but manual classification work by the community is needed to get accurate results. Furthermore, it appears that only a limited amount of Wikipedia sites have this service enabled \citep{mwores}.

\subsection{Problem statement} \label{sec:introduction-problemstatement}
The work by \citeauthor{Adler:2008:MAC:1822258.1822279} is a promising starting point on the impact measure topic, due to their proposal of several formulas that calculate an author ranking based on the impact of an author's edit. \citeauthor{Adler:2008:MAC:1822258.1822279} call those formulas ``contribution measures`` \cite{Adler:2008:MAC:1822258.1822279}. The introduced terminology will be adapted to reduce confusion and credit their work. Unfortunately the authors don't discuss how to efficiently analyze the around 2.6 billion\footnote{\url{https://tools.wmflabs.org/wmcounter}} edits.

This is the bachelor thesis' starting point. 
It will introduce and discuss an open source framework, which prepares Wikipedia's edit history and page views and facilitates the development of distributed impact measures and rankings for Wikipedia authors. Such a framework could lead to better reproducibility of author rankings. The thesis will try to provide reasonable information to keep all tests and results reproducible.

Therefore the following research objective was defined: 

\begin{addmargin}[4em]{4em}
\textbf{Implement and evaluate a framework for distributed calculation of impact measures for Wikipedia authors.}
\end{addmargin}
The following tasks were set to achieve this objective:
\begin{itemize}
	\item \textbf{Task 1}: Analyze the input datasets in regard to their layout and format.
	\item \textbf{Task 2}: Design and implement the distributed framework.
	\item \textbf{Task 3}: Implement \citeauthor{Adler:2008:MAC:1822258.1822279} contribution measures as described in \cite{Adler:2008:MAC:1822258.1822279}.
	\item \textbf{Task 4}: Evaluate the framework's processing speed by comparing it to \citeauthor{Adler:2008:MAC:1822258.1822279} WikiTrust program \cite{wikitrgh}.
\end{itemize}
The first problem to address is the parsing of Wikipedia XML dumps (see \ref{sec:introduction-definitions-wikidatasets}; \cite{wikipediadumps}). 
For example, the English Wikipedia is more than 0.5 TB compressed and extracts to multiple (>= 10) TB of XML data. The student cluster (see \ref{sec:introduction-setup-studentcluster}) does not have enough disk space to store such amounts of data, thus processing the data in its compressed format will be explored in section ``\nameref{sec:implementation-compresseddatasets}``.
This amount of data has to be transformed from its XML state into an usable representation for further processing, for example objects with the attributes as member variables and references between correlated objects, so that distributed algorithms can operate on them.
That also covers the calculation of differences between two revisions of a page.
Different types of processing techniques will be tested on multiple datasets and evaluated based on the execution time.

Wikipedia's page views (see \ref{sec:introduction-definitions-pageviewdatasets}; \cite{mwpvleg, mwpvnew}) is an additional source of information which was not used by to \citeauthor{Adler:2008:MAC:1822258.1822279}, but might be an important factor to rate authors. 
The data needs to be parsed as well and incorporated into the framework's representation of a Wikipedia page. 

The key functionality of the framework is to hide the initial Wikipedia XML or page views data processing from the user, who uses the framework to calculate author rankings for Wikipedia sites.
It should also try to hide the parallel data processing mechanisms as far as it is possible.
Another important point is the extensibility which will make the framework customizable in mostly every aspect.

\subsection{Keywords and abbreviations} \label{sec:introduction-definitions}
This section explains and defines important abbreviations and keywords which will be used throughout the thesis.

\subsubsection{H-index} \label{sec:introduction-definitions-hindex}
The \verb|h-index|:
An author $A$ has a non-empty set $P_A = \{p_1,\dots,p_n\}$ of $n \in \mathbb{N}$ publications. 
Let $c_A : P_A \rightarrow \mathbb{N}$ be a function which returns a publication's citation count. 
Let $P_{A_i}$ be the tuple of publications, sorted by decreasing citation count.
Formally it is $P_{A_i} = (s_1,...,s_n)$ where $\{s_1,...,s_n\} = P_A$ and $c_A(s_i) \geq c_A(s_{i+1})$ for all $i<n$.
The h-index is the biggest $i$ for which $c_A(s_i) \geq i$ holds. A higher value depicts a better performance \cite[p. 16569]{hirsch2005index}.

{\bfseries{Example:}} An author with the publications c(A)=8, c(B)=10, c(D)=5, c(C)=3, c(E)=4 has $P_A = \{A, B, C, D, E\}$ and $P_{A_s} = (B, A, D, E, C)$. 
The resulting h-index is 4. 

\subsubsection{Wikipedia datasets}\label{sec:introduction-definitions-wikidatasets}
Wikipedia\footnote{\url{https://www.wikipedia.org}} is a free encyclopedia. Every language, specific topic or project has its own Wikipedia site.
A \verb|wiki|: Wikipedia sites will be abbreviated to \verb|wiki| with their domain prefix optionally, e.g. \verb|en-wiki| for the English\footnote{\url{https://en.wikipedia.org}} or \verb|de-wiki| for the German\footnote{\url{https://de.wikipedia.org}} version of the Wikipedia. 

Wikipedia creates backups of all their pages, the so called data dumps\footnote{\url{https://meta.wikimedia.org/wiki/Data_dumps}}. Depending on the size of a Wikipedia site, backups are created up to three or more times a month \citep{wikipediadumps}. Such backups contain the full edit history, consisting of revisions which are created when an edit to a page is made. The latest dumps can be downloaded from \cite{mwdumpdl}.

The \verb|dumps|: Throughout this thesis those edit history backups will be called \verb|dumps|.    
A \verb|dump|'s filename, e.g. 'aawiki-20160111-pages-meta-history.xml', is holds multiple pieces of information. It always follows the structure of \verb|<WIKITAG>wiki-<DATE>-pages-meta-history.xml<.FORMAT>|, where \verb|<WIKITAG>| is the wiki's short tag, e.g. \verb|aa|, \verb|<DATE>| is the backup's creation date, e.g. \verb|20160111| in the format of YYYMMDD and \verb|<.FORMAT>| is either \verb|.7z| or \verb|.bz2| depending on the used compression algorithm. When referring to \verb|<WIKITAG>-<DATE>-pages-meta-history.xml<.FORMAT>| the suffix \verb|-pages-meta-history.xml<.FORMAT>| may be omitted. 
Furthermore we define a wiki's name to be \verb|<WIKITAG>wiki|, allowing to reference a dump by its wiki name when the creation date and the format is given by the context. 

\begin{figure}[htbp]
	\centering
	\includegraphics[height=5cm]{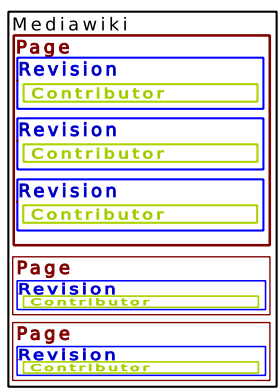}
	\caption{Simplified sketch of a dump's XML structure}
	\label{fig:xmlstructure}
\end{figure}

All dumps follow a specific XML structure\footnote{\url{https://www.mediawiki.org/wiki/Help:Export\#Export_format}}. Figure \ref{fig:xmlstructure} gives a brief and very simplified overview it. 
It begins with a \verb|<mediawiki>| tag and its attributes giving information about it. A wiki consists of multiple pages with Wikipedia content or meta information. All pages are encapsulated within the mediawiki tag, belong to a specific namespace, a title and a list of revisions. Each \verb|revision| tag represents an edit to its associated page and contains the new content in its \verb|text| attribute. A revision is done by exactly one author, which is further specified in the \verb|contributor| attribute.
In the context of wikis the terms \verb|author| and \verb|contributor| can be used interchangeably. That is, because \cite{Adler:2008:MAC:1822258.1822279} use the term ``author`` in their papers, but the dumps and code use the term ``contributor``. To prevent confusion, we try to prefer the former term.
A contribution can either be done anonymously or as an authenticated user. In the former case only the \verb|IP| address is saved, the \verb|username| and user's \verb|id| otherwise. 

At the time of writing, the full XML specification for wiki dumps describing more elements and attributes is available at \cite{mwdumpxml}. The thesis will focus on the \verb|page|, \verb|revision| and \verb|contributor| tags with their most relevant attributes, but extending the parsers to account for more information will be possible and covered in section ``\nameref{sec:implementation-planofimplementation-dataprocessing}`` and ``\nameref{sec:implementation-planofimplementation-codestructure}``.

%
\subsubsection{Pageview datasets} \label{sec:introduction-definitions-pageviewdatasets}
The Mediawiki Analytics Team\footnote{\url{https://www.mediawiki.org/wiki/Analytics}} creates hourly statistics of a Wikipedia page's traffic. This includes the amount of non-unique visits (\verb|COUNT|) as well as the request's size (\verb|SIZE|) in bytes. Those statistics follow a line-wise structure of \verb|<WIKITAG><WIKIPROJECT> <PAGETITLE> <COUNT> <SIZE>| for every page that was requested. \verb|<WIKIPROJECT>| defines the project name. It is either empty for Wikipedia projects like the common wikis or one of the following abbreviations for other projects \cite{wikimdpr}:

\begin{smallitemize}
	\item wikibooks: ``.b``
	\item wiktionary: ``.d``
	\item wikimedia: ``.m``
	\item wikipedia mobile: ``.mw``
	\item wikinews: ``.n``
	\item wikiquote: ``.q``
	\item wikisource: ``.s``
	\item wikiversity: ``.v``
	\item mediawiki: ``.w``
\end{smallitemize}
\verb|<PAGETITLE>| is a visited page's title, which is not normalized, meaning that special characters are url-encoded\footnote{\url{https://www.w3schools.com/tags/ref_urlencode.asp}}. After decoding, it should match with at least one title in the dumps if traffic was observed. 
A \verb|pageview|/\verb|pagecount|: A page view is either one line in a page views file or the number of page views for a specific page. The latter is also called \verb|pagecount|.
All pageviews within an hour are gathered in a gzip-compressed\footnote{\url{http://www.gzip.org/}} file.

The \verb|pageviews|: Throughout this thesis these pageview datasets or files will referred to as \verb|pageviews|.
	
The files from 2007 until December of 2015 follow the naming convention \verb|pagecounts-<YEAR><MONTH><DAY>-<HOUR>0000.gz|, where \verb|<YEAR>| is the full year, \verb|<MONTH>| and \verb|<DAY>| are month and day with a leading zero respectively. The \verb|<HOUR>| placeholder uses the 24-hour format with leading zeros. Since December 2015 the file names have changed to \verb|pageviews-<YEAR><MONTH><DAY>-<HOUR>0000.gz|.
	
The prefix \verb|pagecounts-| or \verb|pageviews-| as well as the suffix \verb|0000.gz| might be omitted if the enough context is given.

In 2015 a new method for collecting pageview statistics was developed by the Wikimedia team, which provides more human statistics, because "spiders" are detected and the data is not sampled anymore \cite{wikimap}. Detecting and filtering automated or robotic traffic can lead to more accurate statistics and therefore serve as better weights for the impact calculation of a wiki edit, because the aspect of visibility and outreach of an edit can be incorporated.

The ``legacy Pageviews`` \cite{wikimap} can be downloaded from \cite{mwpvleg} whereas the newer pageviews data is available at \cite{mwpvnew}.

\subsection{Setup}\label{sec:introduction-setup}
To keep all tests and results reproducible, the used systems and software versions will be listed below. If not stated otherwise or clear from the context, the student cluster was used for the tests.

\subsubsection{Student cluster} \label{sec:introduction-setup-studentcluster}
The DIMA group at the TU Berlin gives certain students access to a shared ten node cluster. Time slots are requested and assigned using a dedicated Slack\footnote{\url{https://dimatub.slack.com/}} channel. 

Each of those ten nodes, except the first node, is equipped with the following hardware and software:

\begin{itemize}
	\item CPU: 1x IBM POWER7 8231-E2B pSeries CPU with 47 cores @ 3720 MHz, revision 2.1 (pvr 003f 0201), as reported by \verb|cat /proc/cpuinfo|
	\item RAM: 65288960 kB (\textasciitilde 62GB) of memory as reported by \verb|cat /prof/meminfo|
	\item Swap: 71685888 kB (\textasciitilde 68GB) of swap with priority -1 and a swappiness of 60, as obtained by \verb|free -k| or \verb|cat /proc/sys/vm/swappiness|
	\item Network: 1x active 1000baseT/Full Ethernet port, reported by \verb|ethtool|
	\item Disks: 
		\begin{itemize}
			\item 4x Toschiba MBF2600RC (600GB)
			\item 2x Fujitsu MBC2073RC (73GB)
		\end{itemize}
	\item OS: Fedora 24
\end{itemize}
The nodes 2 - 10 are the slaves on which the parallel computing takes place. The exception is the first node (\verb|ibm-power-1|) which is the main node where Apache Flink's job manager runs. It has less RAM (50478720 kB, \textasciitilde 48GB), but more swap (143372224 kB, \textasciitilde 136GB) and one of the Fujitsu drives is a Toschiba MBE2147RC (147GB). 

Out of all ten nodes, eight were used during the thesis. Apache Flink's dashboard reported eight task managers with 464 available task slots. If not otherwise specified 400 task slots were used. For the evaluation in section \ref{sec:evaluation} the nodes 2 and 6 were removed due to showing faulty behavior too often. This decision reduced the number of task managers to 6 and the available task slots from 464 to 384, of which 380 were used. The final configuration had the following important settings configured:
\begin{itemize}
	\item \verb|jobmanager.heap.mb = 1024|
	\item \verb|taskmanager.heap.mb = 49152|
	\item \verb|taskmanager.memory.fraction = 0.7|
	\item \verb|taskmanager.network.numberOfBuffers = 262144|
	\item \verb|taskmanager.network.bufferSizeInBytes = 262144|
\end{itemize}

\subsubsection{DigitalOcean Virtual Machine} \label{sec:introduction-setup-digitalocean}
DigitalOcean\footnote{\url{https://digitalocean.com/}} is a cloud-based server hosting provider. Due to WikiTrust's incompatibility with the student cluster's PowerPC architecture, a virtual machine was rented at this provider. The downside of virtualized hardware is that the performance might vary due to the shared underlying host system. Nevertheless, we decided to utilize cloud services, because other suitable dedicated hardware was not available. Furthermore, only one instance with the following specification was rented:

\begin{itemize}
	\item CPU: 12 cores on a Intel(R) Xeon(R) CPU E5-2650L v3 @ 1.80GHz CPU as reported by \verb|cat /proc/cpuinfo|
	\item RAM: 33016556 kB (\textasciitilde 31GB) of memory as reported by \verb|cat /prof/meminfo|
	\item Swap: None
	\item Network: 1x shared 1Gbit/s connection as explained by a moderator\footnote{\url{https://www.digitalocean.com/community/questions/upload-and-download-speed-of-a-droplet}}
	\item Disk: 21.5GB virtual SSD-based disk\footnote{\url{https://www.digitalocean.com/products/storage/}}. 
	\item OS: Debian 8
	\item Datacenter: FRA1, Frankfurt, Germany
\end{itemize}

\subsubsection{Apache Hadoop HDFS} \label{sec:introduction-setup-apachehadoophdfs}
The scalable and fault tolerant Apache Hadoop HDFS\footnote{\url{https://hadoop.apache.org/docs/r2.7.1/}} (2.7.1) distributed file system was used. One NameNode stores the file system's metadata and the remaining DataNodes store the actual data \cite{hadap}. This has the advantage of storing big datasets, e.g. the English wiki, on multiple nodes and allowing them to read different parts of a file simultaneously, thus preventing a single hard drive from becoming a bottleneck. The HDFS data is stored on the nodes 2, 3, 4, 5, 7, 8, 9.

\subsubsection{Apache Flink} \label{sec:introduction-setup-apacheflink}
Apache Flink\footnote{\url{https://flink.apache.org/}} with Hadoop 2.7.0 support was used. Its Java API allows to write distributed MapReduce-programs which can be executed on the cluster, e.g. by providing \verb|DataSet| classes, which are similar to list data structures suitable for distributed processing. Common MapReduce operations like map, filter, reduce and others can be applied to such DataSets. Additionally, Flink handles the orchestration and coordination of the nodes when a distributed task needs to be run. The framework's build uses Flink 1.1.3 as a dependency, but the server had Flink 1.0.3 deployed as the runtime environment. This discrepancy in the auto-generated \verb|pom.xml| was unfortunately overseen until the end, but re-running several tests did not show any significant differences or bug resolutions.

\subsubsection{Apache Maven} \label{sec:introduction-setup-apachemaven}
All code was compiled and executed using either OpenJDK Java\footnote{\url{http://openjdk.java.net/}} 1.8 or Oracle Java\footnote{\url{http://www.oracle.com/technetwork/java/javase/downloads/jre8-downloads-2133155.html}} 1.8 and Apache Maven\footnote{\url{https://maven.apache.org/}} 3.3.0. The framework's dependencies are defined in the \verb|pom.xml| file and fetched automatically by Apache Maven before the build process. It produces a runnable jar file after testing the project with the implemented JUnit tests.

\begin{myboxi}[Key points of section ``\nameref{sec:introduction}``]
\begin{itemize}
	\item Calculating author rankings based on the edit history is an interesting topic that might benefit Wikipedia in various problems.
	\item \verb|dump|, \verb|pageview| and other related keywords and abbreviations facilitate the thesis' understanding.
	\item For better reproducibility, the software versions and hardware specifications are described.
\end{itemize}
\end{myboxi}

\section{Implementation} \label{sec:implementation}
Before starting with the implementation of the framework, the actual basis for it - the datasets - have to be investigated. The explanations in section \ref{sec:introduction-definitions} give a first overview of the dump and pageview datasets, but both are compressed due to their sheer size. There are two questions which immediately arise:
\begin{enumerate}
	\item What is the compression factor? Is it feasible to decompress the data first?
	\item What performance impact would decompression-on-the-fly have?
\end{enumerate}
Investigating the initial dataset's format and structure will help to understand how to operate on those datasets and how to maximize the performance when doing so. 

\subsection{Investigating compressed datasets}\label{sec:implementation-compresseddatasets}
This section will answer both questions from above. The dumps are available in two different compression formats, namely bzip2\footnote{\url{http://www.bzip.org/}} and 7z\footnote{\url{http://www.7-zip.org/}}. 
To get an impression of the compression factor of both algorithms, a selection of different sized dumps were downloaded from \cite{mwdumpdl}. The bzip2 files were decompressed as a preparation for the performance tests and to furthermore reveal their decompressed size.
\subsubsection{Size aspect} \label{sec:implementation-compresseddatasets-sizeaspect}
\begin{figure}[H]
	\includegraphics[width=\linewidth]{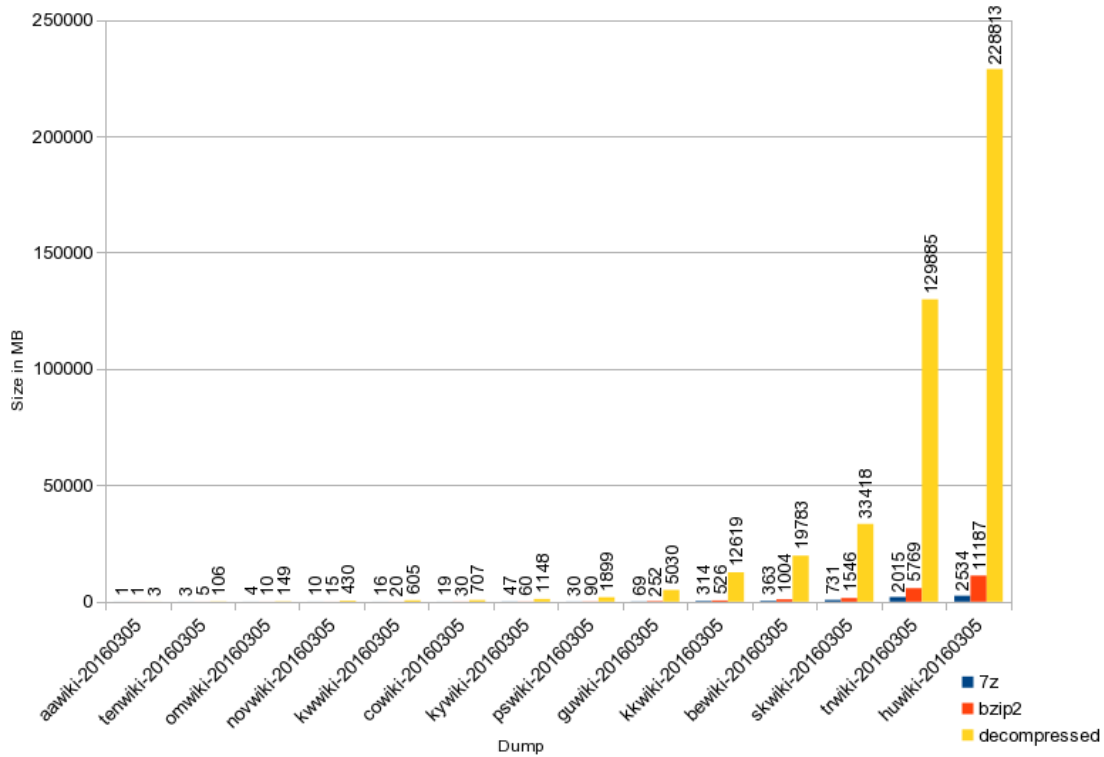}
	\caption{The chart shows sizes of compressed (7z, bzip2) and uncompressed dumps. }
	\label{fig:compression-sizes}
\end{figure}

Figure \ref{fig:compression-sizes} shows the 14 dumps and their 7z, bzip2 and decompressed size. The first thing that catches one's eye is the steep growth of the decompressed data whereas both compression algorithms significantly lower the slope.

\begin{figure}[htbp]
	\includegraphics[width=\linewidth]{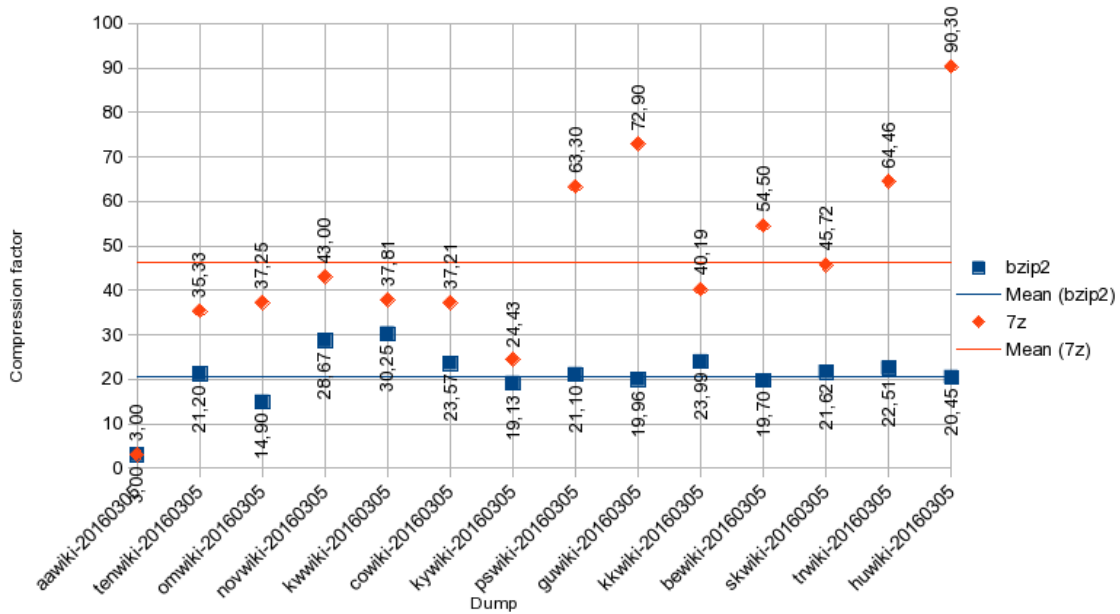}
	\caption{The dumps' bzip2 and 7z compression factors ($F_c$) and an average per compression type.}
	\label{fig:compression-factor}
\end{figure}

The resulting compression rates are plotted in figure \ref{fig:compression-factor}. The formula is: $$ F_c = \frac{S_d}{S_c}$$ where:
\begin{itemize}
	\item $F_c$ is the resulting compression factor
	\item $S_d$ is the decompressed size in MB
	\item $S_c$ is the compressed size in MB
\end{itemize}
There are two significant outliers: The first dump on the far left of the x-axis with a factor of \verb|3|, but that can be explained by the relatively small amount of compressible data. The second outlier is the factor \verb|90.30| for the \verb|huwiki| dump in this test. This might be explainable by the fact that all metadata, like the XML-structure or the contributor's name, as well as the whole page's content are saved for every single revision. This means when only minor changes are introduced in consecutive revisions, the majority of their content will stay the same. Recurring data is usually the optimal input for compression algorithms and 7z compresses it well.

With a mean factor of around \verb|45|, 7z compresses the data more than twice as much as the bzip2 codec with a mean value of \verb|20|. Nevertheless, a compression rate of at least 20 can save a lot of hard disk space. This helps especially when working with huge dumps like the English Wikipedia, which is around 680 GB bzip2 compressed and thus a decompressed size of about twelve terabytes could be expected.

To answer the first question, we can conclude that good compression factors of \verb|20| with bzip2 or \verb|45| with 7z can be achieved. This, however, leads to the fact that decompressing the dumps prior to using them is not really feasible. It might work for smaller dumps, but definitely is a problem for comprehensive wikis when not enough disk space is available.

Those compression achievements only hold for the dumps, not for the pageviews, which are gzip\footnote{\url{http://www.gzip.org/}} compressed. Table \ref{tab:gzip-compression} shows the smallest, an average and the biggest pageviews dataset from May 2016\footnote{\url{https://dumps.wikimedia.org/other/pagecounts-raw/2016/2016-05/}}. With an average factor of around 4 the compression is not nearly as good as bzip2 or 7z, but still helps in the context of using months or years of pageview data. 


\begin{table*}
	\centering
	\caption{Sizes of (de-)compressed gzip Pageviews and the resulting compression ratio.}
	\label{tab:gzip-compression}
	\begin{tabular}{cccc}\toprule
		Pageviews & compressed (MB) & decompressed (MB) & factor \\
		\midrule
		20160504-05 & 65 & 273.38 & 4.21 \\
		20160519-06 & 87 & 386.65 & 4.44 \\
		20160531-19 & 110 & 460.69 & 4.19 \\
		\midrule\midrule
		Average  & 87.33 & 373.57 &  4.28 \\
		\bottomrule
	\end{tabular}
\end{table*}
\subsubsection{Time aspect} \label{sec:implementation-compresseddatasets-timeaspect}
This conclusion directly leads to the second question. What performance impact or trade-off can be expected when using compressed datasets over the uncompressed ones? 
The downside of using compressed data is that not all codecs support splitting the dataset to parallelize the work. Gzip, for example, uses a continuous sliding window to find recurring data to produce an output stream and is therefore not splittable \cite{gziporg}. However, 7z and bzip2 decompression can be parallelized. Bzip2 compresses files in blocks which are delimited by a specific 48-bit pattern and can thus be handled independently \cite{bziporg}. Parallelized systems can therefore read from the same bzip2 file simultaneously at  different block offsets. 7z supports multiple compression methods, but uses the multi-threadable LZMA by default \cite{7ziporg}. Both gzip and bzip2 are natively supported by Apache Hadoop with the \verb|BZip2Codec| and \verb|GZipCodec| \cite{hdaporg}, but the GZipCodec will only run with a parallelism of one for the reason explained before. Since 2014 Apache Flink has been compatible with Apache Hadoop's MapReduce interfaces \cite{flaporg} allowing to re-use the already implemented \verb|CompressionCodecs| and \verb|InputFormats|.
\begin{figure}[htbp]
	\includegraphics[width=\linewidth]{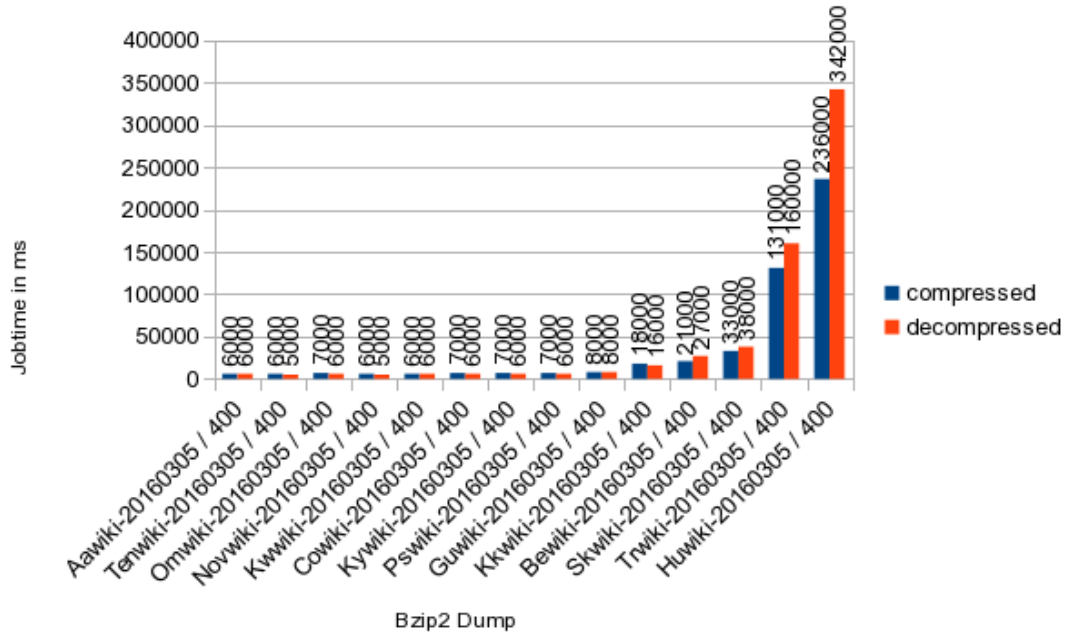}
	\caption{The chart shows that the bzip2 job performance is higher for compressed datasets.}
	\label{fig:bzip2-job-performance}
\end{figure}

\begin{figure}[htbp]
	\includegraphics[width=\linewidth]{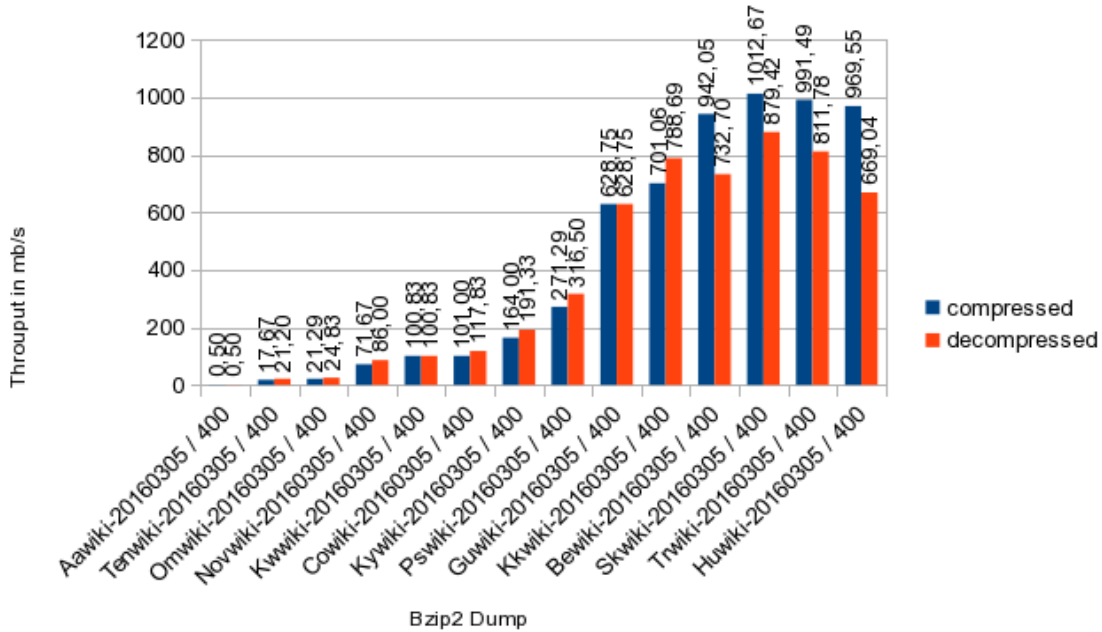}
	\caption{A higher throughput is possible with Bzip2 compressed datasets by processing more data in less time.}
	\label{fig:bzip2-compression-throughput}
\end{figure}

To answer the question a small Apache Flink application was developed, whose job it is to read a (de-)compressed file and count the number of lines in it. Apache Hadoop's \verb|TextInputFormat()| reads data and transparently handles the decompression of compressed files. 
For all fourteen test dumps the application was run on the decompressed XML and bzip2 version with 100, 200, 300 and 400 slots each. Figure \ref{fig:bzip2-job-performance} shows the processing times in ms using 400 slots. The time for the other runs follows a similar pattern. Interestingly, the dumps up until a size of around 300 MB compressed and 12 GB decompressed are processed more or less in the same time. After that, the decompressed dumps take longer to process. Using compressed dumps does not only save disk space, but also increases the performance. 
A possible explanation for that might be a storage bottleneck, e.g. HDFS or the network. Figure \ref{fig:bzip2-compression-throughput} shows the throughput calculated with the following formula: $$T_d = \frac{s_d \cdot 1000}{t_d} \cdot c_d$$ where 
\begin{itemize}
	\item $T_d$ is the resulting throughput in MB/s
	\item $s_d$ is the dump's size in MB
	\item $t_d$ is the dump's execution time in ms
	\item $c_d$ is the dump's compression factor. It is $1$ for decompressed dumps.
\end{itemize}
One can see that the extracted dumps' throughput does not cross 900 MB/s. Only the compressed dumps exceed this limit, because as previously discovered only around one twentieth of the data needs to be transferred and the decompression only increases the CPU load, but does not slow down due to the network or storage limitations. 

\begin{figure}[htbp]
	\includegraphics[width=\linewidth]{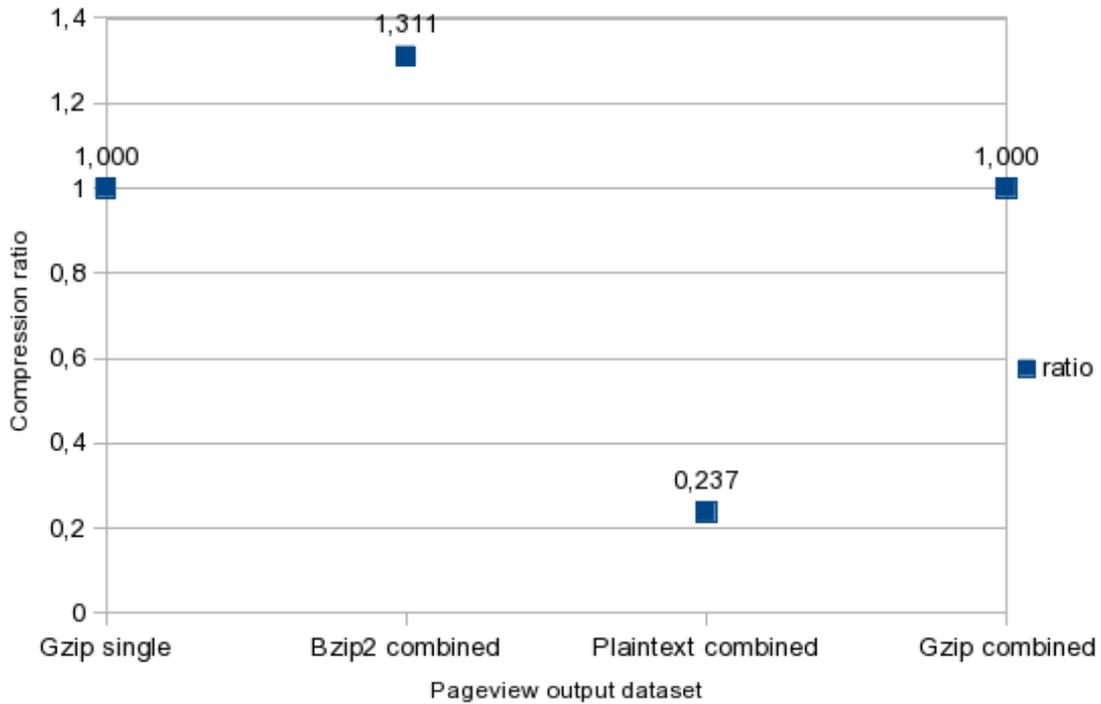}
	\caption{Merging the separate pageview files into a single dataset improves the compression ratio for bzip2, but does not change for gzip.}
	\label{fig:gzip-combined-ratio}
\end{figure}

\begin{figure}[htbp]
	\includegraphics[width=\linewidth]{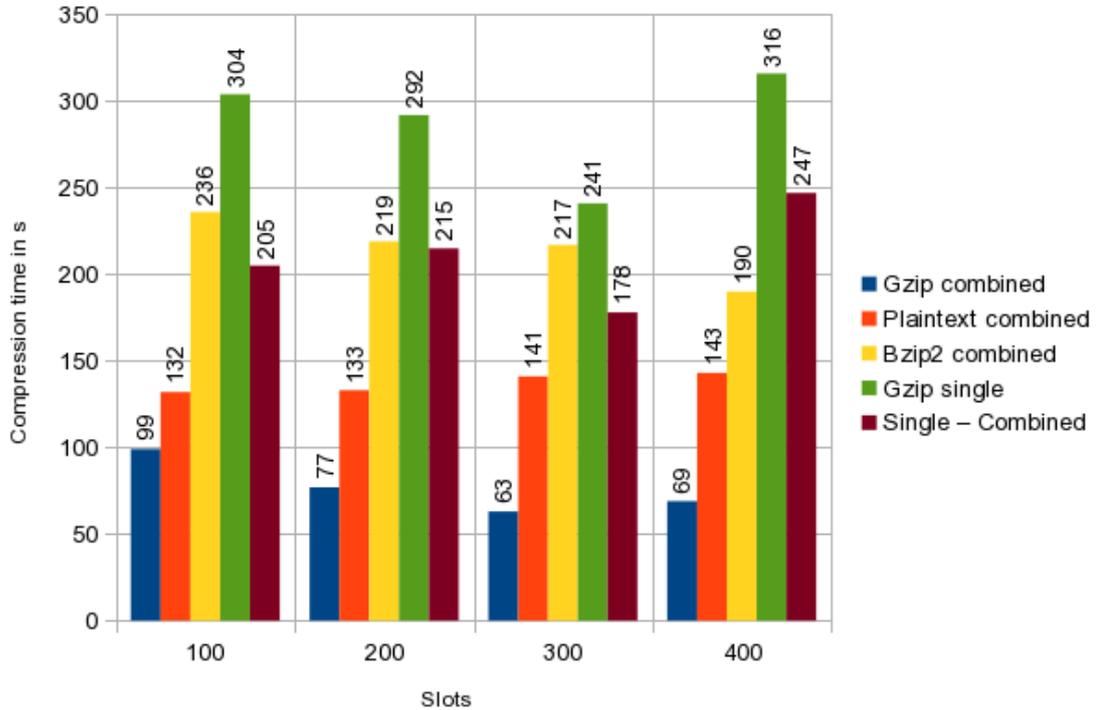}
	\caption{The bar chart shows the time needed to process the single merged and compressed dataset.}
	\label{fig:gzip-combined-jobtimes}
\end{figure}

Another probable cause of slowness might be that the pageviews data is packaged into hourly compressed gzip files. This leads to a dataset consisting of multiple files when a longer time frame needs to examined. Because of the rather small sizes of those files, the compression algorithm does not improve the processing speed. In fact, it slows it down for a single file, which can be seen in figure \ref{fig:gzip-job-performance}. However, the issue can be tackled by concatenating all pageviews files into one huge file. This was concluded by developing and using a preprocessing tool (FileMerger), which reads all files from a folder, decompresses them and then writes one combined, optionally compressed, output dataset. The tool merged 696 separate pageview files from February 2016, accumulating 59.5 GB in total, into single output datasets of the following formats:
\begin{figure}[htbp]
	\includegraphics[width=\linewidth]{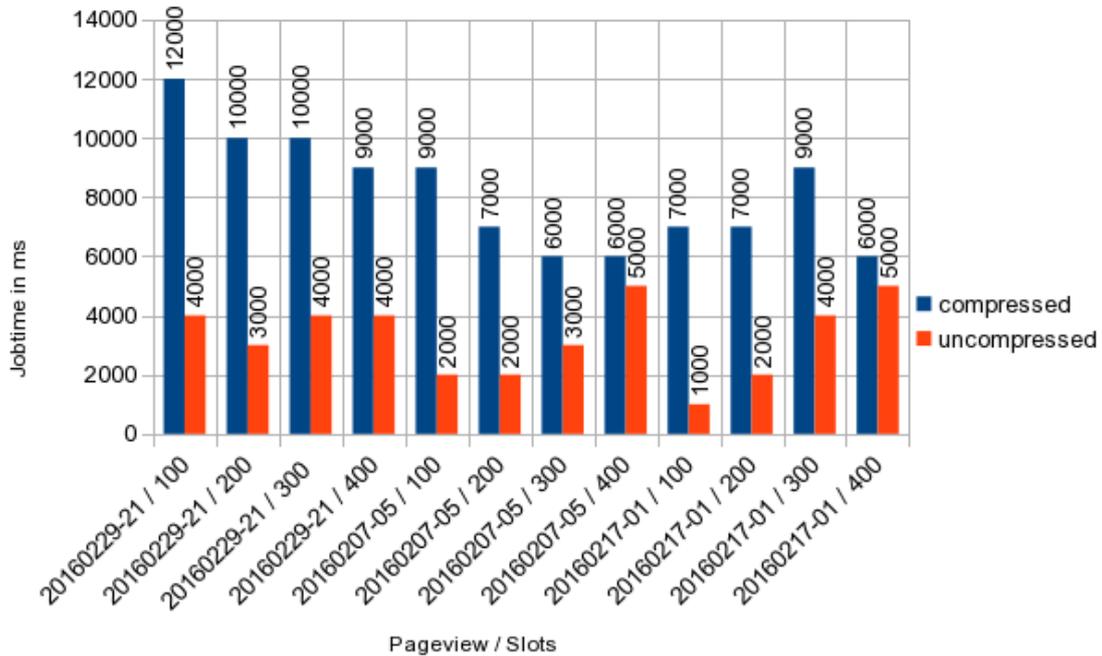}
	\caption{Single pageview files are processed faster when they are not (Gzip) compressed.}
	\label{fig:gzip-job-performance}
\end{figure}
\begin{itemize}
	\item Plaintext
	\item Gzip
	\item Bzip2
\end{itemize}
Figure \ref{fig:gzip-combined-ratio} shows the resulting compression ratios. A ratio greater one means that the size was further reduced. A slightly better ratio can be achieved for bzip2, but it doesn't change for gzip and even drops below 1 for plaintext outputs, because it is not compressed anymore.
Surprisingly the performance test revealed that using the combined gzip output dataset is much faster than plaintext, bzip2 or the separated gzip files. Figure \ref{fig:gzip-combined-jobtimes} shows that a speedup up over 450\% can be reached and more than 240 seconds of computation time can be saved.  
This preprocessing step is worth additional time and effort, because the data only needs to be preprocessed once, but can then be read multiple times with higher performance. This methodology is desired when the dataset does not change over time, but the calculations with the framework on it are repeated often. It takes the following amount of time to fully process the pageview dataset:

\begin{itemize}
	\item Plaintext: 876s
	\item Gzip: 652s
	\item Bzip2: 802s
\end{itemize}
With these preprocessing times and the gained speedup of up to 240 seconds per run, using the combined gzip dataset will pay off when used more than three times. Given the fact that the pageviews datasets contain information about the traffic to all wikis, it is more likely to be static when a longer time frame in the past, e.g. last year, is chosen to accompany author rank calculations for different wikis in that time frame.

After investigating the storage and time aspects of compressed versus uncompressed datasets, the following conclusion can be drawn: Using compressed datasets leads to significant gains by having to store less data and also improves the overall computational performance by having to read and transfer less data from the HDFS. Further speedups can be achieved by preprocessing the pageviews datasets.

\begin{myboxi}[Key points of section ``\nameref{sec:implementation-compresseddatasets}``]
	\begin{itemize}
		\item Compression factors: bzip2 \textasciitilde 20 and 7z \textasciitilde 40 for dumps; gzip \textasciitilde 4 for pageviews.
		\item Shorter processing times and higher performance are achievable with bzip2 for dumps.
		\item Preprocessing the pageviews dataset improves its performance.
	\end{itemize}
\end{myboxi}

\subsection{Plan of implementation} \label{sec:implementation-planofimplementation}
Now that we are familiar with the format and layout of the input datasets, we can focus on the implementation and how to work with those datasets. Before implementing the framework for author impact measure calculations, it has to be designed in a way that allows for flexibility and extensibility of its functionality, e.g. the parsing and the representation of the input data, the data processing steps and finally the result's output format.

The framework's core dependency is the Apache Flink framework for distributed computation, which had been introduced in section \ref{sec:introduction-setup-apacheflink}. Its API is available for the programming languages Java and Scala \cite{flapord-example}. Apache Maven is a dependency manager for Java, so using Java as the framework's language seemed reasonable. Furthermore Java builds are, unlike C or C++, not tied to a specific CPU architecture or platform, because ``[w]hen compiled, the Java code gets converted to a standard, platform-independent set of bytecodes, which are executed by a Java Virtual Machine (JVM). A JVM is a separate program that is optimized for the specific platform on which you run your Java code.`` \cite{docsorcaleorg}

\subsubsection{Data representation and transformation} \label{sec:implementation-planofimplementation-datarepresentation}
The very first step is to model the input data into Java objects with all the necessary attributes and relations. In section \ref{sec:introduction-definitions-wikidatasets} and \ref{sec:introduction-definitions-pageviewdatasets} the rough layout was discussed. The reference implementation of the framework focuses only on the necessary data used to reproduce and implement \citeauthor{Adler:2008:MAC:1822258.1822279} contribution measures, but should be easily modifiable and extensible as explained in section \ref{sec:implementation-planofimplementation-codestructure} and showed in section \ref{sec:implementation-functionalityextensibility}.

In general there are three important pieces of information which can be directly translated from the XML structure of the dumps into equivalent Java classes:
\begin{itemize}
	\item Page
	\item Revision
	\item Contributor
\end{itemize}

These and the following further classes responsible for the framework's data representation are part of the \verb|it.neef.tub.dima.ba.imwa.impl.data| package:

\begin{itemize}
	\item DataHolder
	\item DoubleDifference
	\item PageRelevanceScore, RevisionRelevanceScore, ContributorRelevanceScore, \& DifferenceRelevanceScore
\end{itemize}

\begin{figure}[htbp]
	\includegraphics[width=\linewidth]{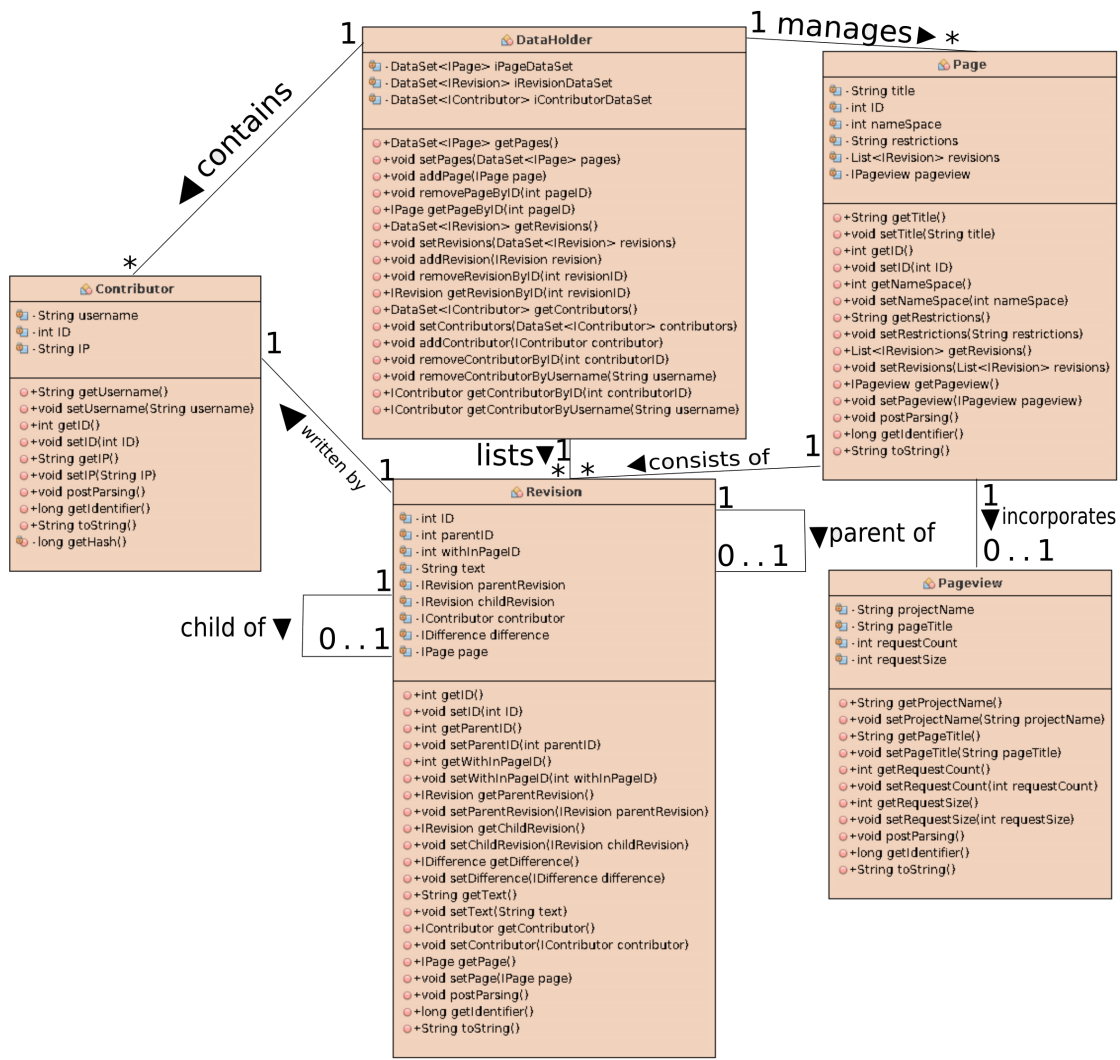}
	\caption{This class diagram shows the relationships, methods and attributes of the major data representation classes.}
	\label{fig:class-diagram-data-structure}
\end{figure}

Figure \ref{fig:class-diagram-data-structure} gives a brief overview of the data structures, which will be described in the following paragraphs.
\paragraph{Page}
In a dump the \verb|<page>| XML tag introduces a new Wikipedia page. Important information to extract are the page's title (\verb|<title>| tag) and the namespace (\verb|<ns>| tag) in which it was published. The title is necessary to later match the pageview data and the namespace a good filter, because \citeauthor{Adler:2008:MAC:1822258.1822279} only considered pages from the main/article namespace, identified by the value \verb|0| \citep{Adler:2008:MAC:1822258.1822279}. A page's ID was parsed from the \verb|<id>| tag for better debugging and testing purposes. The aforementioned attributes are directly represented in the \verb|Page| class. Additionally, a reference named \verb|pageview| to a \verb|Pageview| instance is introduced. When no pageview data is available, the reference is \verb|null|. The page's \verb|<revision>| tags are represented in a list (\verb|revisions|) of \verb|Revision| objects.

\paragraph{Revision}
A \verb|<revision>| XML tag in a Wikipedia page captures its content at a given time. When a change is made, a new revision with its author is appended to the end. The latter is modeled in the \verb|Contributor| class and is referenced by the \verb|contributor| attribute.  The contribution measures by \citeauthor{Adler:2008:MAC:1822258.1822279} use the revisions to eventually calculate a contributor's ranking. Each \verb|Revision| object has a back-reference to its associated \verb|Page|. The model's attributes \verb|ID| (\verb|<id>| tag), \verb|parentID| (\verb|parentid| tag) and \verb|text| (\verb|<text>| tag) are populated with the values of their correspondent XML tags.
Furthermore, a \verb|withInPageID| is assigned during the parsing process to represent a revision's relative position in the page's history. The first revision always has index \verb|1|.
During the realization of the contribution measures, it turned out that implementing the revisions as a double linked list facilitates the calculation of differences between the revisions. Therefore, every \verb|Revision| instance has a reference to the revision before (\verb|parentRevision|) and after (\verb|childRevision|) itself, where ``before`` refers to the revision with $withInPageID-1$ and ``after`` to $withInPage+1$. The first revision has no parent and the reference is therefore \verb|null|. Similarly, the last revision's child is \verb|null|.
The resulting differences between two revisions are wrapped in a \verb|IDifference| object, which is accessible by the \verb|difference| attribute.

\paragraph{Contributor}
Each revision has an unique author defined in the \verb|<contributor>| tag. When it was authenticated while editing a Wikipedia page, the username and its id are saved in correspondent tags. Otherwise the only identifying information is the IP address (\verb|<ip>| tag) \citep[p. 10]{adler2012wikitrust}. 
Due to this fact, when an anonymous contributor is encountered, the username will be set to an imaginary one, which cannot clash with a real contributor, because ``for technical reasons, usernames containing the forbidden characters \# < > [ ] | { } / @ are not possible.`` \cite{wikinaming}. Code listing \ref{code:contributor-setIP} outlines how anonymous authors are mapped to a single non-existent contributor. 
\begin{figure}
	\begin{lstlisting}[caption={Map anonymous authors to an non-existent, ``illegal`` username while parsing.}, label={code:contributor-setIP}]
public static final String invalidUsername = "##<<__-=ANONYMOUS=-__>>##";
//[...]
case "ip":
	this.cContributor.setIP(cValue);
	this.cContributor.setUsername(SkipXMLContentHandler.invalidUsername);
	break;
	\end{lstlisting}
\end{figure}

\paragraph{Pageview}
The pageview data comes from a separate source, namely the pageviews datasets described in section \ref{sec:introduction-definitions-pageviewdatasets}. Each line of a pageview file informs about the request count  (\verb|requestCount|) and request size (\verb|requestSize|) of one specific Wikipedia page. This information can be useful to weight the importance of an edit, e.g. edits to a frequently accessed wiki are more important than others. To transform a raw datapoint into an instance of the \verb|Pageview| class, it has to be separated into its four parts. In contrast to the dumps, the page title is still URL encoded and not normalized, so that needs to be reverted before setting the \verb|pageTitle| attribute. The code listing \ref{code:pageviewparser-parsePageviewData} from the PageviewParser\footnote{it.neef.tub.dima.ba.imwa.impl.parser.PageviewParser} shows how this is achieved. 
Due to the fact that this data is recorded and published in hourly batches, multiple \verb|Pageview| instances with the same page title might exist after parsing. To satisfy the one-to-one relationship with a \verb|Page| object, duplicate Pageview objects are merged into a single one by summing up their separate values. The \verb|projectName| attribute is used in a filter step to reduce the pageview data to a specific Wikipedia project before summing and associating them with their respective pages, both to increase the performance and the accuracy of the pageview data by preventing objects with equivalent page titles from difference wikis to clash. 
\begin{figure}
	\begin{lstlisting}[caption={When parsing pageview data, it must be separated and decoded line by line.}, label={code:pageviewparser-parsePageviewData}]
	// The format is: [wiki shorttag] [title] [count] [size], so split at a whitespace.
	String[] parts = s.split(" ");
	// Get new pageview instance
	IPageview pv = Framework.getInstance().getConfiguration().getPageviewFactory().newPageview();
	pv.setProjectName(parts[0]);
	// The page title is not normalized and urlencoded. We have to revert that, so that it
	// matches the wikipedia (pages) DataSet.
	String pageTitle = parts[1].replace("_", " ");
	pageTitle = java.net.URLDecoder.decode(pageTitle, "UTF-8");
	pv.setPageTitle(pageTitle);
	pv.setRequestCount(Integer.valueOf(parts[2]));
	pv.setRequestSize(Integer.valueOf(parts[3]));
	\end{lstlisting}
\end{figure}

\paragraph{DataHolder}
The framework needs to manage the eventually parsed and preprocessed data, so that it can be further processed. For this purpose, the \verb|DataHolder| class was implemented. After the initial parsing process has finished, the DataHolder will be populated with the Page, Revision and Contributor DataSets. It then offers convenience methods for accessing and manipulating those three major datasets, for example adding, removing or accessing objects by their ID or by certain attributes. Its main purpose is to be the central DataSet distribution point within the framework.

\paragraph{DoubleDifference}
The \verb|DoubleDifference| class implements the \verb|IDifference|\footnote{it.neef.tub.dima.ba.imwa.interfaces.data.IDifference} interface, the idea of which is to hold information about the difference between two objects. The difference itself is calculated by a class which implements the \verb|IDiffer|\footnote{it.neef.tub.dima.ba.imwa.interfaces.diff} interface. In the case of \citeauthor{Adler:2008:MAC:1822258.1822279} most contribution measures use a numeric value to denote the size of the difference between two Revision objects, e.g by determining how many words have changed. To better account for calculations with fractions, the \verb|Double| type was chosen and the  default value was set to \verb|null|.

\paragraph{PageRelevanceScore, RevisionRelevanceScore, ContributorRelevanceScore, \& DifferenceRelevanceScore}
All of the headlined classes implement the \verb|IRelevanceScore|\footnote{it.neef.tub.dima.ba.imwa.interfaces.data} interface for their respective class type. A \verb|RelevanceScore| object has a reference to an associated object, which implements the \verb|IDataType| interface, and its rating - the so called ``relevance`` score. The score should denote the object's importance for the ranking. The RelevanceScore objects are the result of the \verb|ICalculation|\footnote{it.neef.tub.dima.ba.imwa.interfaces.calc} and thus should be used as the actual score for the ranking. When applying the terminology to \citeauthor{Adler:2008:MAC:1822258.1822279} contribution measures, the accumulation of the \verb|DoubleDifference|s of all revisions of a particular contributor will be her \verb|ContributorRelevanceScore|. To keep the framework extensible and also usable for the calculation of other rankings, the relevance scores are not only available for contributors, but also for other aspects of a Wikipedia site, like revisions, pages or differences. RelevanceScore DataSets of the same type can be passed to the \verb|RelevanceAggregator| for further processing.

\paragraph{The IDataType}
All the above described classes, except the DataHolder, implement the \verb|IDataType|\footnote{it.neef.tub.dima.ba.imwa.interfaces.data} interface. It is used to mark those Java objects as serializable to allow Apache Flink to serialize them for distribution across the cluster (more about this issue in section \ref{sec:implementation-problemsandsolutions}) and as \verb|IIdentifiable|\footnote{it.neef.tub.dima.ba.imwa.interfaces.data}. The latter requires the implementation of a \verb|getIdentifier()| method, which is necessary to uniquely identify objects in Apache Flink's DataSets when using several operations such as grouping or joining. For example, the \verb|Contributor| class calculates a very basic hash based on its username as seen in line 14-17 of listing \ref{code:contributor-getIdentifier}. For contributors the username is unique and the preferred matching criteria, but returning \verb|this.hashCode()| for other objects might also be a suitable solution.
\begin{figure}
	\begin{lstlisting}[caption={Contributor - getIdentifier() implementation using a simple hash function that helps to uniquely identify and compare Contributor objects.}, label={code:contributor-getIdentifier}]
	long hash = 0;
	int count = 1;
	String data;
	
	if (this.username == null) {
	// If the user is anonymous, only use the IP address.
	// This is not used when the imaginary username is set in setIP()
	data = this.getIP();
	} else {
	data = this.username;
	}
	
	// For every byte in the username multiply it with its index.
	for (byte c : data.getBytes()) {
	hash += c * count;
	count++;
	}
	
	return hash;
	\end{lstlisting}
\end{figure}

\subsubsection{Data processing flow} \label{sec:implementation-planofimplementation-dataprocessing}
\begin{figure}[htbp]
	\includegraphics[width=\linewidth/2,center]{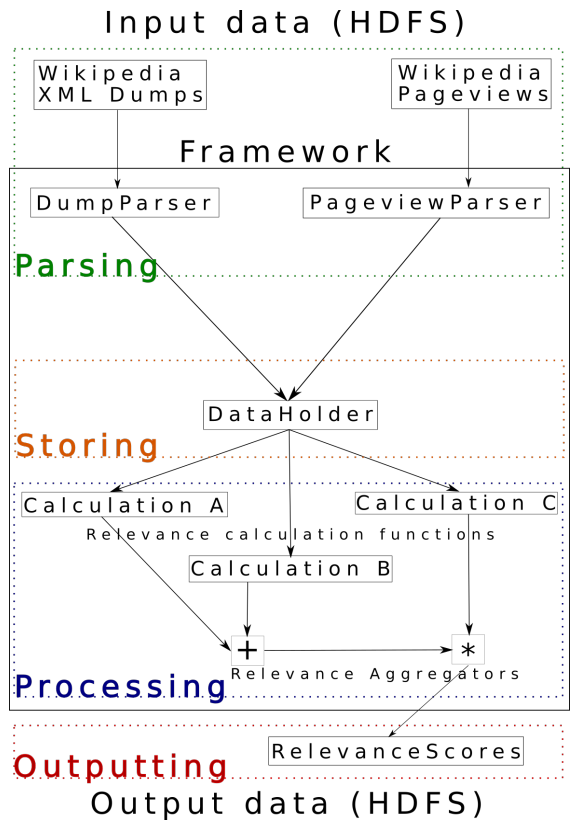}
	\caption{This sketch outlines the four steps of the framework's data processing flow and what components are involved.}
	\label{fig:framework-dataflow}
\end{figure}

This section will explain the framework's data processing flow. That is, the way of the raw input data through all stages of processing until a result is outputted. There are basically four stages that the data passes as shown in listing \ref{fig:framework-dataflow}:

\begin{enumerate}
	\item Parsing
	\item Storing
	\item Processing
	\item Outputting
\end{enumerate}

\paragraph{Parsing}
The first major step is to apply the data transformations described in the previous section. The starting point is dumps and pageviews files in two separate folders. The files should be in their compressed format, because it increases the performance as showed in section \ref{sec:implementation-compresseddatasets}. Instead of saving the files on a hard disk on one node, they should be hosted on a distributed filesystem like HDFS across several nodes on the cluster. Apache Flink supports this data source that will allow parallel access from and to nodes on the cluster to different parts of the data, therefore preventing a single node to become an IO-bottleneck.

\subparagraph{Configuration}
When the framework is invoked, it expects certain configured parameters to function properly. To equip the framework with a globally accessible configuration, the \verb|Configuration|\footnote{it.neef.tub.dima.ba.imwa.impl.configuration} class was implemented. This stores important variables and can be accessed through the framework's \verb|.getConfiguration()| function. It solves the problem of sharing options between different parts of the framework. The default configuration is applied when the framework is instantiated the first time. However, the following three options regarding the input datasets must be set manually, as outlined in lines 6-8 in code listing \ref{code:AdlerContributionMeasuresExample-run}:

\begin{itemize}
	\item The path to the folder containing the dumps (\verb|.setWikipediaDumpPath()|)
	\item The location of the pageviews folder (\verb|.setPageviewDumpPath()|)
	\item The wiki's project name (\verb|.setPageviewDumpShortTag()|)
\end{itemize}

If the dump's variable is not provided or not set before the \verb|.init()| call in line 10, the framework will not be able to read and parse any data. Parsing and merging the pageviews data is optional and skipped if the pageviews path is not set. In the thesis' reference implementation of \citeauthor{Adler:2008:MAC:1822258.1822279} author ranking, the information needs to be supplied on the command line as positional arguments.
\begin{figure}[htbp]
	\begin{lstlisting}[caption={The AdlerContributionMeasuresExample Shows the import steps of initializing, running and outputting the TextLongevity contribution measure by \citeauthor{Adler:2008:MAC:1822258.1822279} with the  framework.}, label={code:AdlerContributionMeasuresExample-run}]
	public void run(String[] args) throws Exception {
	// Initialize the framework and the output objects.
	Framework fw = Framework.getInstance();
	IOutput outputter = Framework.getInstance().getConfiguration().getOutputFactory().newOutput();
	// Configure the framework to use the correct options.
	fw.getConfiguration().setWikipediaDumpPath(args[0]);
	fw.getConfiguration().setPageviewDumpPath(args[1]);
	fw.getConfiguration().setPageviewDumpShortTag(args[2]);
	fw.getConfiguration().getPreFilters().add(fw.getConfiguration().getRegexPreFilterFactory().newIRegexPreFilter("(?is).*<ns>0</ns>.*"));
	fw.init();
	// Instantiate the Calculation object to run.
	ACalculation calculation = new TextLongevityCalculation(); 
	
	// Run and output the calculation.
	fw.runCalculation(calculation);
	outputter.output(calculation);
	}
	\end{lstlisting}
\end{figure}
Further configuration options include setting factories for various classes or defining \verb|PreFilters| and \verb|PostFilters| which will be explained shortly. 

\subparagraph{InputParser}
When the framework knows about the location of the datasets it needs to process, the configured \verb|InputParserFactory|\footnote{it.neef.tub.dima.ba.imwa.impl.factories.parser} will be used to instantiate a new \verb|InputParser|\footnote{it.neef.tub.dima.ba.imwa.impl.parser}. The InputParser's \verb|.run()| method accomplishes three major tasks:
\begin{enumerate}
	\item Instruct the parsing of the dumps and pageviews
	\item Merge the Page and Pageview DataSets and apply the \verb|PostFilter|s
	\item Populate the DataHolder with the resulting data
\end{enumerate}

\begin{figure}[htbp]
	\includegraphics[width=\linewidth]{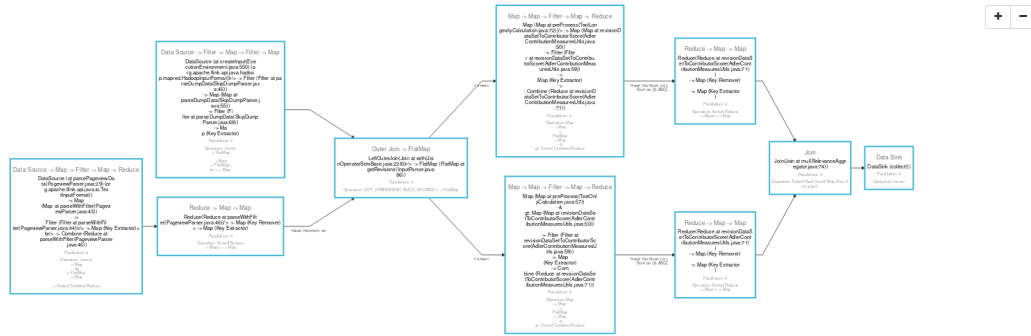}
	\caption{It shows Apache Flink's execution plan for the TextLongevityCalculation, which is generated before the distributed processing starts.}
	\label{fig:apacheflink-executionplan-textlongevity}
\end{figure}

The individual methods for each task need to be called in a specific order. To prevent accidental changes, the \verb|.run()| method is implemented by the abstract class \verb|AInputParser|\footnote{it.neef.tub.dima.ba.imwa.interfaces.parser}. It will call \verb|runDumpParser()| and \verb|runPageviewParser()| for the first task. Those methods then use classes that implement the \verb|IDumpParser| and \verb|IPageviewParser| interfaces (both are in the package\footnote{it.neef.tub.dima.ba.imwa.interfaces.factories.parser}) and are obtained by their respective factories from the configuration to process the two input datasets in parallel. Figure \ref{fig:apacheflink-executionplan-textlongevity} shows Apache Flink's execution plan for a run of the framework's TextLongevityCalculation. On the left side of the centered box, the dump (upper box) and pageviews parsing (lower two boxes) is placed on two different paths, representing two detached data flows.  The box in the middle displays the merging process of the Page and Pageview dataset. The data flow then continues with the usage of the transformed data in calculations as the right side of the figure shows.

After the datasets have been transformed by the parsers, which will be explained in the following paragraphs, the Page and the Pageview objects are still in two separate DataSets. This changes with a call to \verb|mergeDumpAndPageviewData()| right before the abstract class applies the PostFilters in the \verb|applyPostFilters| method. A left outer join of the Page with the Pageview DataSet is performed using the page titles as the key. The \verb|PageAndPageviewJoinFunction| simply assigns a Pageview object to its associated Page object if a match exists. Otherwise it stays \verb|null|. 
At this point there is still only one huge DataSet with Page objects, but in theory it is in a state which can be used for further processing. To speed up the overall performance by removing unwanted objects from this DataSet, it can be refined with \verb|PostFilter|s. The result is then divided into the Page, Revision and Contributor DataSet in the \verb|populateDataHolder()| method, which also takes care of setting the references to those DataSets in the global DataHolder.

\subparagraph{PostFilter}
It should be evident that processing less elements results in less computation time. Therefore the framework offers two primary ways to filter unnecessary objects. One is the filters that can be used to reduce the combined Page DataSet after (``post``) the parsing process has finished are called \verb|PostFilter|s. They must implement the \verb|IPostFilter|\footnote{it.neef.tub.dima.ba.imwa.interfaces.filters.post} interface which defines a \verb|filter(IPage page)| function. It takes a Page object from the merged DataSet as a parameter and returns a boolean decision: If the object fulfills the criterion and should be further processed, then \verb|true| must be returned. A return value of \verb|false| removes the object from the DataSet and any other processing by the framework. The configuration exposes a method to access the list of all PostFilters (line 8) and a new filter should be passed through the \verb|CustomPostFilterFactory|\footnote{it.neef.tub.dima.ba.imwa.impl.factories.filters.post} as shown in listing \ref{code:example-postfilter}. When multiple filters are defined, all will be applied in the same order. There are two methods of filtering: One is using only one \verb|.filter()| call on the DataSet and looping over all PostFilters within it for every element (``inner loop``). The second method is using one loop for all filters and multiple \verb|.filter()| calls on the DataSet (``outer loop``). Listing \ref{code:ainputparser-applypostfilters} is the ``outer loop`` implementation for the PostFilters. An argument for the latter approach is that every subsequent PostFilter will only test against the resulting smaller DataSet of the previous filter. At the time of the framework's implementation, the decision fell on the ``outer loop`` variant. The later conducted comparison of both methods (section \ref{sec:evaluatin-adlerscontributionmeasures-optimizingfilterloops}) does not reveal any significant performance discrepancies on average.
\begin{figure}
	\begin{lstlisting}[caption={Shows how to configure and apply a PostFilter to filter pages from the main/article Wikipedia namespace.}, label={code:example-postfilter}]
fw.getConfiguration().getPostFilters().add(fw.getConfiguration().getCustomPostFilterFactory().newICustomPostFilter(new NameSpacePostFilter()));
// [...] 
static class NameSpacePostFilter implements IPostFilter {
	@Override
	public boolean filter(IPage page) {
		return page.getNameSpace() == 0;
	}
}
	\end{lstlisting}
\end{figure}
\begin{figure}[htb]
	\begin{lstlisting}[caption={Both Pre- and PostFilters are applied subsequently on the whole DataSet to reduce the amount of checked objects.},label={code:ainputparser-applypostfilters}]
final List<IPostFilter> filterList = Framework.getInstance().getConfiguration().getPostFilters();
for (final IPostFilter filter : filterList) {
	filteredIpageDataSet = filteredIpageDataSet.filter(new FilterFunction<IPage>() {
		@Override
		public boolean filter(IPage iPage) throws Exception {
			return filter.filter(iPage);
		}
	});
}
	\end{lstlisting}
\end{figure}
\subparagraph{DumpParser}
Parsing dumps is done by classes that implement the \verb|IDumpParser| interface. There are three different XML-based DumpParsers that come with the framework: 
\begin{itemize}
	\item \verb|DumpParser|: A parser that parses the dumps as they are given.
	\item \verb|SkipDumpParser|: A parser that honors \citeauthor{Adler:2008:MAC:1822258.1822279} condition of skipping subsequent Revision objects by the same author \citep[p 11]{adler2012wikitrust}.
	\item \verb|RegexSkipDumpParser|: A parser that also honors the revision skipping, but uses regular expressions to extract the information.
\end{itemize}
With the focus on \citeauthor{Adler:2008:MAC:1822258.1822279} work, the SkipDumpParser is used as the default.
It starts to read the dumps in the Hadoop compatibility mode with the help of the \verb|PageXMLInputFormat|\footnote{it.neef.tub.dima.ba.imwa.impl.parser}. This class extends the \verb|XmlInputFormat| which was adapted from the Apache Mahout project \footnote{\url{https://github.com/apache/mahout/blob/master/integration/src/main/java/org/apache/mahout/text/wikipedia/XmlInputFormat.java}} to support compressed input files and split the dumps at the \verb|<page>| XML tag. In section \ref{sec:implementation-compresseddatasets-timeaspect} it was discovered that using compressed input files can lead to a significant performance boost. The resulting DataSet contains the raw XML content of all pages in that wiki. What follows is the application of the \verb|PreFilter|s to filter out unnecessary Wikipedia pages before doing the heavy parsing.
In contrast to the implementation of \citeauthor{Adler:2008:MAC:1822258.1822279}, who use regular expressions to parse out the information, the framework's default setting is to use Java's ``Simple API for XML`` (SAX) from the \verb|org.xml.sax| package. A map function on the raw page XML will pass each page into the static \verb|SkipXMLContentHandler.parseXMLString| along with a small configuration wrapper (\verb|DumpParserConfig|) which provides the necessary Page, Revision and Contributor factories. Before returning the resulting Page DataSet to the InputParser for further processing, invalid (\verb|null|) items are filtered from it. 
The \verb|RegexSkipDumpParser|, however, follows Adler et al their idea of using regular expressions for information retrieval from the raw XML page content. It uses the similar \verb|RegexXMLPageParser.parseXMLString| function and lead to further bug discoveries in WikiTrust, but turned out to be slightly faster than a native XML parser for larger datasets in our evaluation section \ref{sec:evaluatin-adlerscontributionmeasures-optimizingdumpparser}.
After an object has been successfully parsed its \verb|.postParsing()| method must be called. This will allow the object to perform some operations before it is added to the DataSet.

\subparagraph{SkipXMLContentHandler}
The static \verb|parseXMLString| function (listing \ref{code:skipxmlcontenthandler-parseXMLString}) first instantiates a new \verb|XMLReader| (line 5), an \verb|InputSource| (line 6) from the given pageXML string and an instance of the SkipXMLContentHandler itself (line 1). 
Both latter instances are then connected to the xmlReader instance to start the parsing process (lines 9-10). The function will either return all successfully parsed pages as Page objects or \verb|null| if an exception occurred (lines 12-14). Furthermore the SkipXMLContentHandler implements the \verb|ContentHandler| from the \verb|org.xml.sax| package defining two important methods, namely \verb|startElement| and \verb|endElement|. These functions are called when a new XML tag starts or ends. This allows to track the parser's state using a \verb|MODES| enum and \verb|switch| statements to handle the different attribute tags within a \verb|<page>|, \verb|<revision>| or \verb|<contributor>| tag. The \verb|characters| method provides the content between two tags which can be assigned to an appropriate object when a tag ends. Using the state tracking method and the extracted tag values, the SkipXMLContentHandler can build objects as explained in the previous section ``\nameref{sec:implementation-planofimplementation-datarepresentation}``.
The main difference between the SkipXMLContentHandler and the XMLContentHandler is that the former removes subsequent revisions by the same author. Due to the nature of the XML dump and the sequential parsing, the code has to ``look back`` and check if the author of the previously parsed revision equals the currently parsed one.  If this happens, then certain references need to be updated for a successful removal of the previous revision. This might include updating the parent revision's \verb|childRevision| attribute and the current revision's \verb|parentRevision|.
\begin{figure}
	\begin{lstlisting}[caption={Parsing the raw page XML content using Java's SAX library.}, label={code:skipxmlcontenthandler-parseXMLString}]
// Instantiate the SkipXMLContentHandler
SkipXMLContentHandler xmlHandler = new SkipXMLContentHandler(config);
try {
	// Setup a StringReader to read the stream.
	XMLReader xmlReader = XMLReaderFactory.createXMLReader();
	InputSource iSource = new InputSource(new StringReader(pageXML));
	
	// Use the SkipXMLContentHandler to parse the data.
	xmlReader.setContentHandler(xmlHandler);
	xmlReader.parse(iSource);
	
	return xmlHandler.getAllPages();
} catch (IOException | SAXException e) {
	return null;
}
	\end{lstlisting}
\end{figure}

\subparagraph{RegexXMLPageParser}
The regular expression parser is similar to the SAX approach, but instead of creating an InputSource, a \verb|BufferedReader| is instantiated and passed to the internal \verb|parseInput| method. It then reads the given XML content line-wise and compares each line against a set of regular expressions that match a beginning, ending or attribute XML tag. This approach and the regular expressions were adapted from WikiTrust (in \verb|analysis/do_evil.ml|) and extended by expressions, for example for a page's namespace. Internally, Java's \verb|Pattern| and \verb|Matcher| classes provide the regular expression functionality. Analogous to the previous parsing method, the MODE switching is re-used and depends on matching a line with one of the regular expressions using the \verb|hasTag| method. Lines that do not match are probably part of a revision's text segment and captured in a StringBuilder if the \verb|TEXT| mode is active. To return the same parsed Revision objects, the same skipping strategies are applied. However, the text had to be URL-decoded with the \verb|StringEscapeUtils.unescapeXml| function to match the SkipDumpParser output. 

\subparagraph{PreFilter}
PreFilters are similar in their core idea to the PostFilters, but they are applied before parsing process. A PreFilter's \verb|filter| function gets a string with a page's XML representation. A return value of \verb|true| will preserve the page in the DataSet and \verb|false| will discard it. There are three convenience PreFilters that the framework provides:
\begin{itemize}
	\item \verb|RegexPreFilter| filters with regular expressions\footnote{\url{https://docs.oracle.com/javase/8/docs/api/java/util/regex/Pattern.html}}.
	\item \verb|XpathPreFilter| is based on the XML Path language\footnote{\url{https://www.w3.org/TR/xpath/}} (Xpath).
	\item \verb|XqueryPreFilter| uses the XML query language\footnote{\url{https://www.w3.org/TR/xquery/}} (Xquery).
\end{itemize}
All three classes reside in the same package\footnote{it.neef.tub.dima.ba.imwa.impl.filters.pre} and can be easily instantiated by using the similar named factories from the configuration and passing the query string. Line 9 in code listing \ref{code:AdlerContributionMeasuresExample-run} shows how the pages can be filtered by their namespace with a regular expression PreFilter, the  \verb|RegexPreFilter|. The other two filters do not provide a matching functionality, but return elements that were selected using the query. Therefore, the filters return true when the return value is not null or an empty string.
In section \ref{sec:evaluatin-adlerscontributionmeasures-optimizingprefilters} the processing speed of all filtering methods was measured and analyzed. The results show that the fastest PreFilter for our test dumps is the RegexPreFilter.

\subparagraph{PageviewParser}
Classes that accord for the parsing of pageviews should base on the \verb|IPageviewParser| interface, which will make them compatible with the \verb|PageviewParserFactory| and thus accessible from the InputParser. The default \verb|PageviewParser|\footnote{it.neef.tub.dima.ba.imwa.impl.parser} class reads the pageviews line-wise into a String DataSet. Each line represents one pageview in its raw format as described in the paragraph in section \ref{sec:implementation-planofimplementation-datarepresentation}. It also covers the important parsing aspects, such as the page title normalization and the filtering for a specific wiki project using the project name. If no such filter is set and the configuration's return value of \verb|getPageviewDumpShortTag()| is null, then the String comparisons costs can be saved by using a simplified object filter function.

\paragraph{Storing}
When all input data is parsed, it needs to be stored, so that it can be used in the next processing step. This section will explain the orange box of figure \ref{fig:framework-dataflow} containing the \verb|DataHolder|, which implements the \verb|IDataHolder| interface. The framework's default DataHolder was already shortly introduced in section ``\nameref{sec:implementation-planofimplementation-datarepresentation}`` - in essence it is a wrapper class for providing access to the three DataSets with Page, Revision and Contributor objects including convenience methods for adding, removing or retrieving individual objects from the DataSets. However, using those methods is quite expensive due to the nature of Apache Flink's DataSet API and the need of \verb|.filter| calls to either remove or get elements. Retrieving single objects from DataSets is especially expensive performance-wise, because it uses \verb|.collect()| to obtain an element from the DataSet and this function triggers an execution plan generation in Apache Flink \footnote{\url{http://apache-flink-user-mailing-list-archive.2336050.n4.nabble.com/Retrieving-a-single-element-from-a-DataSet-td9731.html\#a10027}}. A better approach is to rely on the Pre-/PostFilters for removing elements or on the inter-object references to remove objects.
Anyway, the advantage of storing the DataSets in a globally accessible object is that multiple Calculation functions can read and work with the data in parallel what is indicated by the three outgoing edges in the figure.

\paragraph{Processing}
All efforts of parsing the data would be worthless, if it does not get used. The blue box in figure \ref{fig:framework-dataflow} models how three calculations A, B and C process the data. Results are aggregated following the formula \verb|(A+B)*C|. But how does the data flow through this part of the framework?
\subparagraph{Calculation}
Classes that extend the \verb|ACalculation|\footnote{it.neef.tub.dima.ba.imwa.interfaces.calc} abstract class are the most important part of the framework, because that is where the data is processed. Some example \verb|Calculation| functions based on the contribution measures by \citeauthor{Adler:2008:MAC:1822258.1822279} are implemented and available in the \verb|it.neef.tub.dima.ba.imwa.examples.adlerscm| package and are explained in more detail in section \ref{sec:evaluatin-adlerscontributionmeasures-measuringcontributions}. Again an abstract class was used to provide a \verb|run| function that executes the following methods in the correct order: 

\begin{itemize}
        \item \verb|init()|: Initialize the calculation (e.g. dependencies)
		\item \verb|preProcess()|: Initialize or prepare the DataSet
		\item \verb|process()|: Do the main calculation and process the DataSet
		\item \verb|postProcess()|: Finalize the calculation or the result DataSet
\end{itemize}
The above rules are not strictly enforced and theoretically everything could be done in one single of those. Still, the paradigm should be applied for better structuring and readability of the code. Since everything revolves around the framework, a Calculation must be passed to the framework's  \verb|runCalculation| method, where \verb|run| is eventually called. The instantiation and submission of an example Calculation is shown in lines 12-15 of code fragment \ref{code:AdlerContributionMeasuresExample-run}.
For example, the contribution measure Calculations use the \verb|preProcess| method to prepare the revisions by calculating the differences using \verb|IDiffer| classes. The resulting values are then transformed into the actual relevance scores in the \verb|process| method and the final RelevanceScoreDataSet is saved in a variable that should be accessible through the \verb|getResult()| function. Eventually a DataSink implemented by an \verb|IOutput| class will process it, e.g. by writing it to a file or printing it out.
Additional objects or arguments can be passed to a Calculation instance with the help of its \verb|setArguments| method and the \verb|ArgumentsBundle|\footnote{\url{it.neef.tub.dima.ba.imwa.impl.calc}}. \verb|addArgument| or \verb|getArgument| are used to manage objects in the bundle. The only requirement is that those are serializable and the key to access them is known. 

\subparagraph{Differ}
If a class implements the \verb|IDiffer|\footnote{\url{it.neef.tub.dima.ba.imwa.interfaces.diff}} interface, it becomes a ``Differ`` which can be used to calculate differences between two revisions ``\verb|current|`` and ``\verb|next|`` that are passed to the \verb|calculateDiff| function. The emitted \verb|IDifference|\footnote{it.neef.tub.dima.ba.imwa.interfaces.data} object reflects the differences. For the example implementations, they are numeric values of type \verb|Double| (\verb|DoubleDifference|\footnote{it.neef.tub.dima.ba.imwa.impl.data}) which can be instantiated by the configured IDifferenceFactory.

\subparagraph{RelevanceAggregator}
Two or more separate Calculations can be combined to create more complex ones. In such situations the framework's \verb|RelevanceAggregator|\footnote{it.neef.tub.dima.ba.imwa.impl.calc} can be used to combine the Calculations resulting RelevanceScoreDataSets. The class is based on the \verb|IRelevanceAggregator|\footnote{it.neef.tub.dima.ba.imwa.interfaces.calc} interface which enforces the implementation of the following functionality:
\begin{enumerate}
	\item Joining two IRelevanceScoreDataSets using item-wise \verb|+|, \verb|-|, \verb|*| or \verb|/|.
	\item Mathematical operations with a scalar value $\alpha$ on an IRelevanceScoreDataSet, such as \texttt{+ $\alpha$}, \texttt{- $\alpha$}, \texttt{* $\alpha$} or \texttt{/ $\alpha$}
	\item Aggregations such as \verb|min|, \verb|max|, \verb|sum| on an IRelevanceScoreDataSet.
\end{enumerate}
The return values are again IRelevanceScoreDataSets, what makes chaining them possible. 
When joining such DataSets, the \verb|.join| operation identifies equal objects by calling their \verb|getIdentifier()| function.
Therefore the aggregated DataSets should be of the same size and contain the same objects, but that is normally the case because of working on the same data provided by the DataHolder. Otherwise the join will omit unmatched items. 
Scalar operations happen in a standard map process, which seamlessly integrates into an execution plan, and care should be taken to not divide by zero.
Aggregations in the form of finding the minimum, maximum or sum of a DataSet are costly, because these use operations that will trigger an execution plan due to their \verb|.collect| call and that might decrease the performance.

\paragraph{Outputting} 
Finally the data needs to leave the framework as shown in the red box in figure \ref{fig:framework-dataflow}. This can be achieved using the default \verb|ConsoleOutput|\footnote{it.neef.tub.dima.ba.imwa.impl.output} which will print the resulting relevance scores on the console or with other classes that implement the \verb|IOutput|\footnote{it.neef.tub.dima.ba.imwa.interfaces.output} interface. Each such class will have two functions called \verb|output| that will either accept an ACalculation or a DataSet as a parameter. In the former case \verb|getResult| is used to obtain the final result DataSet. Usually those DataSets can be assumed to be of the type IRelevanceScore. The ConsoleOutput is a wrapper around a DataSet's \verb|.print| method and its usage can be seen in lines 4 or 16 of listing \ref{code:AdlerContributionMeasuresExample-run}.

\paragraph{Execution plan}
Before Apache Flink starts any processing it compiles an execution plan that might look similar to the diagram in figure \ref{fig:apacheflink-executionplan-textlongevity}. The plan begins with a DataSource such as our DumpParser or PageviewParser where data is read using the configuration's \verb|getExecutionEnvironment()|. Flink then continues to collect all DataSet-API calls like \verb|.filter|, \verb|.map|, \verb|.reduce| and incorporates them into its plan until it finds a function call that triggers an execution of the plan:
\begin{itemize}
	\item \verb|.print()|
	\item \verb|.collect()| or \verb|.count()|
	\item \verb|.execute()|
\end{itemize}
The first two implicitly trigger an execution of the created plan. Other DataSinks need to be explicitly executed by a call to \verb|.execute()| \cite{flinkdocs010}. When an execution is triggered, Apache Flink will start to distribute and process the program on the configured cluster.

\subsubsection{Code structure}\label{sec:implementation-planofimplementation-codestructure}
To better understand the implementation and usage of the framework, the code structure and its characteristics will be briefly covered in this section.
The whole code is part of the \verb|it.neef.tub.dima.ba.imwa| package, which further contains three packages and two Java files:
\begin{itemize}
	\item \verb|examples|: A package for examples using the framework.
	\item \verb|impl|: A package containing the (default) implementation of certain necessary interfaces and classes.
	\item \verb|interfaces|: A package consisting of all abstract class interfaces that should be used with the framework.
	\item \verb|Framework.java|: The framework's outermost layer.
	\item \verb|Job.java|: It contains Java's entrypoint ``\verb|main|`` which runs the \\ \verb|AdlerContributionMeasuresExample|
\end{itemize}
A singleton instance of the \verb|Framework| object is available through its \verb|getInstance()| function. On its first initialization it will create the Configuration object and set various default settings. Further essential functionality are the \verb|init| for initializing the parsing and numerous methods with a \verb|run| prefix for starting Apache Flink's execution plan.

As listed above, the package \verb|interface| is about the framework's interfaces and it contains packages, with each having a specific topic and optionally more sub-packages:
\begin{itemize}
	\item \verb|calc|: Interfaces for the Calculation, RelevanceAggregator and ArgumentBundle
	\item \verb|data|: Interfaces for all data classes, including all with the RelevanceScores suffix and the DataHolder
	\item \verb|diff|: The IDiffer interface
	\item \verb|factories|: Same package structure (except for ``calc``) with factories for most of the interfaces.
	\item \verb|filters|: Interfaces for Pre- and PostFilters in separate packages.
	\item \verb|output|: Contains the IOutput interface
	\item \verb|parser|: Parser related interfaces
\end{itemize}
The second package, \verb|impl|, implements most of those interfaces. It adheres to the same package structure, except that an additional package ``\verb|configuration|`` with the Configuration class exists. 
To test and evaluate the framework's performance, \citeauthor{Adler:2008:MAC:1822258.1822279} contribution measures were implemented with the help of it. All six measures are placed in separate packages correspondent to their name in the \verb|examples.adlerscm| package. 

Another important piece of code is in the \verb|Job.java|. It contains Java's entrypoint, the \verb|public void main(String[] args)| function that will be called when the packaged .jar-file is submitted using \verb|flink run|. It instantiates an \verb|AdlerContributionMeasuresExample| object, which then uses the functionality provided by the framework. In fact, the framework cannot exist as a program by itself, because it lacks such a main function. Therefore, it only has a supportive role in the sense of providing a frame for other's work. 

Building the framework with the aforementioned structure was considered to improve its readability and comprehensibility. The naming of the packages by their respective topic should ensure that the correct classes and interfaces can be easily found. In a similar fashion, the implemented classes' naming follows the convention of having a suffix of its interface. 

\begin{myboxi}[Key points of section ``\nameref{sec:implementation-planofimplementation}``]
	\begin{itemize}
		\item Several classes represent the dump and pageviews information with their necessary relationships. All calculations are done on those objects.
		\item The four processing steps consist of multiple components that are responsible for the data processing flow.
		\item The framework's source code is separated into packages for better readability and usability.
	\end{itemize}
\end{myboxi}

\subsection{Functionality and extensibility} \label{sec:implementation-functionalityextensibility}
This section will give an overview from the outside perspective of a user. It is assumed that the user wants to use the framework for its primary developed purpose - the calculation of Wikipedia author rankings. When referring to a ``user`` of the framework, it describes a person that uses it to develop software for achieving their desired goal, for example calculating author rankings.
   
\subsubsection{Functionality} \label{sec:implementation-functionalityextensibility-functionality}
The previous section ``\nameref{sec:implementation-planofimplementation}`` covered the framework's internal processes and data flow that is partly hidden from the user. A detailed insight into the framework's functionality was given in that section. 

\paragraph{Abstraction layer}
The whole parsing process, including dump and pageview parsing, is completely invisible to the user if she decides to use the default parsers. Same applies for running and printing out calculation results. In that regard, the framework poses as an abstraction layer to remove complexity from the user. The only part that she needs to implement are the Calculation classes if the implemented contribution measures are not enough. In that case she will need to interact with the DataHolder, which already provides all necessary DataSets, and the DataSet-API to operate on them. 

\paragraph{Automated tests}
To ensure that the flexible components of the framework work as expected, automated tests with JUnit 4 were implemented. By automatically executing them with every build, changes that modify the framework's functionality in a negative way can be identified. Due to Apache Flink's DataSet-API requiring an Apache Flink environment to function, the flink-spector\footnote{\url{https://github.com/ottogroup/flink-spector}} project was necessary to mimic such an environment. At the time of writing, the Apache Maven repository only contained a version for Flink 1.2.1, what hindered the implementation of certain tests due to the newer version's type handling. 

\subsubsection{Extensibility} \label{sec:implementation-functionalityextensibility-extensibility}
Throughout the previous sections it was mentioned that the framework provides default implementations and classes for a lot of functionality. The reasoning behind this is to deliver a framework that can be used with minimal effort by providing appropriate defaults, but it actually is designed to be flexible. The defaults might be biased towards the implementation of \citeauthor{Adler:2008:MAC:1822258.1822279} contribution measures. However, 
the goal was to not limit or restrict it to only work with specific predefined classes, but to allow for extensibility and adaptation to a user's needs. That is realized using features of the Java language and programming patterns:
\begin{itemize}
	\item Interfaces and abstract classes
	\item Factories
	\item Singletons
\end{itemize}

\paragraph{Interface}
Interfaces are a special type that are used as a ``contract between the class and the outside world`` as well as a way to define ``the behavior it promises to provide`` \cite{docsoracleinterface}. Such ``contracts`` are used throughout the framework. All relevant interfaces were described in section ``\nameref{sec:implementation-planofimplementation-codestructure}`` and can be found in the \verb|interfaces| package. The framework uses them to define method signatures. Code listing \ref{code:icalculation-interface} shows the interface that all Calculation classes have to implement.
\begin{figure}[htbp]
	\begin{lstlisting}[caption={The ICalculation interface defines, which methods a Calculation class has to implement.}, label={code:icalculation-interface}]
	public interface ICalculation<T extends IDataType, S extends Double, B extends IDataType, A extends IArgumentsBundle> extends Serializable {
	
	A getArguments();
	void setArguments(A arguments);
	
	DataSet<? extends IRelevanceScore<T, S>> getResult();
	DataSet<? extends IRelevanceScore<B, S>> getBaseResult();
	
	void init();
	void preProcess();
	void process();
	void postProcess();
	void run();
	}
	\end{lstlisting}
\end{figure}
An interface cannot be instantiated by itself. It must be implemented by a class, but an interface can extend other interfaces to combine their behavior definitions \cite{docsoracleinterface2}. This approach has the advantage that the framework's underlying classes can define functions and methods that take objects, which implement specific interfaces, as parameters or return values. Lines 3,4 show such methods.
The type \verb|A| is a ``Generic Type``\footnote{\url{https://docs.oracle.com/javase/tutorial/java/generics/types.html}} that is defined as \verb|A extends IArgumentsBundle| meaning that only objects of a type which extends the IArgumentsBundle are allowed as a parameter for those function. Generic types can make the source code more readable. Additionally they might help with error detection at compile-time, because the compiler can more precisely infer an object's type \cite{docsoraclegenerics}. Anyway, the framework's extensibility benefits from interfaces, because functions do not know the object's exact type, but have a promise that the object implements certain well-defined methods. For example, the DumpParser and SkipDumpParser both implement the IDumpParser interface and must therefor provide the method body for \verb|parseDumpData|. But the InputParser does not know nor care about the actual implementation when calling the method and is satisfied when the following call to \verb|getPages| returns the parsed data. Similarly the XML parser only knows that the objects have certain methods to assign the values, but not their internal structure or implementation.

\paragraph{Abstract class}
On several occasions ``abstract classes`` were mentioned. That particular type is similar to interfaces, but allows to predefine method bodies. The source code in listing \ref{code:acalculation-abstractclass} presents the ACalculation class with its \verb|run| method. It implements the previously discussed ICalculation interface. By setting the framework's runCalculation parameter type to ACalculation all calculations will have to implement all methods from the interface, except the run method, which had already been implemented by the abstract class. This approach aims at hiding the methods from the users, so that the four processing steps happen in the correct order. It was also used in the InputParser with the AInputParser abstract class, so that important processing steps cannot accidentally be changed. Overall this is another beneficial addition, because it increases the ease of use and lowers the barrier of entry for custom implementations.
\begin{figure}
	\begin{lstlisting}[caption={The ACalculation abstract class, which implements the run function to ensure a particular call order.}, label={code:acalculation-abstractclass}]
	public abstract class ACalculation<T extends IDataType, S extends Double, B extends IDataType, A extends IArgumentsBundle> implements ICalculation<T, S, B, A> {

	
		/**
		* Runs all processing steps in the right order.
		*
		* @see ICalculation#run()
		*/
		@Override
		public void run() {
			this.init();
			this.preProcess();
			this.process();
			this.postProcess();
		}
	}
	\end{lstlisting}
\end{figure}
\paragraph{Factory Pattern}
Although the interfaces and abstract classes facilitate the customization of nearly all components of the framework, changing a data structure like a Revision class would require more changes in the code. That is, because it is instantiated during the parsing process in the SkipXMLContentHandler. If this instantiation was to be fixed, e.g. \verb|IRevision rev = new Revision()|, then changing the invocation to e.g. \verb|IRevision rev = new CustomRevision()| would require a re-implementation of the SkipContentHandler. That in turn requires a new implementation of the DumpParser. This ``chain`` of rewriting would continue up until all type declarations match again. It is not far-fetched that this is not feasible nor in the spirit of a good framework. 
To combat this issue, factories and the global configuration (section \ref{sec:implementation-planofimplementation-dataprocessing}) were introduced. The code snippet \ref{code:revisionfactory} shows that factories are normal Java classes whose only task is to instantiate new objects of a specific type. The trick is that the return value's type is the shared interface of all possible objects of this category. The factories are accessible through the global configuration and the appropriate getter-method. This reduces the location where an instantiation needs to be changed to exactly one line in the factory, because all other classes can use \verb|IRevision cRevision =  Framework.getInstance() .getConfiguration().getRevisionFactory().newRevision();| to obtain revisions with the necessary IRevision base type.
If another class is needed, a new rather simple factory must be implemented and then configured globally using the configuration's suitable setter. In this example it would be \verb|setRevisionFactory(new RevisionFactory())|.
Similar to the interfaces, such factory patterns are implemented throughout the framework for most core components. 

With the last-mentioned features, the framework should give enough freedom for extensibility of its functionality and further use cases in the future.
\begin{figure}
	\begin{lstlisting}[caption={The RevisionFactory for Revision objects as an example for an object factory.}, label={code:revisionfactory}]
public class RevisionFactory implements IRevisionFactory<IRevision> {
	@Override
	public IRevision newRevision() {
		return new Revision();
	}
}
	\end{lstlisting}
\end{figure}
\paragraph{Singleton Pattern}
The framework uses a Singleton pattern to ensure that there is only one instance of the Framework class. Listing \ref{code:singleton-framework} shows the crucial lines of code. The reasoning behind it is that otherwise the DataSet operations and Apache Flink's execution plan might mix up between multiple framework instances. To prevent this, the Framework class has a static variable \verb|instance| with an instance of itself that can only be accessed with the \verb|getInstance()| method. As another precaution the constructor is \verb|private|, so that only the class can instantiate itself. 
The \verb|final| keyword of the \verb|instance| and \verb|configuration| variables enforces that they can only be assigned once.
\begin{figure}
	\begin{lstlisting}[caption={The Framework's singleton implementation for global access to the Framework instance.}, label={code:singleton-framework}]
/**
* Singleton instance of the framework.
*/
private final static Framework instance = new Framework();
/**
* The framework's configuration object.
*/
private final Configuration configuration;

private Framework() {
	this.configuration = new Configuration();
	//[...]
}
public static Framework getInstance() {
	return Framework.instance;
}
	\end{lstlisting}
\end{figure}
\begin{myboxi}[Key points of section ``\nameref{sec:implementation-functionalityextensibility}``]
	\begin{itemize}
		\item The parsing is hidden from the user and automated tests verify its correctness.
		\item Extensive usage of interfaces, abstract classes, factory and singleton patterns are the framework's basis for its flexibility and extensibility.
	\end{itemize}
\end{myboxi}

\subsection{Problems and solutions} \label{sec:implementation-problemsandsolutions}
During the development and evaluation process of the framework, several problems and bugs were encountered. Some of them were easy to resolve or work around, but others were rooted deeper in the dependencies. Fixing the latter ones would require too much effort and time and are therefore discussed for future work.

\subsubsection{Object serialization} \label{sec:implementation-problemsandsolutions-objectserial}
When running distributed processes across a cluster with more than one node, data has to be transferred between them. That means the data, e.g. Java objects, must be transformed from memory into a suitable representation, then transferred to other node and re-instantiated. The receiving node then reverts the transformation to rebuild the same object in memory. This process is called serialization and deserialization. Apache Flink tries to handle it transparently and falls back to the Kryo\footnote{\url{https://github.com/EsotericSoftware/kryo}} serializer if it cannot \cite{flinkdocsserialization}. 

In the early stages of the framework's development Apache Flink was not able to serialize the data objects used in the DataSets, most likely due to their private methods and them not extending the \verb|Serializable| interface. This was addressed with the \verb|IDataType| interface. It extends the aforementioned Serializable interface and should be implemented by all objects that are used with Apache Flink's DataSet-API. This will allow Flink to determine that the object can be serialized, even when falling back to Kryo is necessary.

\subsubsection{Incomplete XML parsing} \label{sec:implementation-problemsandsolutions-incompletexml}
One of the dump parsers was implemented using the SAX XML ContentHandler class. When data between two XML tags is encountered, the \verb|public void characters(char[] chars, int i, int i1) throws SAXException| will be called. It was first assumed that the method returns all data in a single chunk. However reviewing the parsed XML dumps lead to the discovery that some revisions contained truncated text. Indeed, the documentation says that ``SAX parsers may return all contiguous character data in a single chunk, or they may split it into several chunks; however, all of the characters in any single event must come from the same external entity so that the Locator provides useful information.`` \cite{saxcontenthandler} To prevent data loss by ignoring continuous chunks, a \verb|StringBuilder| instance concatenates all data chunks into one string. After an entity is parsed and the full data assigned as a property, the StringBuilder's buffer is reset to length zero for the next entity.
This solution allows to capture all content between the XML tags in its full length. 

\subsubsection{Bzip2 decompression} \label{sec:implementation-problemsandsolutions-bzip2}

The biggest problem throughout the development and evaluation of the framework was related to bzip2 decompression. While investigating compressed datasets and their decompression performance in section \ref{sec:implementation-compresseddatasets} the compatibility bridge between Apache Flink and Apache Hadoop allowed to decompress the data. The \verb|TextInputFormat| in \verb|org.apache.hadoop.mapred| returns a \verb|LineRecordReader| that is capable of handling compressed input data. Those classes were used in the decompression performance testing program. 

Parsing the dumps was realized by splitting the XML data at \verb|<page>| tags. The Apache Mahout project implemented a suitable XmlInputFormat\footnote{\url{https://github.com/apache/mahout/blob/master/integration/src/main/java/org/apache/mahout/text/wikipedia/XmlInputFormat.java}} class that splits Wikipedia dumps. However, its \verb|XmlRecordReader| implementation expects uncompressed input data, because it lacks decompression code for handling compressed data. 
The approach that seemed feasible was to extend Apache Mahout's XmlRecordReader with the appropriate compressed data handling wrappers from the LineRecordReader. However, the final source code introduced some bugs that lead to some severe issues with the framework. Unfortunately, even the community at the Apache Flink mailing list could not further explain nor identify the root cause \footnote{\url{http://apache-flink-user-mailing-list-archive.2336050.n4.nabble.com/Reading-compressed-XML-data-td10985.html}}. 

\paragraph{Losing revisions} \label{sec:implementation-problemsandsolutions-losingrevisions}
Tests with dumps, which are several hundred megabytes in compressed size, revealed that the amount of revisions does not match when compared to the same, but decompressed dump. In all tested cases, processing compressed dumps resulted in a lower number. This was not observable with small compressed dumps (<1MB) and their uncompressed version during the development. For example the uncompressed bgwiktionary-20160111 dump has 901286 revisions, but processing the compressed version results in only 898669 extracted revisions. That is a loss of about 0.290 \%. For the acewiki-20170501 dump, a significant loss of \textasciitilde 5.7\% happens.
There might be a mistake when trying to find the correct boundaries of a page and thus partially skipping pages or some revisions within it. Losing revision has a direct effect on the computation of the author ranking. Revisions that might lift the score for a good author or further discredit a vandal won't be processed leading to inaccurate author rankings.

After further investigation of this bug, the problem was identified and a partial fix implemented. The XmlInputFormat indeed fails to correctly determine the boundaries or positions of <page> tags. That results in some page XML strings being emitted into a DataSet multiple times leading to duplicate content and revisions. It was resolved with a \verb|.distinct()| operation on XmlInputFormat's result DataSet. It will remove all duplicates. This might have an impact on the framework's performance, but a higher accuracy in correctly parsing and reproducing \citeauthor{Adler:2008:MAC:1822258.1822279} contribution measures is more important. Otherwise a direct performance comparison is not possible. With the resolution in place, the uncompressed dumps resulted in the same amount of revisions like their compressed counterpart. The reverse conclusion is that the decompression worked flawlessly, but the ``reference`` decompressed processing had flaws.

\paragraph{Memory issues} \label{sec:implementation-problemsandsolutions-memoryissues}
This was another issue that arose after the development when running the framework on the cluster with datasets of moderate size (e.g. warwiki-20170501). Certain compressed dumps lead to an error related to Java's VM limits causing the whole distributed processing job to fail. Listing \ref{code:compresseddump-oom} shows the first part of Java's stack strace which indicates that the \verb|DataOutputStream|'s byte array cannot be extended, because the new size would exceed the allowed limit. 
\begin{figure}[htbp]
	\begin{lstlisting}[caption={Java's ``JVM Out of memory`` exception that occurred for bigger dumps.}, label={code:compresseddump-oom}]
	02/19/2017 22:20:12     Job execution switched to status FAILING.
	java.lang.OutOfMemoryError: Requested array size exceeds VM limit
	at java.util.Arrays.copyOf(Arrays.java:3236)
	at java.io.ByteArrayOutputStream.grow(ByteArrayOutputStream.java:118)
	at java.io.ByteArrayOutputStream.ensureCapacity(ByteArrayOutputStream.java:93)
	at java.io.ByteArrayOutputStream.write(ByteArrayOutputStream.java:135)
	at java.io.DataOutputStream.write(DataOutputStream.java:88)
	at imwa.impl.parser.XmlInputFormat$XmlRecordReader.readUntilMatch(XmlInputFormat.java:114)
	\end{lstlisting}
\end{figure}
The Apache Spark community discusses the same error message in a post\footnote{\url{http://apache-spark-user-list.1001560.n3.nabble.com/java-lang-OutOfMemoryError-Requested-array-size-exceeds-VM-limit-td16809.html}} on their mailing list. Apparently this is a technical limitation of the JVM, because it cannot create arrays that are bigger than their index. For indexes of type signed integer in a 32bit JVM this would equal to arrays with a maximum size of $\lfloor \frac{2^{32}-1}{2} \rfloor = 2147483647$ elements. The same number can be found as a length check in the aforementioned LineRecordReader in the \verb|maxBytesToComsume| function (see listing \ref{code:linerecordreader-lengthcheck}). Taking into account the \verb|byte| array type and the therefore imposed limit of $\frac{2147483647}{1024 * 1024} \approx 2048$ megabytes per XML page, processing pages of such sizes might not be possible with the framework. 
No further investigation was done to see if the memory issue correlates to the loss of revisions.
\begin{figure}[htbp]
	\begin{lstlisting}[caption={Length checks in Apache Hadoop's LineRecordReader that might prevent the observed error.}, label={code:linerecordreader-lengthcheck}]
	private int maxBytesToConsume(long pos) {
	return this.isCompressedInput?2147483647:(int)Math.min(2147483647L, this.end - pos);
	}
	\end{lstlisting}
\end{figure}
After a while a possible cause was identified: The framework's \verb|pom.xml| contained a directive that enforced a compilation with Java 1.7. Changing the value to Java 1.8 did not fully resolve the problem, but shifted it somewhere else. The new error messages were similar to \verb|Cannot write record to fresh sort buffer. Record too large.| or \verb|Thread 'SortMerger Reading Thread' terminated due to an exception: null|. A significant portion of time was invested into analyzing and debugging those issues, but ultimately no sufficiently satisfying solution was found. Further observations and resolution ideas are discussed in section \ref{sec:discussion}.

In general, to prevent memory issues or revision loss, the dumps could be split into smaller datasets at <page> tags in a preprocessing step. After running the calculations on the separated smaller files, the resulting scores could be accumulated, because the calculations happen on a page basis in theory. The downside of this solution is that it would require costly preprocessing steps and most likely overhead for submitting a job to the cluster multiple times. Additionally, it would restrict the impact measures to ones, where accumulation over several separated pages is possible.

\subsubsection{Page-Pageview merging} \label{sec:implementation-problemsandsolutions-pageviewmerging}
Section \ref{sec:implementation-planofimplementation-dataprocessing} explains that the Page and Pageview DataSets are merged so that pages obtain information about their traffic. At first this was realized with a standard \verb|.join| operation on the Page DataSet. The drawback of using \verb|join| is that the resulting DataSet only contains Page objects that matched with at least one Pageview object. It is not guaranteed that all pages of a wiki receive traffic within an hour, especially when a narrow time frame and therefore small pageview dataset is used. Nevertheless, those unvisited pages are created and edited by authors and therefore need to be present for correct ranking calculations. 

That is why a left outer join was implemented as the solution. In opposite to a normal join it will preserve all page objects that did not match and assign a value of \verb|null|. A calculation depending on the pageview values could then decide how to handle such cases. It could either remove those pages with an intermediate filter step or handle the value \verb|null| as zero, meaning that no traffic was generated on a particular page.

\subsubsection{Differ Factories} \label{sec:implementation-problemsandsolutions-differfactories}
The initial framework's planning phase contained only one Differ and a related factory, because it was assumed that the basis for all calculations would be the same differences between two revisions. While implementing the contribution measures by \citeauthor{Adler:2008:MAC:1822258.1822279} it was realized that this approach does not suite well. Most of the contribution measures require different input values - the revision differences - for their calculation. This would require a factory and Differ class change every time another calculation would be used, what could lead to a more complex system with less usability for the user.  
To accommodate for that situation the central difference calculation was put aside and is now a part of each calculation's preprocessing step. It comes with the advantage that a Calculation class stands for itself and can be exchanged easily.

\subsubsection{Cluster configuration} \label{sec:implementation-problemsandsolutions-clusterconfig}
On the student cluster the \verb|Peel Framework|\footnote{\url{http://peel-framework.org/}} is used to manage the Apache Flink configuration on all nodes. It automatically configures and starts all necessary dependencies and instances, thus abstracting from the configuration layer. While working with the framework on the cluster, several adjustments to the cluster's configuration were needed.

\paragraph{Akka timeout}
After implementing all contribution measures and executing them on the cluster, it would eventually fail with an error message saying that an ``ask timed out`` (listing \ref{code:akka-askedtimeout}).
\begin{figure}[htbp]
	\begin{lstlisting}[language={},caption={Important lines of the Akka ``ask timed out`` error message when the job execution fails.}, label={code:akka-askedtimeout}]
	org.apache.flink.client.program.ProgramInvocationException: The program execution failed: Job execution failed. 
	[...] 
	Caused by: org.apache.flink.runtime.client.JobExecutionException: Job execution failed. 
	[...] 
	Caused by: java.lang.IllegalStateException: Update task on instance a69314c8c1892b9bc52c23ef675202fc @ ibm-power-3 - 64 slots - URL: akka.tcp://flink@130.149.21.80:6122/user/taskmanager failed due to: 
	[...]
	Caused by: akka.pattern.AskTimeoutException: Ask timed out on [Actor[akka.tcp://flink@130.149.21.80:6122/user/taskmanager#689599963]] after [10000 ms]
	[...]
	\end{lstlisting}
\end{figure}
From the error message one learns that it is related to \verb|Akka|\footnote{\url{https://cwiki.apache.org/confluence/display/FLINK/Akka+and+Actors}}. Akka helps Apache Flink to communicate with all distributed nodes. Peel configures a value of 10,000 milliseconds per default, but that was not enough for successful operation. Increasing the value to 50,000 by setting \verb|akka.ask.timeout| resolved the timeout issue. 

\paragraph{Network buffer}
Another issue that was resolved by adjusting the cluster configuration was the amount of network buffers. It correlated with the \verb|paralellism| value. When executing the framework with a parallelism higher than 100 a \verb|Insufficient number of network buffers| exception was thrown. Trial and error adjustments of the options \verb|taskmanager.network.numberOfBuffers| and \verb|taskmanager.network.bufferSizeInBytes| to a value of 262144 removed the limitations and allowed to use a parallelism up to 400.

\paragraph{Shared access}
The fact that the cluster was shared between students lead to problems that are not immediately related to the implementation, but of organizational nature that influenced the development and testing of the framework.
For reliable results and performance only one person was allowed to use the cluster at a time. A slack\footnote{\url{https://slack.com}} channel was used to request and assign a time slot. The fair-use rule applied, so prolonged occupation of the cluster was not welcomed, what made tests on larger datasets difficult due to the needed amount of processing time. Therefore, smaller dumps had to be used and one had to have a well thought out plan on which tasks to run.
Consistency of those datasets throughout the development was hard to maintain, because they were deleted several times from the Apache HDFS filesystem. Either by a student by accident or by the administration without prior notice. In the end re-downloading the datasets was necessary, thus leading to newer and slightly different dumps due to more revisions, when the older ones were not available anymore. This is the main reason why the datasets change throughout the thesis and why smaller ones were preferred not only due to shorter download times.
In rare cases individual nodes of the cluster went offline and had to be manually restarted by an administrator, what could take up to several days. Without all nodes, the cluster's resources decrease and render the comparison of repeated performance tests impossible.

\subsubsection{Differences to \citeauthor{Adler:2008:MAC:1822258.1822279}} \label{sec:implementation-problemsandsolutions-differencestoadler}
During the preparation for the evaluation part of the thesis, both the framework and \citeauthor{Adler:2008:MAC:1822258.1822279} WikiTrust system were run on different dumps. Surprisingly the resulting scores did not match, what caused a deeper investigation of the parsing process. Adjusting the tools to output which pages and revisions were analyzed allowed to resolve all of the following mismatches:

\paragraph{Pages}
A handful of revisions that the framework included, but WikiTrust omitted belonged to pages with a \verb|<redirect>| tag. After removing all pages with that characteristic, the framework was missing revisions that WikiTrust preserved. Further analyzing all details revealed that only pages with the aforementioned tag and a colon in the page's title were removed. The responsible dump parsers were modified to abort the whole parsing of a page, if it matches the criteria.

\paragraph{Revisions}
Further analysis showed that the framework didn't handle certain cases very well. For example a page may have revisions where the author has been deleted and the XML tag is shortened to \verb|<contributor deleted="deleted" />|. The parser could not find a username nor an IP address and assumed them to be null. This in turn prevented successful skipping, when two such consecutive revisions appeared. 
WikiTrust's revision skipping implementation is based on the author's ID, whereas the framework checks usernames. The latter method should yield the same results, because the ``Create account``\footnote{\url{https://en.wikipedia.org/wiki/Special:CreateAccount}} page refuses duplicate usernames with the error message ``Username entered already in use. Please choose a different name.`` and ``once a username has been changed, existing contributions will be listed under the new name in page histories, diffs, logs, and user contributions.`` \cite{wikinaming} In theory one username maps to one ID and vice versa. However, the acewiki-20170501 dump contains revisions by different authors, but with the same ID of 0. Listing \ref{code:dump-different-usernames-ids} shows an excerpt from the dump. WikiTrust only compares the IDs and therefore removes revision 1050 from further analysis. It is not clear which approach the right one is, but for better approximation of the WikiTrust program, the Framework now checks the IDs first. 
Another matching problem was based on the IP-addresses for anonymous users. The assumption was that those authors are treated the same and two consecutive anonymous revisions lead to a skipping step. But at this stage, WikiTrust still separates all revisions, even taking IP-addresses individually and only skipping a revision if its successor has the same IP. The framework was modified to respect this behavior, too. 
\begin{figure}
	\begin{lstlisting}[caption={The listing shows parts of the acewiki-20170501 dump, where different usernames map to the same ID.}, label={code:dump-different-usernames-ids}]
    <revision>
		<id>1050</id>
		<parentid>1049</parentid>
		<timestamp>2008-08-26T04:14:43Z</timestamp>
		<contributor>
			<username>Lam Tamot</username>
			<id>0</id>
		</contributor>
	[...]
	</revision>
	<revision>
		<id>1051</id>
		<parentid>1050</parentid>
		<timestamp>2008-08-26T05:04:25Z</timestamp>
		<contributor>
			<username>Keuramat</username>
			<id>0</id>
		</contributor>
	[...]
	</revision>
	\end{lstlisting}
\end{figure}
All bugs were resolved by changing the SkipXMLContentHandler and RegexSkipDumpParser to correctly skip revisions. Those bugs began to surface with  larger dumps like acewiki-20170501. In contrast the aawiki-20170501 dump, which was extensively used during the development phase, does not have enough revisions or pages for all edge cases to occur. Additionally, bgwiktionary-20170501 served as an isolated verification instance after adjusting the framework. 

Another misbehavior that we believe to be a bug in WikiTrust, was discovered while implementing the RegexXMLPageParser. The regular expression for the start of a revision's text segment is \verb|<text xml:space="preserve">|. However, when downloading the history for a single page, like ``David Siegel (screenwriter)``\footnote{\url{https://en.wikipedia.org/wiki/David_Siegel_(screenwriter)}} through Wikipedia's ``Export pages``-tool\footnote{\url{https://en.wikipedia.org/wiki/Special:Export}}, the text segments will contain another attribute indicating the amount of bytes \verb|<text xml:space="preserve" bytes="540">|.
Therefore, WikiTrust's regular expression does not match and it fails to process this single-page dump. We believe that parsing such dumps should also be possible and adjusted the regular expression to match \verb|<text xml:space="preserve.*">|. It allows any other attributes as long as the last one ends with a quotation mark.

Finally both programs emitted the same list of revisions. Comparison was realized by writing all revision IDs into two separate files and then using the \verb|comm| command line tool to compare both lists \verb|comm -2 -3 <(sort acewiki-revs-fw.txt ) <(sort acewiki-revs-wikitr.txt) &&  comm -2 -3 <(sort acewiki-revs-wikitr.txt ) <(sort acewiki-revs-fw.txt)|. It outputs lines that are in the first file, but not in the second. Doing this check in both directions without any discovered mismatches implies identical files and thus parsed revisions and pages.

\paragraph{Authorship tracking}
An important piece of \citeauthor{Adler:2008:MAC:1822258.1822279} contribution measures is text authorship tracking ``on words as the fundamental unit, to more closely approximate how people perceive edits.`` \citep[p. 3]{Adler:2008:MAC:1822258.1822279}. Their algorithms implement that using recursion, because ``the attribution of words in some version of the document depends on the attribution of matching words from earlier versions.`` \citep[p. 21]{adler2012wikitrust} A problem of the chosen map-reduce approach of implementing the contribution measures on a distributed platform is recursion. It would require recursive subtasks to return values before the task is finished, what is in direct contradiction to a distributed system's ``blocking property`` \citep[p. 1]{Afrati:2011:MER:1951365.1951367}.
Based on this limitation the simpler text tracking algorithm described by \citeauthor{Adler:2007:CRS:1242572.1242608} was implemented. It only considers three revisions to track words. From the consecutive revisions $r_{i-1}, r_i, r_{i+1}$, the first two $r_{i-1}$ and $r_i$ are used to compute the text added in $r_i$. In a second step the algorithm checks how much of $r_i$ is still present in $r_{i+1}$ \citealt[cf.][p. 266]{Adler:2007:CRS:1242572.1242608}.
On the downside this recursion-free algorithm is vulnerable to manipulation. \citeauthor{Adler:2007:CRS:1242572.1242608} model a situation in \citep[p. 266]{Adler:2007:CRS:1242572.1242608} where restoring the wiki from a vandal's or spammer's damage leads to the restored text being attributed to the author who repaired the wiki. She, however, didn't contribute the text herself and might easily be abused. Given a malicious attacker with two accounts $A_d$ for deleting and $A_r$ for restoring, she firsts deletes all text with $A_d$ and immediately restores the previous version with $A_r$. By sacrificing the reputation of a throwaway account $A_d$,  $A_r$'s reputation might increase, because the revision's text is now associated with $A_r$. This attack could as well be repeated with different pages or within reasonable time intervals.

This was considered acceptable risk, because developing a workaround or equivalent implementation of their original text authorship tracking algorithm would be beyond this thesis' scope. In general it might be realizable when the framework's parallelization is limited to map processes on whole pages and not on single revisions. But this is not the case with the current example and reference implementation, thus some deviations from WikiTrust are expected.

\subsubsection{Relevance Aggregation}\label{sec:implementation-problemsandsolutions-relevanceaggregation} \label{sec:implementation-problemsandsolutions-relevanceaggr}
Given a simple formula $s(a) = \sum_{r \in R_a} f(r) + g(r)$ that calculates the author's relevance score $s(a)$, from the set $R_a$ of her revisions $r$, by addition of the individual functions $f(r)$ and $g(r)$, can be approached with the framework in two different ways. The first possibility is to implement only one Calculation and Differ class, where the latter one implements $s(a) = \sum_{r \in R_a} f(r) + g(r)$ and the Calculation emits it. $f$ and $g$ are then bound to that Differ and probably not easily reusable by other Differs or Calculations. The second more portable way is to implement $s'(a) = \sum_{r \in R_a} f(r)$ and $s''(a) = \sum_{r \in R_a} g(r)$ as separate Differ/Calculation classes. A third class implements $s'''(a) = s'(a) + s''(a)$ using the framework's RelevanceAggregator to combine the results of $s'(a)$ and $s''(a)$. The calculation order does not matter due to the associative law of \verb|+|, therefore $s(a) \Leftrightarrow s'''(a)$ holds. It, however, changes when multiplication or division is an operator between $f$ and $g$. In such a case the previous equality does break and special care needs to be taken if one wants to calculate $s_*'''(a) = s'(a) * s''(a)$, because internally $s'(a)$ and $s''(a)$ sum the values of $f$ respectively $g$ for each author. 
To resolve the problem, $s_*'''(a)$ needs access to the results of $f(r)$ and $g(r)$ before they are summed, so that it can multiply all elements pair-wise first. This was achieved by adding a method \verb|getBaseResult|, which should return an IRevelevanceScore DataSet of revisions with their associated rating ($f(r)$ or $g(r)$), to the ICalculation interface. Those DataSets can then be passed to the RelevanceAggregator as usual. 

\begin{myboxi}[Key points of section ``\nameref{sec:implementation-problemsandsolutions}``]
	\begin{itemize}
		\item During and after the implementation, slight adjustments to the framework's design were necessary to remedy problems.
		\item Processing datasets with several millions of revisions raised bugs within WikiTrust and the framework that were unnoticed before, of which most were resolvable.
		\item Fine-tuning the cluster configuration and analyzing Apache Flink's internal classes helped to reduce job failures.
	\end{itemize}
\end{myboxi}

\section{Evaluation} \label{sec:evaluation}

After finishing the implementation, the resulting framework needs to be evaluated in regards to its correctness and performance; answering the question if distributed processing can lead to faster author ranking calculations. 

The next section will cover the following three aspects and focus on an objective execution and measurement of all performance tests. The results' implications and the previous question will be discussed and answered in the following section ``\nameref{sec:discussion}``.
\begin{enumerate}
	\item Comparison of the framework vs. WikiTrust
	\item Performance differences between the framework's computation measures and possible optimizations.
	\item Impact of the Pageview processing on the performance.
\end{enumerate}

For the first point, the WikiTrust program implemented by \citeauthor{Adler:2008:MAC:1822258.1822279} will be used as the reference system for the evaluation. The performance and results of the framework will be compared to those of WikiTrust. In a second step the individual performance of the framework's example contribution measures is evaluated. 
Last but not least the performance impact of incorporating the pageview information is analyzed.

For all performance related comparisons the raw execution time reported by the GNU time\footnote{\url{http://man7.org/linux/man-pages/man1/time.1.html}} command is considered. That means the cluster's setup and tear down time will count towards its performance. To account for sporadic anomalies due to hardware or software delays, each execution is repeated five times on the same dataset and the average is calculated.
\subsection{WikiTrust vs. framework} \label{sec:evaluatin-wikitrust}
\citeauthor{Adler:2008:MAC:1822258.1822279} implemented a ``modular tool [...] [that] processes XML dumps from the Wikipedia [...] which we instrumented to calculate the various contribution measures we have defined`` \citep[p. 4]{Adler:2008:MAC:1822258.1822279}. This tool is open source and was originally published at \url{http://trust.cse.ucsc.edu}, but was offline when accessed. In an email from one of the authors, Luca de Alfaro, it was clarified that the program called ``WikiTrust`` is now available on GitHub \cite{wikitrgh}. At the time of writing, the last change was more than three years ago with code pieces dating back no less than seven years. Most of the program is written in OCaml\footnote{\url{https://ocaml.org/}}, ``an industrial strength programming language supporting functional, imperative and object-oriented styles`` \cite{ocamlorg}.

For the framework's evaluation, the WikiTrust program should serve as a baseline and direct indicator of a possible performance improvement. The source code\footnote{\url{https://github.com/collaborativetrust/WikiTrust/blob/master/util/batch\_process.py}} indicates that Python's \verb|multiprocess| package splits and distributes the work onto all cores for parallelization. However, it is limited by the physical amount of CPU cores of the system it runs on.
In theory higher parallelization by distributing the work across multiple systems should increase the performance and reduce the processing time. This is what we want to achieve with our framework.

\subsubsection{Setup} \label{sec:evaluatin-wikitrust-setup}
WikiTrust's repository contains several README files and instructions on how to compile and use the tool. Unfortunately there were a lot of problems with software version incompatibilities, for example due to updated dependencies. Missing version information about the systems and software used to run WikiTrust lead to intensive debugging and testing until a satisfying working configuration was found. The newest OCaml release 4.04.0\footnote{\url{https://ocaml.org/releases/}} is not compatible with all of WikiTrust's dependencies. For example the package ``sexplib`` lacks essential files for a successful compilation that are not necessary with OCaml up to version 4.02.3. Furthermore, the recommended package management tool GODI\footnote{\url{http://godi.camlcity.org/archive/godi/index.html}} was discontinued in 2014. Any attempt to use it anyway to install the necessary dependencies was not successful.
Luckily, another package manager, Opam\footnote{\url{https://opam.ocaml.org/}},  is still under development and helped to finally build an important part of WikiTrust. Instructions on how to build it are described in section ``\nameref{sec:appendix-buildingwikitrust}``. 

Comparable performance results are obtainable through equal circumstances and environments for both programs. Ideally the student cluster would allow to run WikiTrust and the framework on one single node, followed by running the latter one on the whole cluster. There was one reason why this ideal environment could not be created: OCaml does not fully support\footnote{\url{https://caml.inria.fr/ocaml/portability.en.html}} the IBM-PowerPC-architecture, thus not running or compiling programs for it. Even trying to manually backport a pull request\footnote{\url{https://github.com/ocaml/ocaml/pull/225}} and recompiling a custom OCaml version did not help, because it was intended for a newer OCaml release. Eventually, we came to the conclusion that another server with a suitable architecture was necessary. 

We dealt with this dilemma by renting a powerful virtual machine at DigitalOcean, which is not ideal due to shared hardware, but is better than no comparison. The server's specification is detailed in section \ref{sec:introduction-setup-digitalocean}.

A README-batch file advises to run a python script \verb|batch_process.py| from the \verb|util/| folder. Regarding the instructions, it is a wrapper for the following 5 processing steps:
\begin{enumerate}
	\item \verb|do_split| is responsible for splitting the dump into smaller chunks. It saves them gzip compressed in the \verb|split_wiki/| folder.
	\item \verb|do_compute_stats| calculates statistics and creates ``stat`` files, which are stored in \verb|stats/| and compressed as well.
	\item \verb|do_sort_stats| applies a sorting algorithm to the stat files. Results will be written into the \verb|buckets/| directory.
	\item \verb|do_compute_rep| is where the user reputation scores are calculated and then emitted into a \verb|user_reputations.txt| file.
	\item \verb|do_compute_trust| computes text trust.
\end{enumerate}
All steps, except for \verb|do_sort_stats|, are multi-threaded.
Before running the tool on the test dumps, the source code was reviewed resulting in several conclusions and modifications. The reference paper \cite{Adler:2008:MAC:1822258.1822279} sets the amount of ``judges`` of a revision to 10, but the script's default is 6. This was changed to match the paper. 
The \verb|do_compute_trust| step calculates text trust values and utilizes further external programs, but we learned that it does not affect the user reputation calculation and was therefore removed.
It was also done to make the performance more comparable by focusing on the important tasks.
Afterwards, a variable called \verb|end_time| with a default value of \verb|Timeconv.time_to_float 2008 10 30 0 0 0;| was discovered in the \verb|analysis/generate_reputation.ml| file. From the paper we have learned that \citeauthor{Adler:2008:MAC:1822258.1822279} used to limit the amount of considered revisions for their computations by time. Our framework does not limit the processed revisions by time, so this restriction had to be removed by setting the date into the future (e.g. 1st of January 2020). After each change to the OCaml code, the WikiTrust program had to be recompiled.

Three Wikipedia dumps were preferably used during the development of the framework, because of their rather small size and relatively fast processing speed on the author's system:
\begin{enumerate}
	\item \verb|aawiki-20170501|: 3.5 MB / 229 KB bzip2 / 189 KB 7z
	\item \verb|acewiki-20170501|: 299 MB / 14 MB bzip2 / 9.4 MB 7z
	\item \verb|bgwiktionary-20170501|: 1384 MB / 49 MB bzip2 / 61 MB 7z
\end{enumerate} 
The first dump has 80 processable revisions, the second one 56749 and the biggest test dataset has 899122. ``Processable`` revisions are those that will be further processed that means consecutive edits by the same author have already been filtered out. 
In the beginning we had difficulties to reproduce the same processable revisions with the framework. It turned out that WikiTrust does some additional checks and filtering on pages and revisions. Modifying the methods \verb|add_revision| and \verb|eval| in the \verb|reputation_analysis.ml| to print out revisions \verb|r'| before they are added to the \verb|revs| vector gave valuable insights into which revisions WikiTrust processes. The patch \verb|wikitrust-debug-patch.diff| can be applied to get the same debug output, but it breaks WikiTrust's further processing.  Most bugs and their fixes were already described in ``\nameref{sec:implementation-problemsandsolutions}``, like excluding pages with a specific \verb|<redirect>| tag or the differences in detecting consecutive revisions, where WikiTrust matches using a contributor's ID, but the framework uses the username. With larger dumps more edge cases and bugs appeared, but resolving all those issues was important for basing the comparison on the same parsed data. 
Ideally, this would lead to similar user reputation values in the end, making a pure performance comparison possible.

\begin{myboxi}[Key points of section ``\nameref{sec:evaluatin-wikitrust}``]
	\begin{itemize}
		\item WikiTrust is written in OCaml, which does not compile for the student cluster's PowerPC architecture. Therefore, a x86\_64 virtual server was rented.
		\item To ensure that both programs operate on the same input data, debugging and patching WikiTrust was required.
		\item WikiTrust's last ``text trust`` processing step was omitted, because it did not influence the user reputations, but increased its overall processing time.
	\end{itemize}
\end{myboxi}

\subsection{Performance comparison}\label{sec:evaluatin-wikitrust-perfcomp}
On the DigitalOcean server a Bash script (see section \ref{sec:appendix-wikitrloop}) executed the WikiTrust program fives times on each dump while monitoring its execution time using the GNU time command. Table \ref{tab:wikitrust-perf} shows the resulting processing times. One can easily see that the bigger the dump, the longer it takes for WikiTrust to compute the user reputation scores, while it reaches a maximum throughput of around 100 revisions per second. It is noteworthy that WikiTrust operates on the 7z compressed dumps and gzip compresses the intermediate data between its processing steps.

\begin{table*}
	\centering 
	\caption{Time needed by WikiTrust to calculate the user reputations on the virtual machine without the last processing step.}
	\label{tab:wikitrust-perf}
	\begin{tabular}{cccc}\toprule
		Round & aawiki-20170501 & acewiki-20170501 & bgwiktionary-20170501 \\
		\midrule
		$\#1$ & 23.18s & 09:12.47m & 02:44:54h \\
		$\#2$ & 23.09s & 09:08.70m & 02:45:56h \\
		$\#3$ & 23.20s & 09:07.66m & 02:45:29h \\
		$\#4$ & 23.21s & 09:01.80m & 02:45:04h \\
		$\#5$ & 23.41s & 08:55.17m & 02:44:16h \\
		\midrule\midrule
		AVG  & 23.22s & 09:05.16m&  02:45:07h\\
		\midrule\midrule
		Revisions/s  & 3.45R/s & 104 R/s &  90.75R/s \\
		\bottomrule
	\end{tabular}
\end{table*}

As for the framework, it does not have one special formula to calculate user reputations, but offers functionality to implement various ranking algorithms. Therefore, all contribution measures described in \cite{Adler:2008:MAC:1822258.1822279} were implemented using our framework. A more detailed explanation will be given in the next subsection ``\nameref{sec:evaluatin-adlerscontributionmeasures}``, but for now the focus is on the performance in the aforementioned environment. 
Similar to WikiTrust, each contribution measure was run exactly five times on each of the three dumps. To improve the reproducibility of our tests, the framework's source code at the time of this evaluation can be found on the \verb|wikitrust_comp| branch. As a reference for all further tests, a RegexPreFilter with the regular expression \verb|(?is).*<ns>0</ns>.*| filtered pages from the initial Page DataSet that did not belong to the main namespace. 
Apache Flink 1.0.3 was configured to use 12 task slots and a total allocation of 24576 MB for the taskmanager's heap memory. Apache Flink setup steps are explained in the appendix (``\nameref{sec:appendix-javaflinkfw}``) likewise the nested for-loop to run the framework on all three datasets (``\nameref{sec:appendix-frameworkloop}``). Due to only having one node, there was no need to setup the shared file system HDFS, therefore both tools read their data directly from disk.

Table \ref{tab:framwork-perf} shows the processing times for each contribution measure on each dump. Entries that are marked with '*' indicate that the framework reported an error and the job was canceled while computing and the resulting time is not counted towards the averages. Some calculations end with suffix ``Single``. Those implement the same contribution measure as their suffix-free equivalent, but bundle all computational steps in one Calculation and Differ class, similar to the given example in section ``\nameref{sec:implementation-problemsandsolutions-relevanceaggregation}``. Looking at the column in the middle, we see that the Flink Job fails for the EditLongevity-Calculation, but succeeds for its equivalent, bundled EditLongevitySingle-Calculation. The crashes can be attributed to the relatively small disk. Apparently Apache Flink caches the intermediate results on the disk when multiple separate calculations are combined. Around 15 GB of temporary space was not enough and therefore the whole Job got canceled after the first task encountered an I/O exception. From the bgwiktionary column we learn that with the optimized ``Single``-Calculations less crashes occur while improving the performance significantly. 
\begin{landscape}
	\thispagestyle{empty}
	\ra{1.8}
	\begin{table*}
		\caption{Framework processing times for all contribution measures on the DigitalOcean server. Entries marked with '*' indicate an error and are not included in the averages.}
		\label{tab:framwork-perf}
		\begin{adjustwidth}{-3cm}{}
		\begin{adjustbox}{max width=1.14\linewidth}
			\begin{tabular}{@{}lrrrrrrcrrrrrrcrrrrrrr@{}}\toprule
				
				& \multicolumn{6}{c}{aawiki-20170501} & \phantom{abcdef}& \multicolumn{6}{c}{acewiki-20170501} & \phantom{abcdef} & \multicolumn{6}{c}{bgwiktionary-20170501}& \phantom{abcdef}\\
				\cmidrule{2-7} \cmidrule{9-14} \cmidrule{16-21} & \#1 & \#2 & \#3 & \#4 & \#5 & AVG && \#1 & \#2 & \#3 & \#4 & \#5 & AVG && \#1 & \#2 & \#3 & \#4 & \#5 & AVG &\\ 
				\midrule
				NumEdits & 2:04.07 & 0:47.64 & 0:26.46 & 0:19.25 & 0:20.72 & 0:47.63 &&  0:58.02 & 0:44.20 & 0:42.95 & 0:38.47 & 0:39.92 & 0:44.71 &&  3:02.94 & 3:14.91 & 3:04.54 & 3:07.83 & 3:08.31 & 3:07.71 &\\
				TextOnly & 0:27.03 & 0:19.32 & 0:17.52 & 0:22.25 & 0:13.98 & 0:20.02 && 0:40.40 & 0:19.66 & 0:40.73 & 0:36.18 & 0:40.89 & 0:35.57 && 2:55.77 & 3:31.73 & 3:22.24 & 3:14.42 & 3:26.96 & 3:18.22& \\
				EditOnly & 0:22.91 & 0:17.24 & 0:18.70 & 0:18.80 & 0:18.18 & 0:19.17 &&  0:45.97 & 0:44.30 & 0:50.31 & 0:47.31 & 0:46.34 & 0:46.85 &&  3:11.67 & 3:23.16 & 3:09.68 & 3:22.93 & 3:13.56 & 3:16.20 &\\
				\addlinespace
				TextLongevity  & 0:24.34 & 0:20.46 & 0:20.51 & 0:25.64 & 0:20.00 & 0:22.19 &&  2:16.74* & 2:03.44* & 1:49.83* & 2:09.09* & 0:29.91* & 1:45.80* &&  6:20.61 & 5:39.54 & 5:44.05 & 5:53.18 & 6:03.13 & 5:56.10 &\\
				TextLongevitySingle  & 0:29.00 & 0:22.45 & 0:15.57 & 0:21.81 & 0:18.28 & 0:21.42 &&  1:33.43 & 1:24.81 & 1:37.87 & 0:23.61* & 1:36.17 & 1:33.07 &&  4:08.71 & 4:01.03 & 4:03.36 & 3:55.59 & 3:48.82 & 3:59.50 & \\
				\addlinespace
				EditLongevity & 0:30.33 & 0:22.57 & 0:19.03 & 0:24.55 & 0:19.32 & 0:23.16 && 2:34.00* & 2:20.61* & 2:17.94* & 2:20.38* & 2:20.25* & 2:22.64* && 0:29.42* & 0:28.70* & 6:55.79 & 6:34.50 & 6:13.19 & 6:34.49 &\\
				EditLongevitySingle & 0:28.44 & 0:28.08 & 0:16.28 & 0:18.41 & 0:17.23 & 0:21.69 &&  6:04.80 & 6:23.03 & 6:22.40 & 7:04.07 & 7:32.95 & 6:41.45 &&  4:56.44 & 0:32.58* & 4:04.41 & 4:14.45 & 4:00.99 & 4:19.07 &\\
				\addlinespace
				TenRevisions & 0:21.09 & 0:27.69 & 0:22.42 & 0:25.08 & 0:22.17 & 0:23.69 &&  1:40.23* & 1:38.87* & 1:34.64* & 1:26.61* & 1:34.21* & 1:34.91* &&  6:31.77 & 5:46.26 & 5:27.18 & 6:11.24 & 5:59.62 & 5:59.21 &\\
				TenRevisionsSingle & 0:34.48 & 0:19.74 & 0:13.35 & 0:13.94 & 0:16.38 & 0:19.58 &&  1:24.40 & 1:18.23 & 1:22.83 & 1:13.19 & 1:17.56 & 1:19.24 &&  3:29.53 & 3:38.01 & 3:44.30 & 3:37.68 & 3:27.37 & 3:35.38 &\\
				\addlinespace
				TextLongevityWithPenalty & 0:24.31 & 0:20.36 & 0:26.12 & 0:20.33 & 0:19.50 & 0:22.12 &&  1:34.55* & 1:30.46* & 1:37.21* & 1:26.12* & 1:28.11* & 1:31.29* &&  9:39.89 & 8:24.89 & 8:23.29 & 8:30.48 & 8:24.41 & 8:40.59 &\\
				TextLongevityWithPenaltySingle & 0:37.27 & 0:20.88 & 0:14.99 & 0:19.96 & 0:18.36 & 0:22.29 &&  1:24.26* & 1:33.21* & 1:26.53* & 1:32.58* & 1:30.14* & 1:29.34* &&  7:48.97 & 7:16.85 & 7:16.92 & 6:57.44 & 6:58.45 & 7:15.73 &\\
				\hhline{======================}
				AVG & & & & & & 23.90 && & & & & & 1:56.92 && & & & & & 5:05.65 &\\
				\bottomrule
			\end{tabular}
		\end{adjustbox}
		\end{adjustwidth}
	\end{table*}
\end{landscape}

The averages over all calculations for one dump can be used to approximate the revisions per second:
\begin{itemize}
	\item \verb|aawiki|: 3 R/s
	\item \verb|acewiki|: 485 R/s
	\item \verb|bgwiktionary|: 2941 R/s
\end{itemize}

%
When looking at a dump's rows in both tables \ref{tab:wikitrust-perf} and \ref{tab:framwork-perf}, one can see that the measured execution time usually fluctuates. Averages of all successful runs were calculated to minimize the impact of the comparison.

The following formula describes the speed-up calculation mathematically. Given a dump $D$ and Calculation $C$, let $t_{FW_C}(D)$ be a function that returns the framework's measured processing time in seconds for the Calculation $C$ on the dataset $D$.  Similarly, let $t_{WT}(D)$ be a function that returns WikiTrust's measured processing time in seconds on the dataset $D$. The speed-up in percent is then defined as $$S_{C}(D) = \left( \frac{t_{WT}(D)}{t_{FW_C}(D)} -1\right) * 100$$
$S_{C_{AVG}}(D)$ shall return the speed-up based on the average computation times.
A value of $S_{C}(D) > 0$ denotes faster processing by the framework with a speed-up of $S_{C}(D)$ percent. Negative values indicate slower performance. 

For the smallest dump, WikiTrust is on-par with the framework and even slightly faster on average. An initial speed-up can be observed for the mid-sized acewiki. On average WikiTrust needs 9 minutes to finish, whereas the framework's fastest calculation ends after 35 seconds ($S_{TextOnly_{AVG}}(\text{``acewiki``}) \simeq 14.32$) and its slowest is still $S_{EditLongevitySingle{AVG}} (\text{``acewiki``}) = 35.8$ faster with an execution speed of 6 minutes and 41.45 seconds. 
The biggest performance differences can be seen for the bgwiktionary dump. WikiTrust needs approximately three hours, whereas the framework only needs up to nine minutes, resulting in a speed-up value of $S_{TextLongevityWithPenalty_{AVG}}(\text{``bgwiktionary``}) = 1803.03$, beaning around 19 times faster.

While conducting WikiTrust's performance tests, the single-threaded statistics sorting step was responsible for a major share of whole processing time. A second performance test with only the first two processing steps, namely \verb|do_split| and \verb|do_compute_stats| was done. Table \ref{tab:wikitrust-perf-statistics} shows the resulting processing times without the \verb|do_sort_stats| and \verb|do_compute_rep| steps. Comparing these average speeds with the framework's from table \ref{tab:framwork-perf}, we see that WikiTrust definitely outperforms it for the smallest dump by factor 3. Ignoring acewiki's the outliers EditLongevitySingle and the crashed ones, all other contribution measures perform faster than WikiTrust. The speed-up is even more visible for the bgwikitionary, where the framework performs around three times faster on average. 
\begin{table*}
	\centering
	\caption{WikiTrust's raw performance for parsing and calculating the statistics file by omitting the last three computation steps.}
	\label{tab:wikitrust-perf-statistics}
		\begin{tabular}{cccc}\toprule
		Round & aawiki-20170501 & acewiki-20170501 & bgwiktionary-20170501 \\
		\midrule
			$\#1$ & 07.28s & 1:43.27m & 15:00.68m \\
			$\#2$ & 07.15s & 1:44.99m & 14:54.54m \\
			$\#3$ & 07.00s & 1:43.79m & 15:00.13m \\
			$\#4$ & 07.05s & 1:46.47m & 14:53.47m \\
			$\#5$ & 07.21s & 1:47.84m & 15:00.97m \\
			\midrule\midrule
			AVG  & 07.14s & 1:45.27m &  14:57.96m \\
			\midrule\midrule
			Revisions/s  & 11R/s & 539 R/s &  1001R/s \\
			\bottomrule
		\end{tabular}
\end{table*}

\begin{myboxi}[Key points of section ``\nameref{sec:evaluatin-wikitrust-perfcomp}``]
	\begin{itemize}
		\item For certain combinations of large dumps and calculations, job failures occur on the DigitalOcean server.
		\item WikiTrust's \verb|do_sort_stats| takes significantly longer than \verb|do_compute_stats|, but the framework is as fast as the latter. 
		\item The framework is more reliable and faster when only a single Differ and Calculation class without a RelevanceAggregator is used. 
	\end{itemize}
\end{myboxi}

\subsection{Framework optimizations} \label{sec:evaluatin-adlerscontributionmeasures}

\subsubsection{Measuring contributions} \label{sec:evaluatin-adlerscontributionmeasures-measuringcontributions}
Throughout the thesis, the so called contribution measures by \citeauthor{Adler:2008:MAC:1822258.1822279} were used without introducing them in detail. In the paper \citetitle{Adler:2008:MAC:1822258.1822279} they define and analyze seven formulas that, when applied to all revisions and summed for each author, shall return an author's score that she earned with her contributions to a wiki. All contribution measures rate contributions with more emphasis on the quality than the quantity of an edit, while still trying to differ in specific aspects, such as preventing malicious abuse of the formula \cite[p. 4-5]{Adler:2008:MAC:1822258.1822279}.
A contribution measure is a function of type $\mathbb{A} \mapsto \mathbb{R}$, with $\mathbb{A}$ being the set of all authors of a wiki and $\mathbb{R}$ the rational numbers. It returns a score for an author $a \in \mathbb{A}$ based on all her authored revisions from all pages of a wiki.
\begin{itemize}
	\item \verb|NumEdits| simply counts an author's edits and returns the total number. 
	\item \verb|TextOnly| counts the amount of words that were added by the author.
	\item \verb|EditOnly| calculates the edit distance between a revision and its immediate predecessor. It measures the size of the change by accounting for inserted, moved or deleted words. 
	\item \verb|TextLongevity| puts emphasis on the quality of an edit by not only counting the amount of words added, but also if and how those decay over the following revisions.
	\item \verb|EditLongevity| is similar to TextLongevity, but represents the edit distance of an edit multiplied with the average edit quality of the following revisions. 
	\item \verb|TenRevisions| measures the usefulness of insertions by checking how much of the text is still present in the next ten revisions. 
	\item \verb|TextLongevityWithPenalty| is a combination of TextLongevity and EditLongevity with the goal of rewarding authors of new content, but at the same time, punishing vandals, who delete or insert inappropriate content. 
\end{itemize}
A more detailed explanation and the exact formulas and thoughts behind those contribution measures are described in the reference paper \cite{Adler:2008:MAC:1822258.1822279}. The framework's implementation is based on the paper's definitions and formulas and was carried out as closely as possible, except for circumstances that were discussed in the previous chapter. 

So far, despite the different results in comparison to the WikiTrust system, a huge speed-up was measured for the individual contribution measures. The observed execution times (see table \ref{tab:framwork-perf}) show that some take up to a factor of 2.5 longer depending on the calculated measure. Furthermore the prolonged processing time belongs to the more complex contribution measures, which are composed of several rating functions, such as the TextLongevityWithPenalty. However, as previously discovered, those compound contribution measures' performance can be increased by implementing all calculations in a single Differ class, thus forgoing the RelevanceAggregator or caching overhead. 

\subsubsection{Cluster performance} \label{sec:evaluatin-adlerscontributionmeasures-clusterperf}
To see how a higher parallelization with more task slots further improves the speed-up, the same datasets and calculations were executed on the student cluster. During the initial testing phase, node 6 had some problems leading to irregular job failures. Therefore, the faulty node was excluded from the configuration and all tests continued with 6 task managers and 380 instead of 400 task slots. 

In table \ref{tab:framwork-perf-cluster} are the processing times for the same tests that were done in the previous section, but this time on the student cluster. The resulting average throughput based on revisions per second is:
\begin{itemize}
	\item \verb|aawiki|: 2.37 R/s
	\item \verb|acewiki|: 560 R/s
	\item \verb|bgwiktionary|:  R/s
\end{itemize}
It is observable that the average speeds for aawiki's NumEdits are nearly the same compared to the around ten times bigger acewiki, which even outperforms the aawiki on average by one second for the TextOnly and EditOnly measures. Another observation is that the overall average processing time for the biggest dump, the bgwiktionary, is shorter than for the acewiki, which is more than four times smaller. The contribution measures that increased the acewiki's total average due to being slower than the bgwiktionary are TextLongevity, EditLongevity, EditLongevitySingle, TenRevisions, TextLongevityWithPenalty and TextLongevityWithPenaltySingle. Comparing those contribution measures with the results in table \ref{tab:framwork-perf} introduces a new perspective: most of them failed on the 12 core virtual machine due to temporary storage limitations. Calculations like the TenRevisionsSingle or TextLongevitySingle that finished flawlessly on the virtual machine are also significantly faster and do not seem to be impacted by the slowness. 

Although an exact one-on-one comparison of the framework's execution times on the virtual machine at DigitalOcean.com (table \ref{tab:framwork-perf}) to its performance on the student cluster (table \ref{tab:framwork-perf-cluster}) is not possible due to different hardware specifications, rough tendencies are visible: except for the aawiki, that was discussed earlier, the other dumps were processed more quickly and without failures. 

\begin{landscape}
	\thispagestyle{empty}
	\ra{1.8}
	\begin{table*}
		\caption{Measured framework processing times for all contribution measures on the student cluster.}
		\label{tab:framwork-perf-cluster}
		\begin{adjustwidth}{-3cm}{}
			\begin{adjustbox}{max width=1.14\linewidth}
				\begin{tabular}{@{}lrrrrrrcrrrrrrcrrrrrrr@{}}\toprule
					
					& \multicolumn{6}{c}{aawiki-20170501} & \phantom{abcdef}& \multicolumn{6}{c}{acewiki-20170501} & \phantom{abcdef} & \multicolumn{6}{c}{bgwiktionary-20170501}& \phantom{abcdef}\\
					\cmidrule{2-7} \cmidrule{9-14} \cmidrule{16-21} & \#1 & \#2 & \#3 & \#4 & \#5 & AVG && \#1 & \#2 & \#3 & \#4 & \#5 & AVG && \#1 & \#2 & \#3 & \#4 & \#5 & AVG &\\ 
					\midrule
					NumEdits & 0:39.72 & 0:29.33 & 0:27.56 & 0:26.14 & 0:26.38 & 0:29.83 && 0:44.57 & 0:29.58 & 0:29.17 & 0:26.07 & 0:26.45 & 0:31.17  && 0:34.00 & 0:36.85 & 0:34.64 & 0:37.27 & 0:37.26 & 0:36.00  &\\
					TextOnly  & 0:30.25 & 0:26.47 & 0:26.95 & 0:27.79 & 0:26.98 & 0:27.69 && 0:26.51 & 0:25.86 & 0:25.36 & 0:26.46 & 0:26.66 & 0:26.17 && 0:37.56 & 0:37.30 & 0:34.33 & 0:37.95 & 0:36.03 & 0:36.63 & \\
					EditOnly   & 0:29.20 & 0:26.91 & 0:27.26 & 0:27.20 & 0:27.36 & 0:27.59 &&  0:26.12 & 0:25.22 & 0:25.63 & 0:27.63 & 0:27.14 & 0:26.35 && 0:40.81 & 0:35.89 & 0:38.13 & 0:35.38 & 0:35.38 & 0:37.12  &\\
					\addlinespace
					TextLongevity   & 0:36.91 & 0:33.66 & 0:36.05 & 0:29.74 & 0:36.06 & 0:34.48  &&  1:31.87 & 1:42.66 & 1:31.30 & 1:32.09 & 1:30.86 & 1:33.76 && 0:48.43 & 0:49.87 & 0:50.37 & 0:49.38 & 0:49.96 & 0:49.60  &\\
					TextLongevitySingle  & 0:27.85 & 0:26.13 & 0:26.63 & 0:28.10 & 0:25.55 & 0:26.85   &&  0:37.50 & 0:38.78 & 0:39.79 & 0:39.45 & 0:39.00 & 0:38.90
					&&  0:41.10 & 0:40.79 & 0:42.79 & 0:40.17 & 0:52.43 & 0:43.46 &\\
					\addlinespace
					EditLongevity  & 0:37.96 & 0:35.02 & 0:35.88 & 0:35.12 & 0:41.19 & 0:37.03
					&& 3:17.55 & 3:42.47 & 3:04.55 & 2:57.41 & 3:44.86 & 3:21.37  && 0:58.60 & 1:18.81 & 0:55.09 & 0:57.66 & 0:59.47 & 1:01.93  &\\
					EditLongevitySingle  & 0:27.65 & 0:25.02 & 0:25.71 & 0:25.34 & 0:23.15 & 0:25.37 && 2:01.24 & 2:42.62 & 2:38.88 & 2:26.82 & 2:33.58 & 2:28.63  && 0:41.11 & 0:41.85 & 0:45.82 & 0:42.61 & 0:40.26 & 0:42.33 &\\
					\addlinespace
					TenRevisions  & 0:37.72 & 0:38.37 & 0:38.07 & 0:35.29 & 0:33.25 & 0:36.54  && 1:26.82 & 1:45.42 & 1:29.97 & 1:32.53 & 1:44.56 & 1:35.86  && 0:49.19 & 0:46.70 & 0:49.54 & 0:50.10 & 0:49.97 & 0:49.10  &\\
					TenRevisionsSingle  & 0:30.41 & 0:24.16 & 0:24.56 & 0:26.37 & 0:23.87 & 0:25.87  &&  0:35.67 & 0:38.35 & 0:39.69 & 0:38.42 & 0:36.72 & 0:37.77 && 0:41.68 & 0:42.92 & 0:41.38 & 0:41.84 & 0:41.25 & 0:41.81  &\\
					\addlinespace
					TextLongevityWithPenalty  & 0:53.81 & 0:58.46 & 0:52.26 & 0:49.23 & 0:47.23 & 0:52.20 && 3:37.18 & 4:14.38 & 3:40.28 & 3:37.86 & 3:49.45 & 3:47.83  && 1:16.34 & 1:15.42 & 1:14.64 & 1:16.53 & 1:16.55 & 1:15.90  &\\
					TextLongevityWithPenaltySingle  & 0:46.32 & 0:42.45 & 0:52.93 & 0:52.26 & 0:45.26 & 0:47.84 && 2:46.85 & 3:32.14 & 2:54.46 & 2:49.16 & 3:29.45 & 3:06.41  && 1:04.75 & 1:04.72 & 1:07.06 & 1:10.50 & 1:06.80 & 1:06.77  &\\
					\hhline{======================}
					AVG & & & & & & 0:33.75 && & & & & & 1:41.29 && & & & & & 0:49.15 &\\
					\bottomrule
				\end{tabular}
			\end{adjustbox}
		\end{adjustwidth}
	\end{table*}
\end{landscape}

\subsubsection{Optimizing with filtering methods} \label{sec:evaluatin-adlerscontributionmeasures-optimizingprefilters}
The Pre- and PostFilters were introduced as means to filter pages from a dump that should be parsed and processed by the framework. In section \ref{sec:implementation-planofimplementation-dataprocessing} we introduced three types of PreFilters that are either based on regular expression, Xpath or Xquery queries. Additionally, PostFilters that operate on the fully parsed Page DataSet could be used to limit the pages for further processing. All described PreFilters and the custom PostFilter from listing \ref{code:example-postfilter} were implemented to filter pages that belong to the namespace with ID 0. They were individually enabled for five test runs on the aawiki-20170501, acewiki-20170501, bgwiktionary-20170501 dumps on the student cluster, with a parallelism of 380. To measure the effective filtering time, no calculation was instrumented, but a simple \verb|.count()| operation on the Page DataSet after the filter was applied.

\ra{1.8}
\begin{table*}
	\caption{Page filtering performance by measuring processing times for counting the parsed pages on the student cluster.}
	\label{tab:perf-filters}
	\begin{adjustwidth}{0cm}{}
		\begin{adjustbox}{max width=1\linewidth}
			\begin{tabular}{@{}lrrrrrrcrrrrrrcrrrrrrr@{}}\toprule
				
				& \multicolumn{6}{c}{aawiki-20170501} & \phantom{a}& \multicolumn{6}{c}{acewiki-20170501} & \phantom{b} & \multicolumn{6}{c}{bgwiktionary-20170501}& \phantom{c}\\
				\cmidrule{2-7} \cmidrule{9-14} \cmidrule{16-21} & \#1 & \#2 & \#3 & \#4 & \#5 & AVG && \#1 & \#2 & \#3 & \#4 & \#5 & AVG && \#1 & \#2 & \#3 & \#4 & \#5 & AVG &\\ 
				\midrule
				RegexPreFilter & 0:25.35 & 0:23.16 & 0:29.08 & 0:27.36 & 0:23.00 & 0:25.59 && 0:21.62 & 0:21.43 & 0:20.97 & 0:21.30 & 0:22.20 & 0:21.50 && 0:31.60 & 0:31.36 & 0:30.24 & 0:32.01 & 0:31.70 & 0:31.38 &\\
				XpathPreFilter  & 0:25.96 & 0:22.77 & 0:27.68 & 0:25.76 & 0:23.46 & 0:25.13 && 0:22.68 & 0:21.52 & 0:27.30 & 0:21.53 & 0:21.96 & 0:23.00
				&& 0:48.56 & 0:46.94 & 0:48.31 & 0:48.18 & 0:46.37 & 0:47.67 & \\
				XqueryPreFilter   & 0:25.02 & 0:23.71 & 0:28.99 & 0:27.98 & 0:23.50 & 0:25.84 && 0:21.15 & 0:21.98 & 0:29.61 & 0:21.64 & 0:22.07 & 0:23.29 && 1:24.14 & 1:22.60 & 1:27.37 & 1:24.11 & 1:19.35 & 1:23.51 &\\
				CustomPostFilter   & 0:25.47 & 0:26.75 & 0:22.97 & 0:23.28 & 0:31.06 & 0:25.91 && 0:21.52 & 0:31.60 & 0:20.96 & 0:30.53 & 0:21.24 & 0:25.17 && 0:30.48 & 0:32.63 & 0:32.58 & 0:32.39 & 0:31.28 & 0:31.87 &\\
				\addlinespace
				\hhline{======================}
				AVG & & & & & & 0:25.62 && & & & & & 0:23.24 && & & & & & 0:48.61 &\\
				\bottomrule
			\end{tabular}
		\end{adjustbox}
	\end{adjustwidth}
\end{table*}

The results in table \ref{tab:perf-filters} show no obvious tendencies, but rather mixed results. The aawiki can be considered an outlier, because all average times are the same except for the milliseconds and these might not be as accurate. The second column indicates that the performance decreases in the order of which the tests were run, with RegexPreFilter being the fastest and CustomPostFilter being the slowest. This observation is disproved by the last column, where the filtering times for the RegexPreFilter and CustomPostFilter are on the same level, but the Xpath- and XqueryPreFilters are significantly slower.
With the impression that with larger dumps more accurate results can be obtained, the same test was re-run on the warwiki. It has more than 2 million pages in its main namespace and a bzip2 compressed size of 456 MB. The following results were observed:

\begin{itemize}
	\item RegexPreFilter: 1:36.78
	\item XpathPreFilter: 2:32.54
	\item XqueryPreFilter: 4:01.21
	\item CustomPostFilter: 1:46.29
\end{itemize}

The performance differences are in the range of several seconds and more obvious, resulting in the final order: RegexPreFilter < CustomPostFilter < XpathPreFilter < XqueryPreFilter

\subsubsection{Optimizing with filtering loops} \label{sec:evaluatin-adlerscontributionmeasures-optimizingfilterloops}
While conducting the previous filtering performance tests, the discussion about the method of applying the filters from the PostFilter paragraph in section \ref{sec:implementation-planofimplementation-dataprocessing} was recalled. A decision was made to apply the PreFilters and PostFilters with the ``outer loop`` method, like shown in code listing \ref{code:ainputparser-applypostfilters}. With the following performance test, the given assumption of preferring this method should be tested. Therefor, both loop types were implemented on the branch \verb|filterorderperf| in the \verb|AInputParser| class. To force the loops to iterate more than one time, three PostFilters were implemented and added to the configuration's PostFilter list in the following order:

\begin{enumerate}
	\item \verb|TestPostFilter|: A filter that filters for the main namespace with id 0.
	\item \verb|TestPostFilterMin|: A filter for pages with equal or greater than 2 revisions.
	\item \verb|TestPostFilterMax|: A filter for pages with equal or less than 10 revisions.
\end{enumerate}

Pages that pass all three filters will only have 2 to 10 revisions and belong to the wikis namespace with id 0. 
\ra{1.8}
\begin{table*}
	\caption{Comparing the ``inner loop`` and ``outer loop`` filtering times by counting revisions after parsing by applying three different PostFilters. }
	\label{tab:perf-filtering-loops}
	\begin{adjustwidth}{0cm}{}
		\begin{adjustbox}{max width=1\linewidth}
			\begin{tabular}{@{}lrrrrrrcrrrrrr@{}}\toprule
				& \multicolumn{6}{c}{``inner loop``} & \phantom{c}& \multicolumn{6}{c}{``outer loop``} \\
				\cmidrule{2-7} \cmidrule{9-14} & \#1 & \#2 & \#3 & \#4 & \#5 & AVG && \#1 & \#2 & \#3 & \#4 & \#5 & AVG \\ 
				\midrule
				aawiki-20170501 & 0:29.78 & 0:21.82 & 0:28.04 & 0:27.92 & 0:23.80 & 0:26.27 && 0:26.16 & 0:30.12 & 0:25.58 & 0:35.88 & 0:23.05 & 0:28.16\\
				acewiki-20170501  & 0:20.89 & 0:21.24 & 0:20.72 & 0:21.42 & 0:21.47 & 0:21.15 && 0:19.65 & 0:20.85 & 0:21.47 & 0:20.56 & 0:21.70 & 0:20.85\\
				bgwiktionary-20170501   & 0:31.51 & 0:31.82 & 0:32.96 & 0:31.58 & 0:32.33 & 0:32.04 && 0:31.78 & 0:30.26 & 0:32.06 & 0:31.13 & 0:31.25 & 0:31.30\\
				warwiki-20170501   & 1:37.27 & 1:34.76 & 1:38.33 & 1:48.88 & 1:39.04 & 1:39.66 && 1:35.00 & 1:52.05 & 1:37.45 & 1:41.55 & 1:49.00 & 1:43.01\\
				\addlinespace
				\hhline{==============}
				AVG & & & & & & 0:44.78 && & & & & & 0:45.83\\
				\bottomrule
			\end{tabular}
		\end{adjustbox}
	\end{adjustwidth}
\end{table*}
For the performance measurement, a \verb|.count()| operation was issued on the Revision DataSet after being parsed with the SkipDumpParser on the student cluster. Table \ref{tab:perf-filtering-loops} shows the observed processing times and their averages. Before comparing the times, we noticed that Apache Flink's execution plan changes for both methods. The ``inner loop`` only has one DataSet \verb|.filter()| call, and therefore the execution plan only has this one additional step. The ``outer loop`` approach needs $n$ filtering steps in the execution plan, where $n$ is the amount of filters in the list.
Except for the first row, the average execution times are not significantly different, but the overall average for the ``inner loop`` method is slightly faster. 

\subsubsection{Optimizing with Kryo serialization} \label{sec:evaluatin-adlerscontributionmeasures-optimizingkryo}
When Apache Flink cannot serialize an object by itself, it falls back to the Kryo-serializer. From a more recent documentation one learns that the Kryo-serialization process may get a performance boost if sub-classed objects' types are registered \cite{flkryodoc}. The framework uses inheritance and sub-classing frequently by implementing interfaces and abstract classes. In assumption that Kryo is mostly used for the serialization process, because most classes have private methods or fields that are not compatible with Apache Flink's built-in serializer, a test was conducted:
the following list of classes was registered to Kryo using the framework's ExecutionEnvironment and the \verb|registerType| method immediately after obtaining a Framework object: ArgumentsBundle, RelevanceAggregator, Configuration, Contributor, ContributorRelevanceScore, DataHolder, DoubleDifference, Page, Pageview, Revision, RevisionRelevanceScore, ContributorFactory, DifferenceFactory, PageFactory, PageviewFactory, RelevanceScoreFactory, RevisionFactory, SkipDumpParser, SkipXMLContentHandler, XmlInputFormat, PageviewParser.
TextLongevityWithPenalty is a contribution measure that splits its final result up into three separate calculations, namely EditOnly, TextLongevity and EditLongevity, meaning that a lot of objects are likely to be sent across the cluster. If registering the classes improves the performance, then it should be measurable with that contribution measure. Similar to the previous tests, the program was run five times with a parallelization of 380 on all three dumps. The execution times without the Kryo registration were re-used from table \ref{tab:framwork-perf-cluster}.
\ra{1.8}
\begin{table*}
	\caption{Measured processing times for the TextLongevityWithPenalty calculation with and without type-registration for the Kryo serializer on the student cluster.}
	\label{tab:perf-kryp-serialization}
	\begin{adjustwidth}{0cm}{}
		\begin{adjustbox}{max width=1\linewidth}
			\begin{tabular}{@{}lrrrrrrcrrrrrr@{}}\toprule
				& \multicolumn{6}{c}{Without type-registration} & \phantom{c}& \multicolumn{6}{c}{With type-registration} \\
				\cmidrule{2-7} \cmidrule{9-14} & \#1 & \#2 & \#3 & \#4 & \#5 & AVG && \#1 & \#2 & \#3 & \#4 & \#5 & AVG \\ 
				\midrule
				aawiki-20170501 & 0:53.81 & 0:58.46 & 0:52.26 & 0:49.23 & 0:47.23 & 0:52.20 && 1:01.82 & 0:52.44 & 0:55.17 & 0:50.09 & 0:44.50 & 0:52.80\\
				acewiki-20170501  & 3:37.18 & 4:14.38 & 3:40.28 & 3:37.86 & 3:49.45 & 3:47.83 && 3:53.38 & 4:23.58 & 4:24.05 & 3:48.77 & 3:50.47 & 4:04.05\\
				bgwiktionary-20170501   & 1:16.34 & 1:15.42 & 1:14.64 & 1:16.53 & 1:16.55 & 1:15.90 && 1:11.35 & 1:11.17 & 1:13.34 & 1:13.49 & 1:10.90 & 1:12.05\\
				\addlinespace
				\hhline{==============}
				AVG & & & & & & 1:58.64 && & & & & & 2:02.97\\
				\bottomrule
			\end{tabular}
		\end{adjustbox}
	\end{adjustwidth}
\end{table*}
The results in table \ref{tab:perf-kryp-serialization} do not show any specific trend in regards to a speedup. For the aawiki, both average similar times, but it gets undefined for the other two dumps. The average processing speed for the acewiki is significantly (\textasciitilde 6.65\%) slower with the Kryo type registration. But on the other hand, the speed improves significantly (\textasciitilde 5.07\%) for the bgwiktionary. Comparing the total averages, no significant performance improvements are evident.

\subsubsection{Optimizing with dump parsing} \label{sec:evaluatin-adlerscontributionmeasures-optimizingdumpparser}
As a last performance optimization resort, the SkipDumpParser was compared to the RegexSkipDumpParser to learn which parsing method performs better. The former uses Java's SAX library, whereas the latter uses regular expressions similar to WikiTrust. The code for this performance measurement is archived on the \verb|regexparser| branch. Due to the framework's extensibility, switching between both parsing techniques was easy and straightforward by setting and adjusting the custom \verb|RegexDumpParserFactory|. To abstract from the contribution measures processing time, the framework's task was limited to parsing all revisions and counting them, like done in previous measurements. 
During the RegexSkipDumpParser's first performance test, the debug \verb|System.out.println| statements were not removed, which resulted in an enormous performance loss, compared to the second run without them. The measured processing times in this section and table \ref{tab:perf-fw-regex-vs-xml-parsing} are from the latter one.
\ra{1.8}
\begin{table*}
	\caption{Performance of the RegexSkipDumpParser vs. the SkipDumpParser measured by their computation time for .count() on the resulting Revision DataSet.}
	\label{tab:perf-fw-regex-vs-xml-parsing}
	\begin{adjustwidth}{0cm}{}
		\begin{adjustbox}{max width=1\linewidth}
			\begin{tabular}{@{}lrrrrrrcrrrrrr@{}}\toprule
				
				& \multicolumn{6}{c}{Regex Parser} & \phantom{c}& \multicolumn{6}{c}{XML Parser} \\
				\cmidrule{2-7} \cmidrule{9-14} & \#1 & \#2 & \#3 & \#4 & \#5 & AVG && \#1 & \#2 & \#3 & \#4 & \#5 & AVG \\ 
				\midrule
				aawiki-20170501 & 0:35.95 & 0:24.35 & 0:23.77 & 0:23.10 & 0:23.21 & 0:26.08 && 0:26.42 & 0:22.78 & 0:22.95 & 0:29.12 & 0:27.65 & 0:25.78  \\
				acewiki-20170501 & 0:33.19 & 0:42.71 & 0:31.46 & 0:30.30 & 0:31.14 & 0:33.76 && 0:30.22 & 0:31.70 & 0:31.04 & 0:30.02 & 0:32.54 & 0:31.10 \\
				bgwiktionary-20170501 & 0:28.06 & 0:26.45 & 0:25.42 & 0:26.65 & 0:24.72 & 0:26.26 && 0:31.77 & 0:32.27 & 0:31.94 & 0:31.64 & 0:33.68 & 0:32.26  \\
				warwiki-20170501 & 1:56.76 & 1:53.80 & 1:54.39 & 1:56.68 & 1:50.74 & 1:54.47 && 2:13.02 & 2:16.09 & 2:20.33 & 2:19.33 & 2:16.97 & 2:17.15  \\
				\hhline{==============}
				AVG & & & & & & 0:50.14 && & & & & & 0:56.57\\
				\bottomrule
			\end{tabular}
		\end{adjustbox}
	\end{adjustwidth}
\end{table*}
Relying on the SkipDumpParser for the best performance can be done for the aawiki and acewiki, because for those dumps it gives the best average performance. However, when comparing the other two dumps, extracting information with line-wise reading and regular expressions outperforms the native XML parser by at least 16.5\% (warwiki) up to 18.5\% (bgwiktionary). Averaging over all dumps, the RegexSkipDumpParser is still more than significantly faster with about 11.37\% performance improvement. 

\begin{myboxi}[Key points of section ``\nameref{sec:evaluatin-adlerscontributionmeasures}``]
	\begin{itemize}
		\item Compared to the virtual server, the cluster processes the contribution measures much faster and without job failures.
		\item The fastest page filtering method is the RegexPreFilter. 
		\item Applying multiple PostFilters in a single filter DataSet-API call improves the performance slightly.
		\item The Kryo type registration does not result in faster processing times.
		\item The RegexSkipDumpParser's regular expression and line-wise reading outperforms the SkipDumpParser's native SAX-library for larger dumps.
	\end{itemize}
\end{myboxi}

\subsection{Pageviews parsing} \label{sec:evaluatin-pageviews} 
WikiTrust does not incorporate pageview information into its calculations and the contribution measures do not use this information. The framework supports parsing and associating pageviews with their respective pages, because we believed that this additional metric might help to develop more advanced contribution measures in the future. For example, it is imaginable that \citeauthor{Adler:2008:MAC:1822258.1822279} contribution measures could be weighted by the amount of traffic that a page receives to reward or punish authors who edit frequently visited wikis due to higher importance and impact of their edit. 

However, processing another source of data comes at the cost of possible performance loss. The impact of providing this information was subject of another performance test. For the pageview dataset, the pageviews for February 2017 were downloaded\footnote{\url{https://dumps.wikimedia.org/other/pageviews/2017/2017-02/}} totaling in 672 files and more than 34 GB of gzip compressed data. In a first step it was further processed with our FileMerger-tool to form one homogeneous dataset, what was discovered in section \ref{sec:implementation-compresseddatasets-timeaspect} to improve the performance. Nearly five minutes (4:58.22) were needed to finish the conversion, but as explained in the referenced section this time investment is returned with multiple framework executions.
\ra{1.8}
\begin{table*}
	\caption{Measured processing times for the TextOnly calculation with and without pageviews on the student cluster.}
	\label{tab:perf-pageviews}
	\begin{adjustwidth}{0cm}{}
		\begin{adjustbox}{max width=1\linewidth}
			\begin{tabular}{@{}lrrrrrrcrrrrrr@{}}\toprule
				& \multicolumn{6}{c}{Without pageviews} & \phantom{c}& \multicolumn{6}{c}{With pageviews} \\
				\cmidrule{2-7} \cmidrule{9-14} & \#1 & \#2 & \#3 & \#4 & \#5 & AVG && \#1 & \#2 & \#3 & \#4 & \#5 & AVG \\ 
				\midrule
				aawiki-20170501 & 0:30.25 & 0:26.47 & 0:26.95 & 0:27.79 & 0:26.98 & 0:27.69 && 5:10.15 & 3:40.96 & 3:41.68 & 3:37.64 & 3:36.26 & 3:57.34 \\
				acewiki-20170501  & 0:26.51 & 0:25.86 & 0:25.36 & 0:26.46 & 0:26.66 & 0:26.17 && 3:45.24 & 3:42.09 & 3:44.53 & 3:43.19 & 3:46.25 & 3:44.26 \\
				bgwiktionary-20170501 & 0:37.56 & 0:37.30 & 0:34.33 & 0:37.95 & 0:36.03 & 0:36.63 && 3:56.96 & 3:48.73 & 3:58.79 & 3:59.17 & 3:50.43 & 3:54.82 \\
				\addlinespace
				\hhline{==============}
				AVG & & & & & & 0:30.16 && & & & & & 3:52.14\\
				\bottomrule
			\end{tabular}
		\end{adjustbox}
	\end{adjustwidth}
\end{table*}
As a reference for comparison, the TextOnly contribution measure was chosen. The left column of table \ref{tab:perf-pageviews} contains the previously measured times from table \ref{tab:framwork-perf-cluster}. 
The contribution measure was modified to be $txt(r) * pv(r)$ where $txt(r)$ is the revision's TextOnly score and $pv$ is a function that returns the pagecount from the revision's associated page. Multiplication of both values happens in the TextOnlyWithPageviewDiffer. The implementation of the TextOnlyWithPageviews-Calculation can be found on the framework's branch \verb|adlercm_pageviews|. The current pageviews implementation only assigns the pagecounts for the given input data time frame to the pages, thus not providing granular statistics like the amount of views a specific revision had. 
The modified TextOnly calculation can be interpreted as a measure that values edits to frequently visited pages, but ignores edits to pages that did not receive any traffic due to multiplication with zero. The framework expects two more positional command line arguments to activate the pageview parsing: 
\begin{enumerate}
	\item Path to the pageview dataset(s)
	\item The wiki's project name as explained in section \ref{sec:introduction-definitions-pageviewdatasets}
\end{enumerate}
For the aawiki and acewiki it is simply the prefix before ``wiki``, but for the bgwiktionary one has to use ``bg.d`` to get the correct pageview information. 

The TextOnlyWithPageview's performance results are listed in table \ref{tab:perf-pageviews}. The first obvious discovery is that all three dumps were processed in a similar time span of 3:52 minutes on average. It is around 2:20 minutes slower than the mean processing time without the pageviews. No correlation between the amount of revisions and the processed time is visible: Acewiki has the smallest processing time, followed by bgwiktionary and aawiki. The very first run using pageviews takes more than 1 minute longer than the other ones and is an obvious outlier.

\begin{myboxi}[Key points of section ``\nameref{sec:evaluatin-pageviews}``]
	\begin{itemize}
		\item The pageviews data should be preprocessed as discussed in section \ref{sec:implementation-compresseddatasets-timeaspect}.
		\item The framework supports the incorporation of pageviews data, but although it makes new contribution measures possible, it impacts the overall performance negatively.
	\end{itemize}
\end{myboxi}

\section{Discussion} \label{sec:discussion}
This section will start with a discussion of the previous section's evaluation results and their implications for the performance and the developed framework. After that the framework's use cases such as the usability to calculate author ranking as well as an outlook at how it could benefit Wikipedia will be discussed. In the last part, future work in the form of modifications or tests that might further improve the framework will be outlined.

\subsection{Performance and ranking results} \label{sec:discussion-performancerankingresults}


Each section from the previous ``\nameref{sec:evaluation}`` chapter revealed some performance and ranking results that need to be discussed under the aspects of what we learn from those and what their implications and meaning are.

\subsubsection{WikiTrust vs. Framework}
The first section dealt with the execution times of \citeauthor{Adler:2008:MAC:1822258.1822279} their WikiTrust program and the framework developed by us. 

\paragraph{WikiTrust performance}
Throughout the performance tests, we noticed that the execution times always fluctuated in the range of several seconds. This was not limited to the framework, but also happened for WikiTrust, when no other work was done on the virtual server and both programs were allowed to exhaust all resources. An explanation is the shared environment where the virtual server does not have guaranteed dedicated resources and other customers can influence the overall hypervisor's performance. It is likely that the virtual hard drive and the shared disk I/O performance is to blame for a major part of the fluctuation. 

\citeauthor{Adler:2008:MAC:1822258.1822279} note in the README-batch file that for larger wikis with more than 100,000 revisions, the processing speed for their first approach can be as low as 20 to 60 revisions per second. Therefore, they advise to use the second approach with the \verb|batch_process.py| that was also used in our tests. Furthermore, \citeauthor{Adler:2008:MAC:1822258.1822279} claim in the aforementioned file that processing the Italian Wikipedia ``took about 1 day on an 8-CPU-core machine``, but without specific information about the amount of revisions, the needed processing time and the hardware that was used, it is impossible to verify, reproduce or put our own values in relation. Within eight years the wiki probably accumulated a lot more revisions, so its actual size was smaller in the past and therefore faster to process. We, on the other hand, can now say that WikiTrust is able to process around 100 revisions per second on a 12-core server (see \ref{sec:introduction-setup-digitalocean} and table \ref{tab:wikitrust-perf}), paving the way for further comparisons. 

It needs to be noted that the performance does not seem to scale linearly. Otherwise the comparatively small aawiki with its 80 revisions would be expected to be processed within 0.8 seconds. But it takes around 23 to finish bringing the speed down to less than 4 revisions per second (see table \ref{tab:wikitrust-perf}). We believe that the increasing processing time stems from the overhead of combining the following two factors: 

\begin{itemize}
	\item Splitting the small dump into smaller files
	\item Calling external programs for processing steps like 7z or gzip to de-/compress dumps, split\_wiki or stats files 
\end{itemize} 

\paragraph{Framework performance}
WikiTrust is not alone with the additional overhead for relatively small dumps. The same observation was made for the framework, because its average processing time on the aawiki is also around 23 seconds (table \ref{tab:framwork-perf}) and 33 seconds (table \ref{tab:framwork-perf-cluster}), whereas it processes the other dumps faster by a significant factor. For the latter one, some measured times for the aawiki were even longer than for the other dumps. Either this is a fluctuation in the cluster's performance, or the circa 1.5s differences come from the overhead of parallelizing 80 items on 380 task slots, therefore slowing down the overall execution time by managing all tasks. We conclude that the smaller the dump and the higher the parallelization is, the more the overhead outweighs the content processing, thus leading to an overall reduced performance.

At first, we were confident that we could gain a significant speed-up from distributing the work onto multiple nodes, in comparison to WikiTrust on single node. However, due to the architecture limitations of OCaml and our budget, we could not create an ideal environment where this comparison was possible. The approximated environment only consisted of a single compatible node. Nevertheless, we were surprised about several observations:

We did not initially expect WikiTrust's sorting step to be the bottleneck leading to the quite long processing times in table \ref{tab:wikitrust-perf}. After removing this particular step and looking at the table \ref{tab:wikitrust-perf-statistics}, we expected it to be much faster than the framework. However, its average processing times on the same node are more or less in the same range. 
A reason could be that WikiTrust's sophisticated authorship tracking takes more computational time, than the framework's simpler implementation, thus bringing the measured time differences closer to each other. 

The general expectation was that the bigger the dump, the longer the execution time, because logically, more data has to be processed. With the results from table \ref{tab:framwork-perf-cluster}, this no longer holds true. Despite the ordering by size and processable revisions being aawiki < acewiki < bgwiktionary, the ordering by total average processing time changes to aawiki < bgwiktionary < acewiki. The contribution measures that increased the acewiki's total average due to being slower than the bgwiktionary are TextLongevity, EditLongevity, EditLongevitySingle, TenRevisions, TextLongevityWithPenalty and TextLongevityWithPenaltySingle. A possible conclusion is that Flink starts to cache data onto the disk, which decreases the performance drastically. 
However, some calculations that finished flawlessly on the virtual machine are also significantly faster and do not seem to be impacted by the (caching) slowness. This is not unexpected, because not only were more nodes involved, but also faster CPU-cores and more RAM. Apparently scaling out an already distributed algorithm by increasing the amount of CPU cores and task slots can improve its performance. That scaling process is furthermore simplified by being based on technology like Java, Apache Flink or HDFS that pose as an abstraction layer for the hardware and operating system. 

\paragraph{Processing failures}
Another surprising observation was that during the framework tests on the virtual server, the acewiki could not be processed successfully, because of not enough caching disk space. The amount of data that needed to be cached was in no obvious relation to the dump's size. Acewiki's pages or revisions must have a different structure or distribution, because the framework did not show similar problems with a several times bigger dump. Possible explanations are either a lot of revisions in a single page, thus creating a long list of them, or huge revisions. Even when only small changes are made, a revision will contain the new complete content. All this data, however, needs to be handled by Apache Flink for processing and the single node was not powerful enough. Furthermore, all crashes also seem related to the compound contribution measures that use the RelevanceAggregator to combine their components' results. Using the single class based implementation of those measures that end with the \verb|Single| suffix prevents the creation of additional RelevanceScore DataSets that need to be cached until they are processed with the RelevanceAggregator, therefore reducing the amount of cached data and successfully finishing the task. Implementing contribution measures in that way is recommended and the RelevanceAggregator should only be used if absolutely necessary. Ostensibly, higher performance is still more probable by implementing all relevance calculations in a single Differ and Calculation class.
Unfortunately, the other tests and optimizations prevented a deeper look into this behavior, so finding an explanation and solution must be considered future work. 

\subsubsection{Sources of error}\label{sec:evaluatin-wikitrust-sourcesoferror}
\begin{figure}[htbp]
	\centering
	\includegraphics[width=\linewidth]{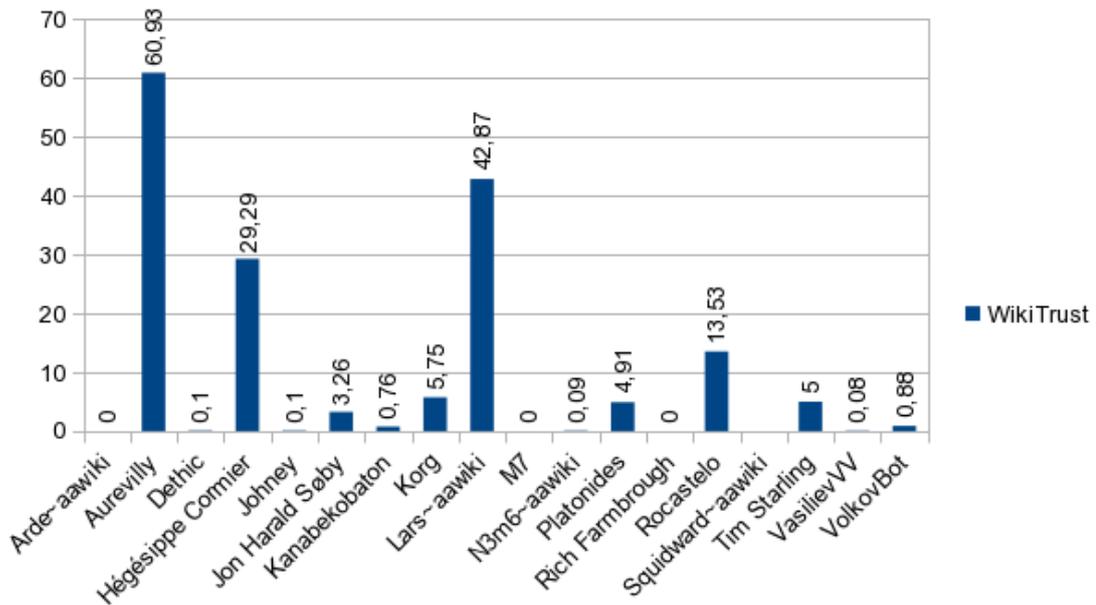}
	\caption{WikiTrust's user reputation results on the aawiki.}
	\label{fig:comparison-wikitr}
\end{figure}

\begin{figure}[htbp]
	\centering
	\includegraphics[width=\linewidth]{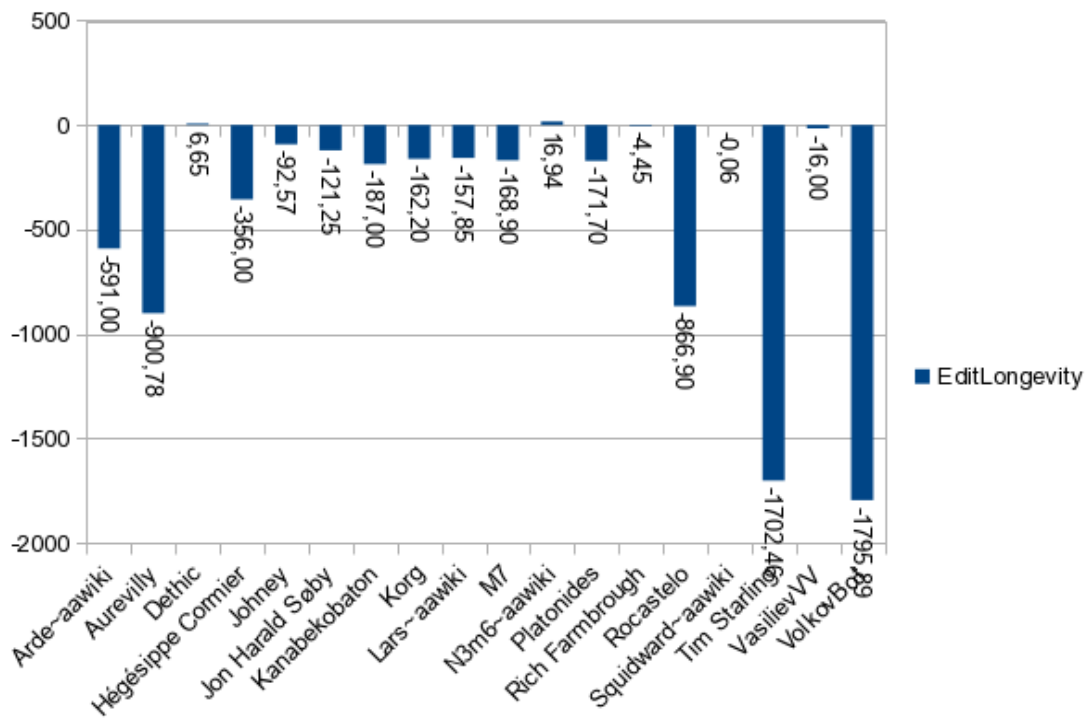}
	\caption{Framework's EditLongevity results}
	\label{fig:comparison-fw-editlongevity}
\end{figure}

\begin{figure}[htbp]
	\centering
	\includegraphics[width=\linewidth]{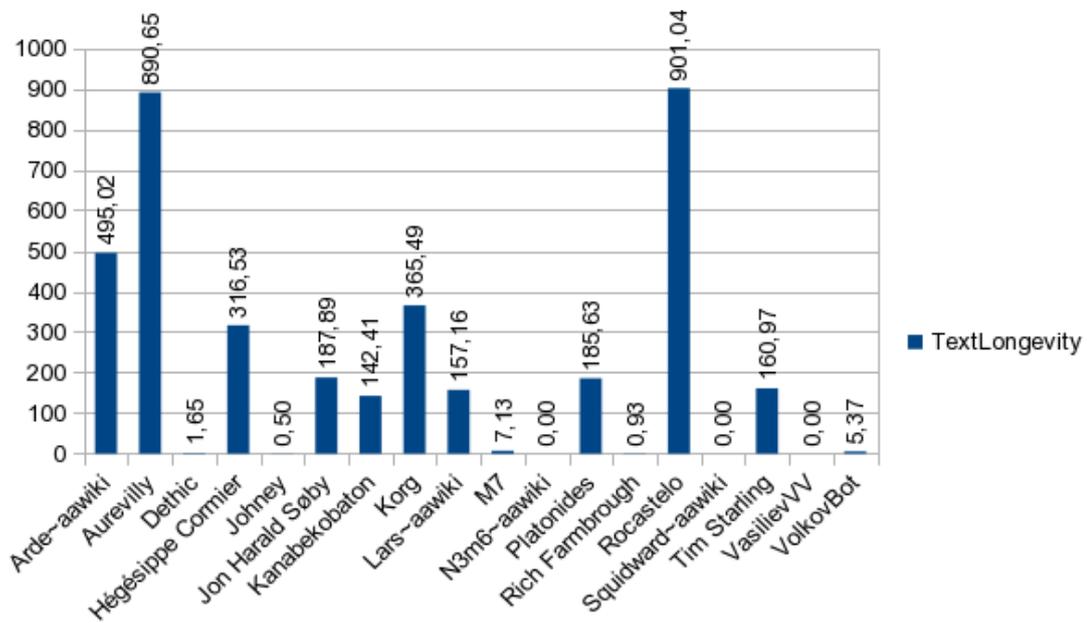}
	\caption{Framework's TextLongevity results}
	\label{fig:comparison-fw-textlongevity}
\end{figure}

\begin{figure}[htbp]
	\centering
	\includegraphics[width=\linewidth]{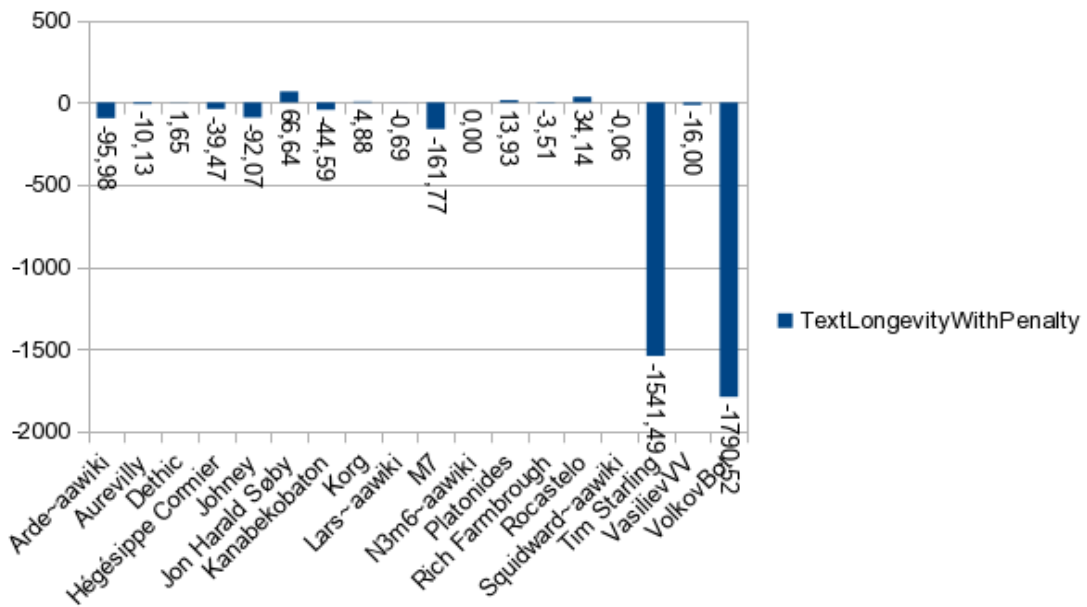}
	\caption{Framework's TextLongevityWithPenalty results}
	\label{fig:comparison-fw-textlongevitywithpenalty}
\end{figure}

\begin{figure}[htbp]
	\centering
	\includegraphics[width=\linewidth]{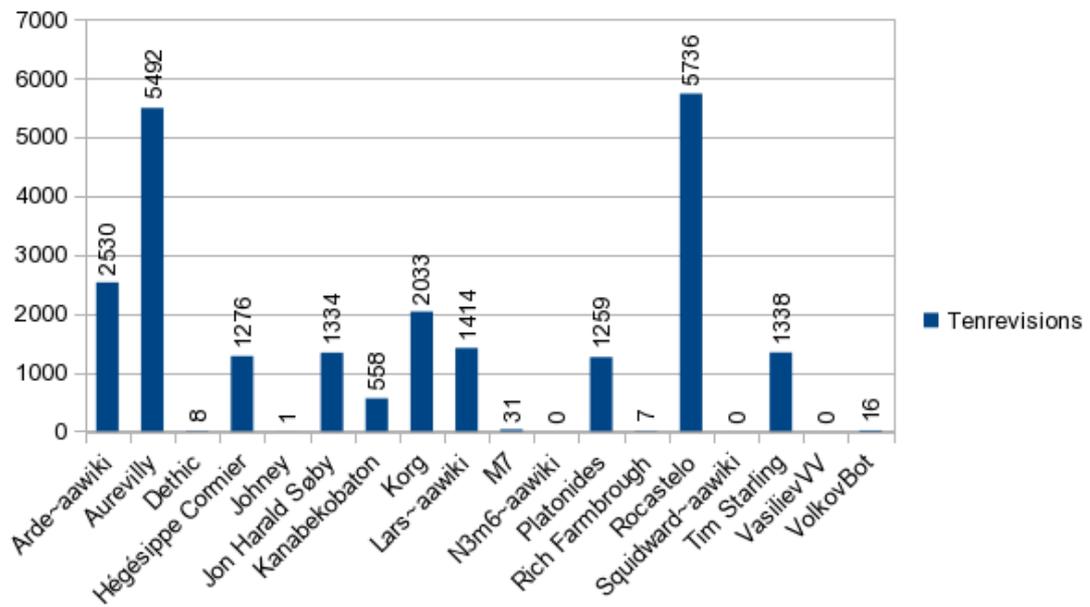}
	\caption{Framework's TenRevisions results}
	\label{fig:comparison-fw-tenrevisions}
\end{figure}

\paragraph{Sources of error}
The speed-up results from section \ref{sec:evaluatin-wikitrust} can only serve as an indicator of performance differences rather than absolute comparison values. The reason is that WikiTrust's user reputation results are not equal to any of the framework's. Figure \ref{fig:comparison-wikitr} shows the former's author ranking scores. Most contribution measures which were implemented with the framework have their results plotted in figures \ref{fig:comparison-fw-editlongevity} to \ref{fig:comparison-fw-tenrevisions}. The NumEdits, TextOnly and EditOnly are not included due to the number of returned authors, which was not within the realm of WikiTrust. As one can see, none of the contribution measures show the exact same bars as WikiTrust. Also no direct similarities or correlation can be recognized. 

Despite \citeauthor{Adler:2008:MAC:1822258.1822279} claiming that a tool was ``instrumented to calculate the various contribution measures we have defined`` \citep[p. 5]{Adler:2008:MAC:1822258.1822279}, none of the framework's results looked similar to WikiTrust's. This might be directly correlated to the long processing times, because WikiTrust's third computation step (``do\_sort\_stats``) accounted for the majority of it and none of the described contribution measures seemed to need sorting of any kind. Therefore, we gave WikiTrust the benefit of the doubt and timed its performance for the splitting and statistics computation steps (see table \ref{tab:wikitrust-perf-statistics}). Although the framework processes small dumps slower than WikiTrust, it starts to outperform its competitor when big dumps are processed. In view of the goal of processing the english Wikipedia with a huge amount of data, ignoring smaller and focusing on bigger dumps seems acceptable.

After investigating the discrepancies, reading WikiTrust's sourecode and further paper's by Adler et al, several sources of error were discovered. Overall the process of trying to successfully reproduce any of WikiTrust's results and calculations was difficult and tedious, ultimately leading to its abortion due to time constraints.

\subparagraph{Continued development}
Since the publication of \cite{Adler:2008:MAC:1822258.1822279} in \citefield{Adler:2008:MAC:1822258.1822279}{year} the WikiTrust's development continued until mid 2014, as can be seen on GitHub's commit history page\footnote{\url{https://github.com/collaborativetrust/WikiTrust/commits/master}}. The tool has plenty of releases\footnote{\url{https://github.com/collaborativetrust/WikiTrust/releases}}, but neither a detailed changelog, nor other useful information, except the paper's publication date, from which a correlation between the paper and their implementation of the contribution measures could be deducted. Most of them were referenced again in a dissertation \cite{adler2012wikitrust} in 2012, what would indicate a re-use of the previously defined algorithms. However, the dissertation explores further ranking methods which might have influenced the development of WikiTrust. 
Even more surprising was the fact that applying WikiTrust on its own test datasets in the \verb|test-data/| folder yielded slightly different results. Using the \verb|wiki1.xml| and its \verb|wiki1.stats| as a reference, the output differs in regard to the output (e.g. TextInc and TextLife is omitted) as well as in its level of detail and values (e.g. other EditInc or EditLife).
That leads to the assumption that maybe not all tools relevant to the paper \cite{Adler:2008:MAC:1822258.1822279} were relocated from the website to GitHub or were modified in other ways. This is unfortunate, because it hinders to correctly implement and reproduce their work.

\subparagraph{Base values}
An important thing to consider is error propagation. If the initial values on which the further algorithms are based differ, then those differences will propagate through all calculations and eventually lead to incorrect results. Some of such errors, like parsing the wrong amount of Page or Revision objects (see section \ref{sec:implementation-problemsandsolutions}) were identified and resolved. However, more such edge cases might exist that are yet to be discovered. We are confident that differences in the resulting user reputations might also stem from the different text tracking approach, which was discussed in section \ref{sec:implementation-problemsandsolutions-differencestoadler}. Without exact attribution throughout all revisions, an author might get rated for changes that were not originally authored by her, thus changing her and the original author's score.
Therefore, slight differences were likely, but expected to be in the same range, especially for the small test dumps. 
An unexpected discrepancy was the list of rated revisions. Reading the formulas in \cite{Adler:2008:MAC:1822258.1822279} creates the impression that the contribution measures are calculated for all authors and revisions. But WikiTrust's generated statistics file for the aawiki-20170501 contains only 68 out of 80 revisions that were rated using EditLife. From the source code in \verb|analysis/wikidata.ml| we learn that ``[this] line gives the "edit longevity" of an edit.`` Analyzing the omitted 12 revisions did not indicate any obvious reasons for their exclusion and were considered perfectly valid. The possibility of bugs within WikiTrust was regarded, due to the fact that the reduced ``statistics file contains information about every version [...]`` \citep[p. 4]{Adler:2008:MAC:1822258.1822279}.
All those discrepancies and uncertainties about the generation of the necessary basis for the author ranking could be partly responsible for the final results' mismatches.

\subparagraph{Implementation details}
Reading WikiTrust's source code uncovered some more of its implementation details. For example the earlier explained variable \verb|end_time|, which limits the amount of parsed revisions to a specific deadline. Two more variables \verb|rep_scaling| and \verb|max_rep| could be used to manipulate the generated author reputation. At first it was thought that disabling the scaling by setting it to a multiplicative neutral value of 1 and increasing the \verb|max_rep| to a sufficiently high value would only affect the final result's scores. Several tests using different combinations of those parameters showed that this was not only the case, but that also the list of rated authors had changed, which was more than unexpected. With a neutral reputation scaling setting, WikiTrust returns 15 instead of 14 authors for the aawiki-20170501. To minimize the confusion about which resulting author lists to compare, all tests were conducted with the initially identified default parameters due to this behavior. 
Furthermore, it did not become clear how exactly the claimed modularity of WikiTrust worked, making it impossible to focus and compare individual contribution measures between both programs. A second attempt to contact WikiTrust's authors to clarify some of its operational flow and configuration, was left unanswered.
We believe that such subtle implementation differences were responsible for further dispersing the results. 

\subparagraph{Result Correlation}
Knowing that the results of WikiTrust are of limited use to validate the correctness of our implementation, other sources of information were searched for. The reference paper itself includes several statistics and results of the presented contribution measures for the ``Wikipedia dump of February 6, 2007`` while considering only ``versions before October 1, 2006`` \citep[p. 4]{Adler:2008:MAC:1822258.1822279}. The used dump was not further specified, but we assume the English Wikipedia. Unfortunately, their analysis is based on a sample of 5 out of 25 million randomly selected revisions due to ``memory limitations of the software package.`` \citep[p. 5]{Adler:2008:MAC:1822258.1822279} Randomizing the sample dataset makes it impossible to reproduce the results without information about the samples, like all revision or page IDs. It also did not allow to put the framework's results in relation with those from the paper. 

\subsubsection{Optimizations}
The second part of the evaluation focused on running the framework on the student cluster and testing various optimizing strategies to further increase the performance. Some expectations were fulfilled, but others were invalidated by showing interesting results.

\paragraph{Filtering methods}
The expectation was that using PostFilters instead of PreFilters to reduce the Page DataSets' size would be significantly slower, because all raw XML pages need to be parsed first. 
The results from table \ref{tab:perf-filters} suggest that while our assumption held true for the aawiki and acewiki, where first parsing the pages and then filtering using a PostFilter is the slowest operation, it relativizes for the bigger dumps.

For the bgwiktionary, the filtering times for the RegexPreFilter and CustomPostFilter are on the same level, whereas the Xpath- and XqueryPreFilters are significantly slower. The latter observation about the PreFilters also holds for the warwiki. We try to explain it with the implementation of the Xpath- and XqueryPreFilter. Both get the full XML string of a page passed to their filter function, which internally needs to parse it before applying the filter. Therefore, a page gets parsed twice: Once in the PreFilter and a second time in the DumpParser. Neither the RegexPreFilter, nor the CustomPostFilter require preprocessing in any form, thus the page will be parsed only once and processed faster. While the difference between bgwiktionary's CustomPostFilter and RegexPreFilter are not significant, the full cost of filtering after parsing is more noticeable with larger dumps, like the warwiki, where a significant difference is measured (\textasciitilde 8.95\%). On the other hand, immediately passing the raw XML strings without prior examining it to the XML parser and checking the namespace afterwards could payoff when most pages belong to the target namespace and no or slight overhead of later removing pages occurs.  
Nevertheless, it surprised us that the CustomPostFilter outperforms the two PreFilters and is only a little slower than the RegexPreFilter. 

We conclude that the Xpath- and XqueryPreFilter should be avoided if performance is important. It proves the assumption that eliminating pages before the parsing step, leads to overall faster processing times. The question that remains is why this behavior is not observable for the two smaller dumps and why the acewiki is overall faster than the aawiki. It is probable that this can be attributed to measurement errors or the previously discussed overhead of processing relatively small dumps.
Based on this outcome, we recommend the RegexPreFilter as a namespace filter, because it provided the best overall performance, but depending on the requirements another filtering technique might perform better.

\paragraph{Filtering loops}
At the time of implementing the Pre- and PostFilters, it was believed that applying the filters in the ``outer loop`` would yield a higher performance, because in each iteration the DataSet would contain less elements to filter, but apparently returning \verb|false| immediately after the first mismatch is at least as fast. The fact that the former method needs $n$ instead of one single filtering operation was not considered back then, and might be a reason for the slightly lower speed. Therefore, we believe that the bigger the list of filters to apply, the more switching to the ``inner loop`` method should be viable and considered. 
Such a change is easy to accomplish for the PreFilters, because those are usually implemented in a dump parser. For the PostFilters, a change in the \verb|AInputParser| abstract class is necessary, which cannot be directly changed by the framework's users. However, the InputParser class extends this abstract class, so overriding the \verb|applyPostFilters| method without calling \verb|.super()| is suggested. That imposes more work on the framework's user, which is against the idea of simple extensibility. This minor design flaw will either be removed in an upcoming version or the ``inner loop`` will become the default. 
We think that the measured performance gains (\textasciitilde 2.29\%) depend on the amount of filters, but that it is nearly the same for only one filter. 

\paragraph{Kryo serialization}
After reading about the possible performance boost by registering the used classes with the Kryo serializer, our hopes were high to see another performance improvement. Unfortunately, the results from table \ref{tab:perf-kryp-serialization} do not look promising and no obvious performance gain is visible. Considering that the quoted claim of a possible speed-up comes from the documentation for Apache Flink 1.2, and an older version is used for the framework, it might only hold for the newer version, due to features or code changes that are not available in the older one. The earlier version's documentation does not explicitly hint towards a speed-up. On the other hand, a significant improvement was observed for the bgwiktionary dump. However, the acewiki's significant performance decrease should not be ignored. It could be a measurement error due to caching, and that Kryo's type-registration only creates a benefit when no caching is involved. 
Nonetheless, based on the limited and mixed results, a confident conclusion cannot be drawn yet, because we do not want to warrant a possible performance loss. We believe that with further tests on bigger dumps, a more clear result can be seen. An update of the framework to the newer version was considered, but later discarded, because it might bring unforeseen side effects and was also not available on the student cluster. 

\paragraph{Dump parsing}
We were surprised that the regular RegexSkipDumpParser outperforms the native XML parsing library for bigger dumps. During the implementation and the prior performance tests, we believed that the library would process the XML tags more efficiently. Another supportive argument for it was that for some lines, the regular expression would be matched twice causing some overhead: First within the \verb|hasTag| method to identify e.g a username XML tag, and then again in the immediately following \verb|getMatch| function to retrieve the matched content. An explanation for the notwithstanding higher performance could be based on the fact that the native library processes every XML tag and parses all information from it, for example present attributes, so that it can provide them in the function calls. Although it facilitates accessing the attributes and prevents bugs with regular expressions, it is not used very often for parsing the dumps and CPU time is lost. 
It is interesting that this observation does not hold for the two smaller dumps, because one could expect that less lines should also be processed faster. However, this is not the case and could be explained with Apache Flink's distribution and orchestration overhead or with unknown measurement errors, but apparently the performance gain starts somewhere between the size of the acewiki and bgwiktionary. 
Putting the smaller dumps aside, a performance gain of more than 1/6th is enormous. We therefore conclude and recommend to use the RegexSkipDumpParser. 

No time was left to implement and test other native Java XML parser libraries or other parsing approaches, for example native string operations like \verb|.startswith|, \verb|.endswith| instead of regular expressions.

\paragraph{Pageviews parsing}
During the implementation of this feature, it was obvious that it would induce longer processing times due to additional input and filter operations, but we did not imagine times to slow down by a factor of five. The processing times of around four minutes are within the same realm as the slowest calculations on the acewiki (see table \ref{tab:framwork-perf-cluster}). It is possible that while the page and pageview merging took place, disk caching was used, but unfortunately that was not verified during the performance tests. 
Another reason for this could be that the whole pageview dataset is read by the framework and then filtered for the correct, matching wiki project tag. This gives the freedom of reusing a preprocessed pageviews dataset for different wikis by simply passing another project tag. The other approach that was discarded, had the idea of the FileMerger-Tool already filtering the entries for the correct project tag. By accepting a slightly longer preprocessing time, the homogeneous pageview dataset would likely be drastically smaller, due to only a small percentage of the entries belonging to the given wiki and probably lead to a faster reading and merging process. However, the reason for discarding this approach was that the target project tag had to be known prior to the preprocessing. If that is not the case, then the preprocessing would need to be repeated several times, thus decreasing the overall performance gain. 
On the other hand, given the apparent slowness of the chosen approach, adjusting the FileMerger-Tool to sort, split and output the pageviews for every available project tag into a separate folder can be discussed and further researched. Another research question that we did not have the time to follow was at how much pageview data the performance starts to decrease. 
Until those question are answered, the additional processing time has to be remembered when such new measures are researched.

\subsubsection{General observations}
Throughout the development and the performance tests, several general observations could be made that need to be discussed or at least outlined for further research. 

\paragraph{Performance fluctuation}
The observed and at several points mentioned fluctuation of processing times during the evaluation can have various reasons, which we were not able to identify satisfactorily. Due to the nature of a distributed system, those factors can be quite diverse and add up. That begins with network latencies and processing/polling timeouts from managing the different task slots. Furthermore caches down from the CPU, HDFS up to Apache Flink's caching methods, can all influence the processing time. As outlined multiple times before, we chose to average over five runs to combat sporadic execution time changes. 
The decision fell on five executions, because it seemed like a acceptable trade-off between the reliability of the average value and the amount of time to obtain this value. It would be interesting to learn if further increasing the amount of runs would yield more stable and reliable average results and lead to better performance comparisons. While conducting the various (de-)compression tests in section \ref{sec:implementation-compresseddatasets}, the usual processing time fluctuation was not noticed and ignored. Those values are based on a single run, but they were usually run several times with different amount of task slots, and still showed the same tendencies. Nevertheless, we believe that re-running those tests for better reliability is necessary.  

\paragraph{Performance differences}
Most of the conducted performance tests had some controversial results, like the processing time of bigger dump being faster than a smaller one. Although we tried to attribute those occurrences to the amount of parallelism compared to the dump's size, no real evidence for this claim was obtained. This often left us with results that were hard to interpret or to draw a conclusion from. 
A further point that needs to be highlighted is that performance comparisons with the aawiki or acewiki usually lead to processing time discrepancies that were either impossible to find due to the aforementioned problems, or too small to reliably identify. Applying those tests to the bigger dumps like bgwiktionary or warwiki improved that a bit and made performance differences more obvious and measurable. 
We suppose that given the framework's purpose of processing large wikis in a distributed manner, the fluctuations and measurement problems with the small wikis should be overlooked. Instead, the focus needs to shift on the bigger dumps, where gaining performance improvements is desirable. The smaller dumps were continuously tested for consistency reasons.
 
\paragraph{Scalability}
With regards to the previous section, we tried to run the framework and its contribution measures on bigger dumps, like warwiki-20170501 (456 MB bzip2 and \textasciitilde 10 GB decompressed) or eowiki-20170501 (1.5 GB bzip2 and \textasciitilde 30 GB decompressed). In section ``\nameref{sec:implementation}`` some problems (see \ref{sec:implementation-problemsandsolutions-bzip2}, \ref{sec:implementation-problemsandsolutions-clusterconfig}) with bigger dumps were highlighted. At first, we thought that scaling the system from a five year old development system, to a virtual server and then to a student cluster would sufficiently increase the processing power to prevent those issues. It did help with the processing times and allowed to process some bigger dumps, but after unsuccessfully trying to run the TextOnlyWithPenaltyCalculation on the warwiki in the evaluation phase, we have to admit the thoughts of possible design flaws. Although the parsing step seems to run without any exceptions, at the end the job was canceled due to errors related to Apache Flink's sort buffer when flatMapping the revisions into the DataHolder's Revision object DataSet: 
\begin{itemize}
	\item \verb|Cannot write record to fresh sort buffer. Record too large.|
	\item \verb|'SortMerger Reading Thread' terminated due to an exception: null|
	\item \verb|'SortMerger Reading Thread' terminated due to an exception: Error at remote task manager|
\end{itemize}
Repeated executions resulted in different error messages, of which some were discussed online  \footnote{\url{http://apache-flink-user-mailing-list-archive.2336050.n4.nabble.com/FlinkML-ALS-matrix-factorization-java-io-IOException-Thread-SortMerger-Reading-Thread-terminated-duel-td8809.html}}\footnote{\url{http://apache-flink-user-mailing-list-archive.2336050.n4.nabble.com/Cannot-write-record-to-fresh-sort-buffer-Record-too-large-td13644.html}}\footnote{\url{https://issues.apache.org/jira/browse/FLINK-1085}}, but applying the suggested changes did not help. Although increasing \verb|taskmanager.memory.fraction| to 0.8 and decreasing \verb|taskmanager.numberOfTaskSlots| to 8 (totaling 40 task slots) lead to more data being processed, ultimately the job failures did not stop to occur. So in essence the framework fails to properly calculate author ranking for dumps surpassing a specific size. There are three major points that are considered limiting factors:
\begin{enumerate}
	\item Data representation
	\item Calculation processing
	\item Dump parsing
\end{enumerate}
\subparagraph{Data representation}
The initial idea on how to represent the wiki dump data within the framework was described in section \ref{sec:implementation-planofimplementation-datarepresentation}. The main idea was to organize the data around a central DataHolder, and then access this data for further processing. That method forces the preservation of all objects that are in the DataHolder, what means that the full content of all revisions is stored somewhere within the Apache Flink Job, until it is processed by a Differ and Calculation class. The revisions are then organized as a double linked list within an ArrayList bound to a Page object. Without knowing the detailed internals about Apache Flink's serialization, one can assume that certain overhead is necessary to maintain all those relationships. The decision to preserve as much information in the basic objects was made as a compromise to allow arbitrary calculations on the data, without e.g. forcing a specific calculation type. 
If it is clear that only revisions are used to calculate author rankings, then the framework could be modified to calculate the revision differences while parsing a page and only storing the resulting RevisionRelevanceScores in the DataHolder. This would drastically reduce the amount of data within the DataHolder and eliminate some processing steps. Due to the fact that less objects flow through the framework, we assume that its resource footprint should become smaller and larger dumps more reliably processable.

\subparagraph{Calculation processing}
Similar to the previous subparagraph, the framework was designed to parallelize the computations for every revision in favor of only parallelizing the processing of pages. The mathematical definition of \citeauthor{Adler:2008:MAC:1822258.1822279} contribution measures would allow to compute ``local`` author rankings based on the revisions of a single page and later summing up all local rankings into a final one for the whole wiki. That method would have also had the advantage of allowing to recursively process a page's revisions before calculating a score, so that the sophisticated authorship tracking could be applied. There would be no need for a DataSet of Contributors or Revisions in the DataHolder, further reducing the amount of information. Nonetheless, we believed that it could lead to single nodes processing huge pages individually, thus reducing the overall performance. By processing DataSets of Revision objects, the decision of distributing the work equally on all nodes would be given to Apache Flink.

Both discussed points are regarded to be part of the processing problem. Nevertheless, we are still confident that initially developing an open framework was the right decision. The implemented openness and extensibility should make the realization of the aforementioned, narrower modifications possible, so that processing big dumps becomes achievable. Furthermore, the suggestion from section \ref{sec:implementation-problemsandsolutions-memoryissues} to split up a big wiki into smaller chunks of pages and executing the framework on them is already possible with the current implementation.

\subparagraph{Dump Parsing}
Further analysis and efforts to identify the root cause narrowed it down to the initial splitting of the raw XML data into \verb|<page>|-separated XML strings. For this test, the only operation was counting all elements in the DataHolder's Page DataSet after parsing and filtering. All jobs on bgwiki, eowiki, arwiki failed with similar errors that were observed before. Removing the \verb|.distinct()| operation merely allowed the eowiki job to finish successfully, but it emitted less data (more items, but lower total size) than the other dumps. We therefore believe that the framework's problems could also be resolved by implementing a proper XmlInputFormat or extending the DelimitedInputFormat with (de)compression support.


\begin{myboxi}[Key points of section ``\nameref{sec:discussion-performancerankingresults}``]
	\begin{itemize}
		\item A credible comparison of WikiTrust's and the framework's performance is not possible, because of discrepancies in the user ranking results. Despite that, the distributed framework performs better. 
		\item An increase in the dump's size does not necessarily imply longer processing times.
		\item The tested optimizations can make the reference framework about 20\% faster, but the constant fluctuation of the processing time made exact conclusions hard.
		\item Applying the framework on large dumps resulted in unexpected job failures and exceptions that we were not able to resolve despite intensive debugging, but suggested several changes that might fix it.
	\end{itemize}
\end{myboxi}

\subsection{Framework use cases} \label{sec:discussion-furtherusecases}
The following section will cover a brief discussion of the developed framework's use cases. 
\subsubsection{Usability for Author ranking} \label{sec:discussion-usabilityauthorranking}

The thesis' problem statement was to explore how and if a distributed framework could process a huge amount of edits efficiently, while at the same time facilitating the development of new author ranking methods. Both aspects were considered while planning and implementing the framework (see section \ref{sec:implementation-planofimplementation}), but stronger focus was put on the functionality and extensibility to create the basis for future work on that topic.
 
Its usability and functionality to calculate author rankings was demonstrated during the evaluation (see table \ref{sec:evaluation}). Although the exact results from the WikiTrust system could not be exactly reproduced, due to technical and human limitations, the contribution measures defined in \cite{Adler:2008:MAC:1822258.1822279} were successfully implemented. On the downside, we have seen and previously discussed problems with applying the framework on dumps that contain several million pages, so that its usability for those bigger Wikipedia pages is limited. However, we provided several ideas on how our reference implementation could be adjusted to remedy those problems. 
On the other hand, we are confident that the following features make it a viable option for author rankings:
\begin{itemize}
	\item \verb|Extensibility|: It allows to easily replace parts of the framework with custom implementations by heavy use of interfaces and factories, thus making it flexible in its functionality and feature set.
	\item \verb|Abstraction|: It abstracts from the Wikipedia and Pageview dump parsing, allowing to concentrate on implementing impact measures.
	\item \verb|Platform independence|: It is written in Java, which provides platform independence as long as the Java Virtual Machine (JVM) runs on it. Furthermore, Java is taught in introductory Computer Science classes - at least at the TU Berlin, so its paradigms are well-known and well-studied. 
	\item \verb|Scalability|: It is implemented using Apache Flink, which allows to scale horizontally from one system onto several without the need of adjusting the implementation.
	\item \verb|Open Source|: Its source code \cite{fwrepo} is public and can be modified by the community or any individual.
\end{itemize}

\subsubsection{Usability for Wikipedia} \label{sec:discussion-usabilitywikipedia}
At least three use cases can be imagined that could benefit the self-organizing and open collaboration style of Wikipedia.
\paragraph{Vandalism detection}
Wikipedia can only function properly as long as all editors contribute edits in good faith and intention. Sadly, so called vandals try to sabotage the work by introducing mal-intended edits, such as inserting inappropriate or deleting the whole content. It is in the communities interest to identify and if possible, prevent harmful activity. 
There are several projects and papers about this topic. For example, the Objective Revision Evaluation Service (ORES) \cite{mwores} applies machine learning techniques to a wiki's edit history. Once activated for a specific wiki, it can help analyzing and classifying edits, so that responsible editors can faster identify and appropriately act to resolve the issue.  However, the table of supported wikis \cite{mwores} is relatively small compared to the list of downloadable wikis\footnote{\url{https://dumps.wikimedia.org/backup-index-bydb.html}}.
The authors of \citep[p. 5]{Adler:2008:MAC:1822258.1822279} used contribution measures to identify non-human editors (bots) and vandals, when added text was immediately removed in the next revision. 
With regards to the aforementioned methods, we believe that the latter one can directly be implemented with the framework, because it re-implements the same contribution measures. For the former point, the framework also takes the edit history as an input. We think that incorporating machine learning algorithms is possible, therefore supporting the ORES system in its work or providing vandalism detection for other wikis.
\paragraph{Adminship}
The community members that actively ``delete copyright violations, protect frequently vandalized pages, block malicious users, move pages when there are name conflicts, exclude bulk vandalism from the recent changes list [...]`` \citep[p. 3441]{Burke:2008:TUM:1358628.1358871} are administrators with additional privileges. To become an administrator, a user has to submit a Request for adminship \footnote{\url{https://en.wikipedia.org/wiki/Wikipedia:Requests_for_adminship}} (RfA) to the community. The candidate's experience as well as her trustworthiness are then discussed by it. One argument to avoid is Editcountitis\footnote{\url{https://en.wikipedia.org/wiki/Wikipedia:Arguments_to_avoid_in_adminship_discussions\#Editcountitis}}, what describes that the number of edits does not allow conclusions on the user's ability to administer the Wikipedia well, and the edits' content should be used as an indicator. This sentiment towards the quality of an edit, however, is what \citeauthor{Adler:2008:MAC:1822258.1822279} try to capture with their contribution measures. 
We therefore believe that implementing the framework with its contribution measures that focus on the edit quality might pose a helping hand and facilitate the decision. It could also serve as a self-evaluation service for checking your own score, before submitting a RfA. 
\paragraph{Gamification}
The word gamification can be defined as ``a process of enhancing a service with affordances for gameful experiences in order to support user's overall value creation.`` \cite[p. 19]{Huotari:2012:DGS:2393132.2393137} Such gameful experiences can be understood as rewards for (repeating) actions by the user, so that she creates additional value. Similar rewards have already been given away by the Wikipedia community in the form of Barnstars\footnote{\url{https://en.wikipedia.org/wiki/Wikipedia:Barnstars}} for hard work and other renowned achievements. 
We suggest that the framework could be used to periodically calculate an author ranking for all wikis and then displaying the various rankings on a user's page. Depending on the frequency of those calculations, a gamified environment could be created, where Wikipedia editors become eager to increase their rank, or fine-tune their individual contribution scores. It is believed by us that introducing gamification in such a form to Wikipedia could benefit it due to the increased engagement and activity. 

\subsubsection{Other use cases}
Two further use cases that stem from the framework's extensibility are calculating rankings for Revision or Page objects.
\paragraph{Revision rating}
The RevisionRelevanceScore and the Differ could be used to calculate scores for single Revision objects in the context of labeling them for further analysis, e.g. for the recent changes page of an article similar to ORES. The calculations could be based on previous revisions or the authors who contributed text. 
\paragraph{Page rating}
With the PageRelevanceScore it is possible to rate Page objects. Calculating relevance scores for pages might lead to building scoring systems that do not focus on authors, but on whole wikis. Such an approach of comparing wikis based on their page quality or page quantity measure might shift some focus on wikis that do not have a big community yet and would be happy to receive some contributions. 

Those are just two further ideas that the developed framework should be capable of handling without extensive modifications, but were not examined in this thesis.

\begin{myboxi}[Key points of section ``\nameref{sec:discussion-furtherusecases}``]
	\begin{itemize}
		\item The framework's open design makes modifications and different use cases possible, which are not limited to author rankings, but general rankings based on the parsed data objects.
		\item It might benefit Wikipedia for vandalism detection, moderator or admin elections, or to increase the intrinsic motivation with methods of gamification.  
	\end{itemize}
\end{myboxi}

\subsection{Future work} \label{sec:discussion-futurework}
As the last part of this section, we want to outline future work that the framework could benefit from. For example during the implementation and evaluation certain bugs, design issues and places for improvement were identified. Additionally, some further research facets about author contribution and Wikipedia rankings were introduced, but cannot be researched due to this work's scope and time constraints.
Facilitating the usability and future work with the framework was one of the core concerns, so documenting the framework's classes well was a part of it. The comments and the detailed description in section \ref{sec:implementation} should give enough insight to extend and improve the framework's functionality over time. Throughout the thesis, different improvable aspects came up.
\paragraph{Parsing improvements}
The XML parsing problems with the custom XmlInputFormat implementation for handling compressed datasets were remedied by a call to \verb|.distinct| (see \ref{sec:implementation-problemsandsolutions}). A flawless implementation of a transparently decompressing XmlInputFormat class, even better natively for Apache Flink, would not only benefit the framework and its performance, but possible other projects that process Wikipedia dumps. Analyzing and implementing other splittable compression formats with a higher compression ratio than bzip2 is also an area of interest.
Earlier in this section, the surprising performance boost by using the RegexSkipDumpParser were debated and other ideas were presented. An implementation of those and other parsing methods might further increase the framework's performance. Right now, the framework only parses the minimal necessary information to calculate \citeauthor{Adler:2008:MAC:1822258.1822279} contribution measures, but more is available. Therefore, adding new fields to the IPage, IRevision and IContributor interfaces and adjusting the parsers to populate them could be useful for new contribution measures.
We think that further performance gains could be achieved if the framework's objects did not have private fields or methods. The goal is them to be classified as simple POJOs by Apache Flink and then no fall back to the apparently slower Kryo serializer is needed. 
\paragraph{Pageviews preprocessing}
In the previous subsection we discussed the huge impact of pageview parsing on the overall performance, and formed some ideas how that could be reduced. It would require a rewrite of the FileMerger and a re-evaluation of the needed preprocessing time versus the final framework time as well as a discussion if the reduced usability is acceptable. Given a mandatory preprocessing step, it could decode the page titles simultaneously to reduce the framework's workload a bit more. The reduce step of summing up the page count and traffic values for all entries could also be moved from the framework to the FileMerger or a dedicated pageviews preprocessor, but then a discussion about splitting the framework into different programs, like WikiTrust's approach, is necessary.  
\paragraph{Text tracking}
One of the reasons for the discrepancies between the results from WikiTrust and the framework was the simpler text tracking algorithm. Section \ref{sec:implementation-problemsandsolutions-differencestoadler} discusses the exact details and technical limitations. On the downside, it does not provide as exact text tracking results and is also vulnerable to malicious authors, who try to game it. Further analyzing and understanding, adjusting and implementing the original algorithm in a distributed manner is considered future work and would most likely result in more reliable and accurate contribution measure results.
\paragraph{Page mapping}
Due to the job failures that were observed in the evaluation, we suggested to refrain from mapping on the Revision object level, and just map on the Page object level. As discussed earlier in this chapter, combining this with the previously mentioned text tracking issue could be an option to research.
\paragraph{Rankings}
Despite future work on the framework's implementation, another topic to research and explore is designing new contribution measures with the help of the framework. For example, by incorporating external information like the pagecounts or other attributes that can be obtained by a Page or Revision object. Analyzing where differences to WikiTrust exist and how those can be removed while maintaining the distributable programming style, is an important task to ensure the ranking's correctness. 
\citeauthor{DeAlfaro:2011:RSO:1978542.1978560} their dissertation analyses and discusses in detail how to design impact measures.
Applying those to different wikis and making the information available to the public periodically.

\begin{myboxi}[Key points of section ``\nameref{sec:discussion-futurework}``]
	\begin{itemize}
		\item Improving the parsing or preprocessing as well as more precise text tracking techniques are considered future work.
		\item Further development of contribution measures with the framework is another important topic.
	\end{itemize}
\end{myboxi}





\section{Conclusion} \label{sec:conclusions}
The previous discussion outlined the strengths and potential places of improvement in the thesis' work. We succeeded in implementing a framework to calculate impact measures for Wikipedia authors in a distributed manner that abstracts from the initial parsing and preparation of the edit history and pageviews input data. The flexible and extensible design allows for the realization of various impact measures that might improve or benefit Wikipedia in topics such as vandalism detection, moderator or admin elections, or increasing the intrinsic motivation with methods of gamification.

Prior to the implementation, we analyzed the input datasets and concluded that bzip2 is the preferred compression algorithm for the edit history in regards to saved disk space and processing time. Preprocessing the pageviews further increases the performance.

Throughout the thesis, we tried to adhere to the important aspect of reproducibility that we believe to have achieved by thoroughly documenting the framework and its source code, the used soft- and hardware as well as the evaluation in as much detail as possible. Therefore, providing code snapshots, important datasets and implementing automated tests seemed natural and is believed to facilitate future work on this topic.

\citeauthor{Adler:2008:MAC:1822258.1822279} their work on the contribution measures and the WikiTrust program was chosen as a reference system for our framework. A re-implementation of their formulas allowed to demonstrate its usability and possible performance gains from distributing the work on multiple systems. Unfortunately, we were not able to exactly reproduce WikiTrust's results nor \citeauthor{Adler:2008:MAC:1822258.1822279} their observations in the evaluation phase due to technical limitations and missing information on their and our side. 

Experiments with several code optimizations indicated that the performance of our framework's reference implementation can be improved by up to 20\%. Although it proved the extensibility and interchangeability of its components, the goal of calculating author rankings for the huge English Wikipedia was not accomplished. However, we also showed that incorporating pageviews information paves the way for more diverse and interesting impact measures.

We are confident that our work is a first step and lays the groundwork for distributed author rank calculations based on impact measures for the Wikipedia. By open sourcing the framework and all related information, we hope to facilitate and involve the Wikipedia community in further development of the framework and research on this topic.

\section{References}
\begingroup 
\renewcommand{\section}[2]{}
\printbibliography[title={}]
\endgroup

\section{Appendix} \label{sec:appendix}
Despite the list of tables and figures, this section covers the distribution and organization of the thesis' tools, source code and the steps necessary to build and run those programs. Figures with additional information that can improve the reader's understanding of the results are part of it. 

\subsection{Source code on CD and GitHub} \label{sec:appendix-sourcecode}
The developed tools, their output and performance data, as well as some important datasets will be published alongside the thesis. To allow for further development, not only by the author, but also the community or otherwise interested people, the data will be published in a repository on the collaborative development platform GitHub\footnote{\url{https://github.com}} at \cite{fwrepo}. A read-only copy will be provided to the thesis' supervisors for correction on a CD-ROM.
\subsubsection{Structure}
The repository consists of two folders. One named \verb|code| with the programs that were mentioned and developed throughout the thesis, whereas the \verb|data| folder contains all data, such as input or output files of those programs to facilitate the reconstruction of the evaluation measurements and calculations.

Contents of the \verb|code| folder:
\begin{itemize}
	\item \verb|BzipTest|: A tool for reading compressed files and used for the evaluation of the (de-)compression performance.
	\item \verb|FileMerger|: A tool that can be used to merge multiple input files into a single DataSink (output) with optional compression.
	\item \verb|Framework|: A framework for implementing and running contribution measures for author ranking in a distributed manner. 
	\item \verb|Test-Tools|: A collection of Python tools to parse and analyze small dumps without multi-threading.
	\item \verb|WikiTrust|: \citeauthor{Adler:2008:MAC:1822258.1822279} their WikiTrust program.
	\item \verb|Wikitrust-framework-setup.txt|: Rough instructions on how to setup WikiTrust, Apache Flink and the framework.
\end{itemize}

Contents of the \verb|data| folder:
\begin{itemize}
	\item \verb|compression-comparison|: All data and results belonging to the compressed datasets and their performance analysis.
	\item \verb|crash-logs|: Stacktraces from failed jobs.
	\item \verb|dumps|: Contains the most important Wikipedia dumps, like aawiki-20170501, acewiki-20170501, bgwiktionary-20170501 in their bzip2 and 7z format. 
	\item \verb|parsing-tests|: Results from analyzing the lists of parsed revisions and pages from WikiTrust and the framework.
	\item \verb|performance-comparison|: Various performance measurements and timings can be found in this folder.
	\item \verb|result-comparison|: Contains WikiTrust's and the framework's calculation results for comparison with each other. 
\end{itemize}
Some folders contain subfolders for better categorization, but almost all have a \verb|README.md| file with further information.

\subsubsection{License}
The GNU General Public License v3 (GPL-3) was chosen as the license for the source code developed during the thesis. It provides freedom in using the code commercially, modifying or distributing it, while at the same time requiring derivative work to be disclosed with a statement of what significant changes were made. We believe that this decision helps to maintain the framework's openness and potentially building a community around it, so that new and better features, such as advanced parsing methods or contribution measures, can be developed by independent individuals. 

\subsection{Java, Flink, WikiTrust and framework installation}\label{sec:appendix-javaflinkfw}\label{sec:appendix-buildingwikitrust}
In this section the necessary steps for a partially successful build of \citeauthor{Adler:2008:MAC:1822258.1822279} their WikiTrust program are described. The operating system of the DigitalOcean server was the Linux based Debian 8\footnote{\url{https://www.debian.org/}} with up-to-date packages. Listing \ref{code:appendix-buildwikitrust} is a listing of all necessary commands.
\begin{figure}[htbp]
	\begin{lstlisting}[language=Bash, caption={It shows the command line steps necessary to build WikiTrust and its dependencies.}, label={code:appendix-buildwikitrust}]
	$> sudo su
	$> apt-get install ocaml git opam pkg-config zlib1g-dev libmysqlclient-dev libpcre3-dev libjson-c-dev camlp4-extra p7zip-full
	$> ocaml -version
	The OCaml toplevel, version 4.01.0
	$> opam --version
	1.2.0
	$> opam init
	$> eval `opam config env`
	$> opam install pcre extlib ocamlfind json-static json-wheel mysql ocamlnet sexplib type_conv xml-light camlzip
	
	$> git clone https://github.com/collaborativetrust/OcamlLdaLibs.git
	$> cd OcamlLdaLibs
	$> make all
	$> cd ../
	$> git clone https://github.com/collaborativetrust/WikiTrust.git
	$> cd WikiTrust
	$> make all
	$> # Copy wikitrust-patch.diff from the repository.
	$> git apply wikitrust-patch.diff
	$> make all
	\end{lstlisting}
\end{figure}
Debian's repository already provides OCaml, Opam and most of the depencies as packages, which can be installed (line 2) after acquiring higher privileges (line 1). The exact package names and versions might differ in the future or on other operating systems. The following four lines show the installed OCaml and Opam versions that were used throughout this thesis.
After its installation, Opam needs to be initialized and configured for the current terminal session, before  it can install all necessary OCaml libraries (lines 7-9).
Once everything is set up, the WikiTrust source code can be obtained with git and then compiled using the "make" command like showed with lines 11-17. The build of WikiTrust might fail with an error similar to \verb|Error: Unbound module Http_client|. Nevertheless, it has already built the required tools in the \verb|analysis/| folder, so we ignored the further complications. 

A patch file from our repository (\verb|wikitrust-patch.diff| in the \verb|code/WikiTrust| folder) should be applied to change the number of edit judges to 10 rather than 6 and to set the revision parsing end time into the future. It also removes WikiTrust's last processing step of computing the text trust.

Installing Apache Flink and our framework on a single system is just a couple of steps. A recent version of Java is required, preferably Java 1.8 from Oracle, as well as Apache Maven for building the framework. Both can be installed using the operation system's package manager once the correct sources are configured (listing \ref{code:appendix-flink}, lines 2-6). To setup Apache Flink an archive must be downloaded and extracted from \cite{flkdlc}. It is crucial to choose a version with at least Hadoop 2.7 if compressed datasets should be supported. Otherwise the decompression might not work.
Afterwards, Flink can be started or stopped locally with the appropriate \verb|start-local.sh| or \verb|stop-local.sh| executables in its \verb|bin/| folder. With all dependencies met, the framework's repository can be cloned from \cite{fwrepo}. Executing \verb|mvn clean package| inside the \verb|Framework/| folder will build it.
\begin{figure}[htbp]
	\begin{lstlisting}[language=Bash, caption={Java and Flink installation instructions on a Debian 8 system.}, label={code:appendix-flink}]
$> sudo su
$> echo "deb http://ppa.launchpad.net/webupd8team/java/ubuntu xenial main" | tee /etc/apt/sources.list.d/webupd8team-java.list
$> echo "deb-src http://ppa.launchpad.net/webupd8team/java/ubuntu xenial main" | tee -a /etc/apt/sources.list.d/webupd8team-java.list
$> apt-key adv --keyserver hkp://keyserver.ubuntu.com:80 --recv-keys EEA14886
$> apt-get update
$> apt-get install oracle-java8-set-default maven

$>  java -version
java version "1.8.0_131"
Java(TM) SE Runtime Environment (build 1.8.0_131-b11)
Java HotSpot(TM) 64-Bit Server VM (build 25.131-b11, mixed mode)

$> wget http://mirror.23media.de/apache/flink/flink-1.0.3/flink-1.0.3-bin-hadoop27-scala_2.10.tgz
$> tar xfv flink*.tgz

$> git clone https://github.com/gehaxelt/thesis-imwa
$> cd thesis-imwa/code/Framework
$> mvn clean package
	\end{lstlisting}
\end{figure}

\subsection{WikiTrust and framework bash-loop}\label{sec:appendix-wikitrloop} \label{sec:appendix-frameworkloop}
On the DigitalOcean virtual machine a screen\footnote{\url{https://www.gnu.org/software/screen/}} session was initiated to conveniently reconnect to it in case the connection terminates. A variable \verb|DUMP| holds the first part of the dump to analyze. A for-loop iterates five times and executes the WikiTrust program while saving the execution time and the resulting user reputation scores.

\begin{figure}[htbp]
	\begin{lstlisting}[language=Bash, caption={Screen session and bash for-loop to execute and time WikiTrust multiple times on different dumps.}, label={code:appendix-wikitrloop}]
	$> pwd
	/root/
	$> screen -S wikitrust
	$> export DUMP="aawiki"
	$> for i in $(seq 1 1 5); do
	cd /root/WikiTrust/
	mkdir "/root/Output/$DUMP-$i";
	echo "=======================  $i  ===============" >> "/root/Output/perf-$DUMP";
	/usr/bin/time -o "/root/Output/perf-$DUMP" --append python2 /root/WikiTrust/util/batch_process.py --cmd_dir /root/WikiTrust/analysis --dir "/root/Output/$DUMP-$i" "/root/Dumps/$DUMP-20170501-pages-meta-history.xml.7z";
	cat "/root/Output/$DUMP-$i/user_reputations.txt" > "/root/Output/reputation-$DUMP-$i";
	rm -rf "/root/Output/$DUMP-$i";
	cd -;
	sleep 5s;
	done
	\end{lstlisting}
\end{figure}

Similar to the \nameref{sec:appendix-wikitrloop}, another for-loop in Bash helped to execute the framework on different dumps multiple times. Listing \ref{code:appendix-frameworkloop} presents two nested for-loops, where the outer loop iterates over the dumps defined in the \verb|DUMPS| variable. The inner loop iterates over a sequence of integers to process each dump five times. The \verb|CALC| variable helps to assign the recorded performance to the executed contribution measure.

\begin{figure}[htbp]
	\begin{lstlisting}[language=Bash, caption={Automated, nested bash for-loop to execute the framework five times on each of several dumps.}, label={code:appendix-frameworkloop}]
$> export CALC="numedits"
$> export DUMPS="aawiki acewiki bgwiktionary"
$> for dump in $DUMPS; do
for i in $(seq 1 1 5); do
cd /root/flink-1.0.3
echo "=======================  $i  ===============" >> "/root/Output/perf-$dump-$CALC";
/usr/bin/time -o "/root/Output/perf-$dump-$CALC" --append ./bin/flink run /root/code/Framework/target/Framework-1.0-SNAPSHOT.jar "file:///root/Dumps/$dump-20170501-pages-meta-history.xml.bz2";
cd -;
sleep 5s;
done;
done;
	\end{lstlisting}
\end{figure}

\subsection{WikiTrust and framework author reputations}
The section contains figures with all results for the aawiki-20170501 dump which were computed with the WikiTrust system and the framework with different contribution measures. Users with a resulting score of zero and the anonymous user were removed before plotting. For better comparability, all figures have a list of all present authors on the x-axis. Therefore, ones without a rating (value 0) were not part of the original result. 


\clearpage
\subsection{List of figures, tables and listings}
\renewcommand\listfigurename{Figures}
\listoffigures
\renewcommand\listtablename{Tables}
\listoftables
\lstlistoflistings

\end{document}